# Recent progress in configuration–interaction shell model

Menglan Liu and Cenxi Yuan*

*Sino-French Institute of Nuclear Engineering and Technology,*
*Sun Yat-Sen University, Zhuhai, Guangdong 519082, P. R. China*
**yuancx@mail.sysu.edu.cn*

Since Mayer and Jensen employed the single-particle shell model to interpret the magic numbers, various microscopic nuclear models have been developed to study the nuclear force and structure. The configuration–interaction shell model (CISM), performed in truncated model space with the inclusion of the residual interaction, is one widely-used nuclear structure model. In the last decade, CISM has progressed in investigating the cross-shell excitation in exotic light nuclei, the similarity and difference in mirror nuclei, and the isomerism and seniority conservation in medium and heavy nuclei. Additionally, researchers have attempted to construct effective Hamiltonians for nuclei near $^{132}$Sn and $^{208}$Pb through a unified way in the CISM framework. In parallel, related models, including the nucleon-pair approximation (NPA) approach, the Monte Carlo shell model (MCSM), the projected shell model (PSM), the Gamow shell model (GSM), etc., have also been extensively developed and validated in the last decade. This paper reviews the recent progress in CISM and some related models.

*Keywords*: Configuration–interaction shell model; nuclear structure; effective Hamiltonian.

PACS Number(s): 21.60.Cs, 21.10. -k, 21.30.-x

## 1. Introduction

The nuclear shell model, assuming that nucleons move independently in a mean field with a strong spin-orbit force, was successfully used to explain the appearance of magic numbers.[1,2] According to the independent-particle model (IPM), nucleons fill the single-particle orbits from low to high energy.[3] Doubly magic nuclei were found to be spherical and more stable than their neighbors due to the complete filling of the shells. However, the spherical mean field described in the IPM cannot represent the complete contribution of the nuclear potential, which plays a fundamental role in studying atomic nuclei.

*Corresponding author.

As early as the 1950s, evidence showed that not all nuclei were spherical.[4] The non-negligible residual interaction among nucleons, even for those close to doubly magic nuclei, highlighted the need for considering configuration mixing beyond single-particle configurations.[5,6] Therefore, the configuration–interaction shell model (CISM) was proposed, considering the residual interaction and mixing single-particle configurations. Generally speaking, CISM has succeeded in describing nuclei at the $\beta$-stability line and near the driplines. While more and more deviations between IPM and observations have been noticed during the development of nuclear experimental facilities, CISM has been further developed and employed to interpret the exotic phenomena, including intruder states,[7,8] halo structure,[9–12] mirror differences[13–15] and new magic number.[16–18]

In a typical CISM calculation, one needs to select a core (typically doubly magic nuclei), choose a model space, construct the effective Hamiltonian (also named effective interaction) and solve the many-body Schrodinger equation through the diagonalization process with a code. Including the residual interaction or constructing the effective Hamiltonian in the model space is a key issue for CISM applications. The effective Hamiltonian is constructed from the nucleon-nucleon interaction (nuclear force). In one case, the nuclear force is deduced from the nucleon-nucleon scattering experimental data,[19] i.e., the realistic nuclear force, resulting in realistic effective Hamiltonians. In another case, the phenomenological effective Hamiltonian is based on the phenomenological nuclear force derived from the observed nuclear structure data.[20] The former one starts in a more *ab initio* way. But to apply the realistic nuclear force, one needs a proper theoretical approach to deal with the hard-core and in-medium effects. On the other side, the phenomenological nuclear force depends significantly on experimental data. Therefore, an accurate phenomenological nuclear force should be fitted from sufficient structure data in the target region.

In 1965, Cohen and Kurath constructed a phenomenological effective Hamiltonian for the $1p$ shell.[21] In 1968, Kuo and Brown built a realistic one for the $pf$ shell.[22] In 1988, the semi-empirical USD family was proposed for the $sd$ shell.[23] Today there are hundreds of phenomenological effective Hamiltonians, many of which are very successful in particular model spaces. While the neutron dripline for F and Ne was not experimentally determined until 2019,[24] the CISM using the effective interaction YSOX reproduced the dripline of C, N and O simultaneously.[25] Furthermore, CISM nicely described the properties of nuclei within the island of inversion[26,27] and well interpreted the anomaly-huge mirror differences in the $sd$ region.[15] In the medium and heavy mass region, the shell evolution reflected by isomeric states was studied with the configuration mixing,[28] which required substantial computation resources. Nonetheless, one has to choose different effective Hamiltonians, with different origins, for different nuclear regions. It is of great significance to construct effective Hamiltonians for several regions from one basis.

It should be noted that from IPM to CISM, the consumption of computation resources increases significantly. Roughly, the dimension of the Hamiltonian matrix increases exponentially with the mass number when approaching the middle of the shell. Thus, only a few low-lying states can be calculated with CISM in nuclei near doubly magic nuclei (with few valence particles considered in compact model space). In the heavy mass region, one unique major shell was already so large that the calculation of mid-shell nuclei was over ability. Fortunately, supercomputer capabilities have been rapidly developed in recent decades. On the other hand, the diagonalization techniques have been largely improved. Shell model codes such as MSHELL[29] and KSHELL[30,31] for massive parallel computation have been proposed. Nowadays, CISM can be systematically applied to nuclei around $^{208}$Pb.[28,32–34] The achievable M-scheme dimension has been improved to the magnitude of $10^{11}$, while some nuclei have a gigantic Hamiltonian matrix, reaching a dimension of $10^{50}$. Overall, the number of observed nuclides is far from prediction.[35] CISM carries the burden of describing unknown nuclides.

Related shell models have been adopted for a smaller computation dimension or a more accurate understanding of nuclear physics. For the former aim, the nucleon-pair approximation (NPA)[36] approach uses collective nucleon pairs as the building blocks of the model space for even systems. The Monte Carlo shell model (MCSM)[37–40] applies the quantum Monte Carlo method to select important configurations. The projected shell model (PSM)[41] uses a deformed basis with a good classification scheme. For the other, the Gamow shell model (GSM)[42] applies the Berggren basis in the complex-energy plane so that it is able to study continuum and resonance states, weakly bound and unbound nuclei. The realistic shell model (RSM)[19,43,44] employs realistic effective interactions. The *ab initio* theories start from relatively strict quantum many-body theory.

This work aims to review the recent progress in CISM and related models. In detail, this review is arranged as follows. Section 2 introduces the theoretical framework of CISM. Section 3 introduces the recent progress in CISM and related models. The first part discusses intruders, mirror differences, isomers, and seniority conservation. The NPA approach, MCSM, PSM, GSM, RSM and *ab initio* theories are given attention in the second part. Finally, a conclusion is made in Sec. 4.

## 2. Theoretical Framework of Configuration–Interaction Shell Model with Effective Hamiltonian

### 2.1. *General introduction*

The nuclear shell model proposed by Mayer and Jensen, also called the independent particle model (IPM), is based on the following hypotheses: (1) Each nucleon in the nucleus moves independently in the spherically symmetric mean field generated by the other nucleons. (2) Spin–orbit coupling part should be considered.[1,45] It concentrates on independent particle movement and ignores the residual interaction. On this basis, the configuration-interaction shell model (CISM) considers the residual interaction. The nuclear states are described by mixed configurations rather than single configurations.

CISM assumes that a part of the nucleons form a frozen core. The rest of the valence nucleons are allowed to occupy specific orbits above the core, which make up the model space. The effective Hamiltonian (effective interaction) is constructed based on the model space. The valence nucleons are subject to the interactions from the core, which correspond to the one-body parts (single-particle energies) of the Hamiltonian. The interactions among valence nucleons are the residual interactions typically considered up to the two-body parts. The eigenvalues and eigenstates can be obtained with the CISM code by solving the many-body Schrodinger equation with the diagonalization process. The details on model space, effective Hamiltonian, and code are introduced in this paper.

### 2.2. *Model space*

The choice of model space is crucial to solving a particular problem. The theoretical calculations will be inaccurate or wrong if the model space is insufficient, either too small or unsuitable. For instance, describing the fluorine and neon isotopes near the neutron dripline would be incorrect without considering the intruder neutron orbits beyond the $N = 20$ shell.[46–48] In contrast, it takes vast computation resources if the model space is too large as the computation time grows exponentially with the model space size.[20] That is also why the systematic investigation of heavy nuclei is still challenging.

Typically, while a doubly magic nucleus is defined as the frozen core, the orbits of one or two adjacent major shells beyond the core are chosen to make up the model space. For example, in the light mass region, there are the *p*-shell model space,[49,50] the *sd*-shell model space[23,51] and the *pf*-shell model space,[52–54] which include one major shell. In comparison, the *psd*-shell model space[55] and *sdpf*-shell model space[8] consist of two major shells.

Due to the spin-orbit splitting, harmonic oscillator shells do not define model spaces in the medium and heavy regions. In contrast, the order of the magic numbers is used for the denomination. In the northeast of $^{56}$Ni, there are the model spaces *jj*44, *jj*45, *jj*46, *jj*55, *jj*56, *jj*57, *jj*66 and *jj*67. The numbers 4, 5, 6 and 7 represent the number of orbits beyond the magic numbers 28, 50, 82 and 126, respectively. For example, the frozen core of the model space *jj*45 is the nucleus composed of 28 protons and 50 neutrons, i.e., $^{78}$Ni. All nuclei, with $28 \leq Z \leq 50$ and $50 \leq N \leq 82$, are included in the *jj*45 model space. Table 1 gives a more detailed description of the mentioned model spaces.

### 2.3. *Effective Hamiltonian*

The knowledge of nuclear force plays a fundamental role in studying atomic nuclei. The nuclear force can often be decomposed into central, spin–orbit and tensor forces.[56–58] The central part contributes the most to the whole nuclear bindings. The inclusion of the spin–orbit term was the key to interpreting the magic numbers larger than 20, making IPM generally acceptable.[1,2] The spin–orbit interaction in the CISM, from a nucleon–nucleon view, should be related to the mean-field spin–orbit term in the IPM from a more fundamental perspective. The tensor force, proposed in the 1940s,[56–58] was known to be important to shell evolution,[59] along with other aspects such as the pairing interaction and the iso-vector spin-orbit terms.

Table 1. Details of the model spaces in the medium and heavy mass region.

| Model space | Frozen core | Orbits | Proton number range | Neutron number range |
|---|---|---|---|---|
| *jj*44 | $^{56}$Ni | $\pi 0f_{5/2}$, $\pi 1p_{3/2}$, $\pi 1p_{1/2}$, $\pi 0g_{9/2}$, $\nu 0f_{5/2}$, $\nu 1p_{3/2}$, $\nu 1p_{1/2}$, $\nu 0g_{9/2}$ | 28–50 | 28–50 |
| *jj*45 | $^{78}$Ni | $\pi 0f_{5/2}$, $\pi 1p_{3/2}$, $\pi 1p_{1/}$, $\pi 0g_{9/2}$, $\nu 0g_{7/2}$, $\nu 1d_{5/2}$, $\nu 1d_{3/2}$, $\nu 2s_{1/2}$, $\nu 0h_{11/2}$ | 28–50 | 50–82 |
| *jj*46 | $^{110}$Ni | $\pi 0f_{5/2}$, $\pi 1p_{3/2}$, $\pi 1p_{1/2}$, $\pi 0g_{9/2}$, $\nu 0h_{9/2}$, $\nu 1f_{7/2}$, $\nu 1f_{5/2}$, $\nu 2p_{3/2}$, $\nu 2p_{1/2}$, $\nu 0i_{13/2}$ | 28–50 | 82–126 |
| *jj*55 | $^{100}$Sn | $\pi 0g_{7/2}$, $\pi 1d_{5/2}$, $\pi 1d_{3/2}$, $\pi 2s_{1/2}$, $\pi 0h_{11/2}$, $\nu 0g_{7/2}$, $\nu 1d_{5/2}$, $\nu 1d_{3/2}$, $\nu 2s_{1/2}$, $\nu 0h_{11/2}$ | 50–82 | 50–82 |
| *jj*56 | $^{132}$Sn | $\pi 0g_{7/2}$, $\pi 1d_{5/2}$, $\pi 1d_{3/2}$, $\pi 2s_{1/2}$, $\pi 0h_{11/2}$, $\nu 0h_{9/2}$, $\nu 1f_{7/2}$, $\nu 1f_{5/2}$, $\nu 2p_{3/2}$, $\nu 2p_{1/2}$, $\nu 0i_{13/2}$ | 50–82 | 82–126 |
| *jj*57 | $^{176}$Sn | $\pi 0g_{7/2}$, $\pi 1d_{5/2}$, $\pi 1d_{3/2}$, $\pi 2s_{1/2}$, $\pi 0h_{11/2}$, $\nu 0i_{11/2}$, $\nu 1g_{9/2}$, $\nu 1g_{7/2}$, $\nu 2d_{5/2}$, $\nu 2d_{3/2}$, $\nu 3s_{1/2}$, $\nu 0j_{15/2}$ | 50–82 | 126–184 |
| *jj*66 | $^{164}$Pb | $\pi 0h_{9/2}$, $\pi 1f_{7/2}$, $\pi 1f_{5/2}$, $\pi 2p_{3/2}$, $\pi 2p_{1/2}$, $\pi 0i_{13/2}$, $\nu 0h_{9/2}$, $\nu 1f_{7/2}$, $\nu 1f_{5/2}$, $\nu 2p_{3/2}$, $\nu 2p_{1/2}$, $\nu 0i_{13/2}$ | 82–126 | 82–126 |
| *jj*67 | $^{208}$Pb | $\pi 0h_{9/2}$, $\pi 1f_{7/2}$, $\pi 1f_{5/2}$, $\pi 2p_{3/2}$, $\pi 2p_{1/2}$, $\pi 0i_{13/2}$, $\nu 0i_{11/2}$, $\nu 1g_{9/2}$, $\nu 1g_{7/2}$, $\nu 2d_{5/2}$, $\nu 2d_{3/2}$, $\nu 3s_{1/2}$, $\nu 0j_{15/2}$ | 82–126 | 126–184 |

There are three ways to understand the nuclear force among nucleons: strong interaction, nucleon–nucleon scattering data, and nuclear structural properties. As a result of the low-energy scale, it is challenging to deduce the nuclear force from the strong interaction. The nuclear force is called realistic when fitted to the nucleon–nucleon scattering data, and it needs to be renormalized by removing the repulsive core effect and considering the influence of outer space. The phenomenological nuclear force is fitted to the spectroscopic properties, such as the binding energy and the energy spectrum. In CISM, both the realistic and phenomenological nuclear forces are employed.

One can construct the shell-model Hamiltonian in the model space based on the nuclear forces. As a many-body problem, the Hamiltonian, in principle, should include the single-particle, two-body, three-body and many-body terms. However, it is challenging to include three-body and many-body terms in practice except for very light nuclei. Though the role of the many-body force was believed to be less critical for the energy spectrum,[44] including the three-body force was shown to be necessary for improving the accuracy of the shell-model description.[19] The effect of the three-body force should be considered when the realistic nuclear forces are used, while the phenomenological nuclear forces fitted to the spectroscopic properties (partially) include the many-body correlation.

One way to express the shell-model Hamiltonian in second quantization is to employ single-particle energies (SPEs) and two-body matrix elements (TBMEs). TBMEs can be decomposed into monopole and multipole components. The extraction of the monopole part of the effective Hamiltonian started in 1964.[60] SPEs and monopole components together govern the evolution of the single-particle structure. In the last two decades, they have been used to explain shell evolution.[7,61] On the other hand, multipole components, such as the quadrupole interaction, have been studied since the 1950s[6] and are crucial for understanding collective states.[62]

Currently, the Hamiltonian used for the shell model depends on the targeted region. In the $p$ region, Cohen and Kurath derived the effective Hamiltonians by fitting observed energy levels.[21] In the $sd$ region, the USD family by Brown *et al.*,[23,51,63] obtained from a least-squares fit of hundreds of states in the $sd$ shell,[63] denotes classical phenomenological Hamiltonians. In the $pf$ region, there are classically Kuo–Brown (KB) interaction,[22] GXPF1 interaction,[54] KB3 interaction[53] and KB3G interaction.[52]

The nuclear shell-model space, including two major shells, is more complex mainly because of the cross-shell part. For the $psd$ model space, the related effective interactions are composed of three parts, i.e., the $p$ shell, the $sd$ shell and the cross-shell. In this region, for the first time, Millener and Kurath added the repulsive component to the central part of the $G$ matrix[64,65] and built the MK interaction[66] for particle-holes. The classical WBT and WBP interactions proposed by Warburton and Brown took the W (also named as USD) interaction[23,51] for the $sd$-shell interaction and fitted binding energies for the other two parts of the interaction.[55] In detail, WBT fitted TBMEs, while WBP fitted potential for the cross-shell component. Suzuki, Fujimoto and Otsuka modified the strength of proton-neutron interaction and SPEs and proposed the SFO Hamiltonian.[67] Yuan *et al.* proposed YSOX[25] which is based on SFO, SDPF-M[68] and $V_{\rm MU}$[69] plus an M3Y spin-orbit force.[70] In the $sdpf$ model space, there are SDPF-M,[68] SDPF-MU,[71] SDPF-U,[72] etc. In addition, the extended pairing-plus quadrupole Hamiltonian combined with monopole corrections (EPQQM) was adopted by Hasegawa and Kaneko[73] and applied in the medium and heavy mass region.[73–77,347] The Kuo–Herling Hamiltonian was modified by Warburton[78] and Brown[79] for nuclei around $^{208}$Pb.

At present, it is anticipated to construct shell-model Hamiltonians for different model spaces in a unified way. For this purpose, the monopole-based universal interaction $V_{\rm MU}$, consisting of the Gaussian central force and the tensor force comprising $\pi$ and $\rho$ meson exchanges, was introduced by Otsuka *et al.*[69] $V_{\rm MU}$ is universal as it can universally describe shell evolution. Based on $V_{\rm MU}$, various Hamiltonians were built for different regions, especially those involving orbits of two major shells. For example, the cross-shell part of the Hamiltonian YSOX in the $psd$ region is based on $V_{\rm MU}$. SDPF-MU is another Hamiltonian using $V_{\rm MU}$ for the cross-shell part and set in the $sdpf$ region, as the name indicates. Besides, a new Hamiltonian based on $V_{\rm MU}$ is suggested to describe the nuclei in the medium and heavy mass region in a unified way.[32]

### 2.4. *Code*

Solving the many-body Schrodinger equation is on schedule after constructing the Hamiltonian in the chosen model space. For this sake, the diagonalization of the Hamiltonian matrix is required to derive the eigenvalues (energies) and eigenvectors (wave functions). Upon the deduction of the eigenstates, other structure and decay properties can be calculated with corresponding operators.

A practical problem typically aims to obtain a few lowest eigenvalues of a large, sparse and real-symmetric matrix. The Lanczos method[80] and the Davidson method[81] were developed with iteration methods to achieve this goal. Employing these iteration approaches, the Glasgow group made the treatment of the giant matrix possible. After that, the code Antoine[82] written by Caurier *et al.* in the M-scheme was generally adopted. Researchers at Drexel University completed the DUPSM code in the J-scheme. The matrix written in M-scheme has a larger dimension than in J-scheme but is less dense and less complex to be calculated in algebra. Therefore, both the two schemes are adopted in practice. For illustration, Caurier *et al.* also wrote a J-scheme code NATHAN.[83] The OXBASH code, widely distributed and constantly updated, was first written in J-scheme and now is in a hybrid scheme. Its version 2004 can carry out shell model calculations with dimensions up to nearly 50,000 in the J-scheme and 2,000,000 in the M-scheme.[84]

Parallel computation has been realized in shell-model codes this century. NuShell and NushellX, written in J-scheme, are successors of OXBASH. With OpenMP parallelization, NuShellX is able to calculate within a dimension of more than 100,000,000.[85] Based on NushellX, Brown introduced the NuShellX@MSU for the pre-treatment of the input files and post-treatment of output files. Figures and tables for energy levels and decay data are direct outputs of NuShellX@MSU, including comparison with experimental data. Meanwhile, researchers at the University of Tokyo developed the M-scheme code MSHELL[29] and published the OpenMP-MPI hybrid code KSHELL.[31] In recent years, another OpenMP-MPI hybrid code Bigstick[86] (in M-scheme), uniquely implementing a second level of factorization, has been published. Thanks to the codes, our computation ability has been increasing, reaching a dimension around $10^{11}$ in M-scheme.

## 3. Recent Progress in CISM and Some Related Models

### 3.1. *CISM with phenomenological Hamiltonians*

The CISM with phenomenological Hamiltonians succeeds in providing spectroscopic properties for nuclei from the $\beta$-stability line to the driplines, including binding energies, levels, electromagnetic moments and transitions, spectroscopic factors and $\beta$-decay strengths. With CISM, one endeavors to search for simple rules to describe complex systems and is encouraged by newly observed out-of-the-ordinary phenomena. CISM helps to understand the emergence and disappearance of the magic effect, the uniqueness and the systematics of specific properties and the conservation and breaking of various symmetries. In recent decades, several phenomenological effective Hamiltonians have been developed to interpret the observed exotic phenomena by figuring out the important terms of the nuclear force, which may have been ignored before due to computational limitations.

In this section, we will introduce the recent progress in CISM. First, the cross-shell excitation in light nuclei; then, the mirror energy difference in $A \sim 20$ nuclei; next, the new isomers and the new nuclides in the medium and heavy nuclear region; besides, the unified description of medium and heavy nuclei with effective nuclear force; finally, the seniority conservation in the medium and heavy nuclear region.

#### 3.1.1. *Cross-shell excitation in light nuclei*

In neutron-rich light-mass nuclei, valence neutrons occupy the higher major shell than the valence protons. The cross-shell proton-neutron interactions play an important role. Meanwhile, the particle excitation across the shell gap is usually prohibited in the CISM calculation, which is reasonable if the excitation energy of the target low-lying state is smaller than the shell gap. However, both common configurations in one major shell and intruder configurations from the next major shell contribute largely to some nuclear states. It occurs when the gap between two major shells is significantly reduced. In such a case, cross-shell excitation needs to be considered, with a complete inclusion of configuration mixing, corresponding residual interaction, and extensive calculation resources. A systematic study on cross-shell excitation is beyond the present computational ability in the medium and heavy mass region. On the contrary, the particle excitation across the $Z$, $N = 8$ and $N = 20$ shells has been investigated in depth in the last two decades. Overall, the number of nucleons for cross-shell excitation has been extended for describing such light nuclei in a unified way because of the significant variation of the shell gap and the large individual differences.

##### 3.1.1.1. C, N and O driplines and YSOX including $(0\text{–}3)\hbar\omega$ cross-shell excitation

One elementary problem of nuclear physics is the location of the driplines, which is related to the saturability of the nuclear force. On the one hand, the production of dripline nuclei is arduous because of their weakly bound property. Figure 1 illustrates the light-mass part of the nuclide chart. It also shows that the dripline nucleus of oxygen isotope $^{24}$O was discovered in 1970,[89] while the dripline nucleus $^{31}$F was first observed nearly 30 years later.[90] Until now, only the neutron drip line up to Ne isotopes has been experimentally located.[24] On the other hand, unexpected structure arising in the driplines challenges the theoretical approaches. The dripline locations of C, N and O isotopes were not simultaneously reproduced by the same effective Hamiltonian until recent decades.[25]

It was shown that the inclusion of $2\hbar\omega$ excitations was important for significant theoretical progress.[25] Using $^4$He as the core, Ref. 25 constructed a new Hamiltonian in the *psd* model space based on three parts of interactions: SFO (with its *p*-shell interaction), SDPF-M (with its *sd*-shell interaction), and $V_{MU}$ plus M3Y spin-orbit force (for calculating the *psd* cross-shell interaction). While the WBT and WBP interactions include only the (0–1)$\hbar\omega$ states, this new Hamiltonian YSOX considers (0–3)$\hbar\omega$ cross-shell excitation during its construction. Owing to the consideration of (2–3)$\hbar\omega$ states, YSOX simultaneously reproduced the C, N and O neutron driplines and correctly assigned the ground-state spins to $^{10}$B and $^{18}$N and located $^{22}$C and $^{24}$O as the dripline nuclei, as illustrated in Fig. 2.

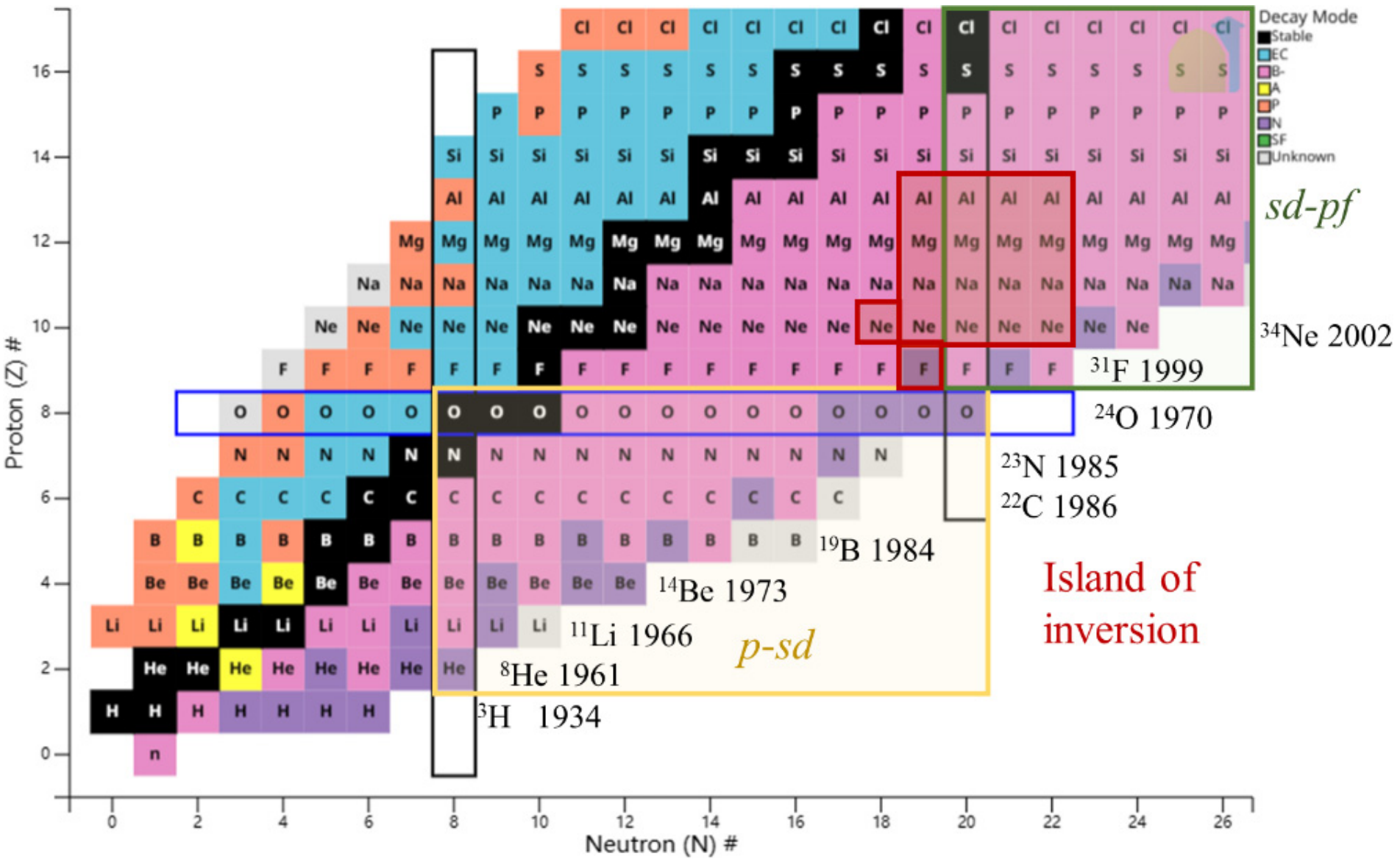

Fig. 1. Nuclide chart in the light mass region. Data are from Refs. 87 and 88. The year of discovery of the most neutron-rich isotopes up to Ne is noted.

#### 3.1.1.2. Application of YSOX in $2 \leq Z \leq 8$ and $8 \leq N \leq 20$

Overall, YSOX nicely described the ground-state and excitation energies, the electromagnetic and Gamow–Teller transitions of isotopes with $5 \leq Z \leq 8$, from the $\beta$-stability line to the neutron drip line. The predicted two-neutron separation energy ($S_{2n}$) in $^{22}$C shown in Ref. 25 is in nice agreement with the successive experimental measurement.[10] The occupation weight in $\nu 1s_{1/2}$ of $^{20}$C measured in Ref. 92 can be well reproduced by YSOX.[93] As it well describes the configuration mixing in C isotopes around the $N = 14$ shell, it has also been used to investigate the relation between the $S_{2n}$ value and halo radius in $^{22}$C.[93] In addition, YSOX has well reproduced the excitation energies and spectroscopic factors in low-lying negative-parity states in $^{11}$Be recently obtained from the $^{12}$B (d, $^3$He) $^{11}$Be reaction,[94] and in $^{18}$N derived from neutron-adding reaction.[95] Its predictions on the quadrupole deformation properties in $^{16}$C, including the $B(E2, 2_1^+ \to 0_{g.s.}^+)$ value and the large deviation between proton and neutron quadrupole matrix elements, are close to the recent experimental data.[96] Successful applications in the last decade have validated the YSOX interaction.

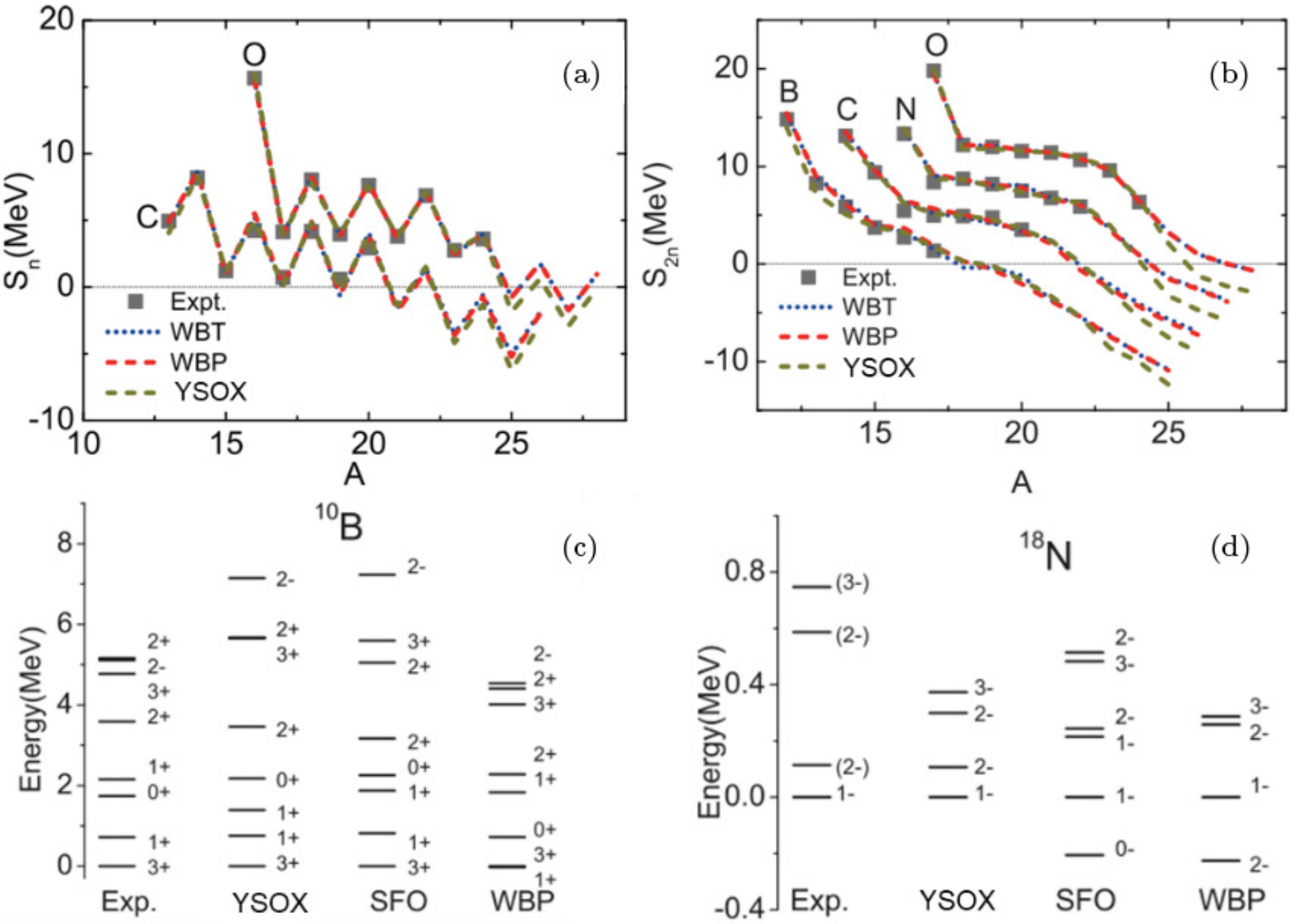

Fig. 2. From Ref. 25. Experimental and theoretical (a) $S_n$ in C and O isotopes, (b) $S_{2n}$ in B, C, N and O isotopes, (c) excitation spectrum in $^{10}$B, (d) excitation energy related to the $1^-_2$ state in $^{18}$N. Experimental data of the ground-state energy are from AME2003[91] and excitation energy are from Nudat2.[88] Interactions of WBP and WBT predict $^{22}$C ($^{26}$O) to be unbound (bound), while YSOX reproduces the locations of the neutron drip line: $^{22}$C and $^{24}$O. YSOX correctly describes the ground state of $^{10}$B and $^{18}$N, while other interactions do not.

The importance of cross-shell excitation in the region with $2 \leq Z \leq 8$ and $8 \leq N \leq 20$ can be illustrated by the complex shell evolution there. $^{24}$O possesses the characteristic of a doubly magic nucleus, $N = 16$ being a new magic number.[16] The $N = 16$ shell closure also exists in the Be and C isotopes, but the $N = 14$ shell closure arising in O isotopes vanishes in Be, C and N isotopes.[97–99] It was suggested that the $N = 16$ shell closure originated from the absence of the interaction between $\pi 0d_{5/2}$ and $\nu 0d_{3/2}$.[16,99] CISM analyses showed that both the cross-shell proton-neutron interaction[97] and the neutron-neutron monopole interaction[100] contributed significantly to the $N = 14$ shell evolution. While a novel magic number emerges, the conventional magic number disappears. The proton excitation may be significant in the low-lying states where the neutron shell gap is closed. Since there are no single patterns for the shell closures, not only should the cross-shell excitation be considered but also significant number of particles crossing the shell gap should be set for universally describing the *psd*-region nuclei. As a result, YSOX which both synthesizes the advantages of existing interactions and considers the extended cross-shell excitations is promising in the *psd* region.

The mid-shell nuclei have less probability of cross-shell excitation globally due to the Pauli-blocking effect. For instance, the ground states in $N = 11$ isotones $^{17}$C[101] and $^{19}$O[102] are illustrated with pure $(\nu 0d_{5/2})^3$ configuration. Between them, $^{18}$N is dominated by the $(\pi 0p_{1/2})^{-1}$ $(\nu 0d_{5/2})^3$ configuration in the ground and several low-lying states.[95] In contrast, one should be prudent for nuclei near the conventional shell closures, especially for the exotic nuclei with enhanced shell evolution. It can be speculated that C, N, and O isotopes near the neutron drip line are more related to the proton cross-shell excitation. When talking about the neutron excitation from $p$ to $sd$ shell, nuclei around $N = 8$ but deviating from the $\beta$-stability line are more relevant. As evidenced by the experimental discoveries,[103–105] the $N = 8$ shell closure vanishes in the extremely neutron-rich nuclei $^{11,12}$Be.

Experiments showed that the ground state in the halo nucleus $^{11}$Be is dominated by the $s$-wave intruder, indicating an inversion of $\nu 1s_{1/2}$ and $\nu 0p_{1/2}$.[103] A recent measurement has revealed the intruder configurations in $^{12}$Be,[104,105] as illustrated in Fig. 3. It has been found that the $d$-wave intruder is dominant in the ground state of $^{12}$Be. CISM calculations with YSOX well reproduced the measured low-lying levels and spectroscopic factors, indicating a lowering of neutron $d$-orbit from $^{11}$Be to $^{12}$Be. Recent experiments, along with CISM reproductions, also focused on the low-lying states in $^{11,12}$Be. It has been recently found that $p$-shell configurations dominate in the negative-parity states of $^{11}$Be.[94] However, the *sd*- and *sp*-configurations were respectively suggested as dominating the $2^+$ and $1^-$ states.[105] In other words, the cross-shell excitation affects the mentioned low-lying state of $^{12}$Be but is less significant in the negative-parity states of $^{11}$Be.

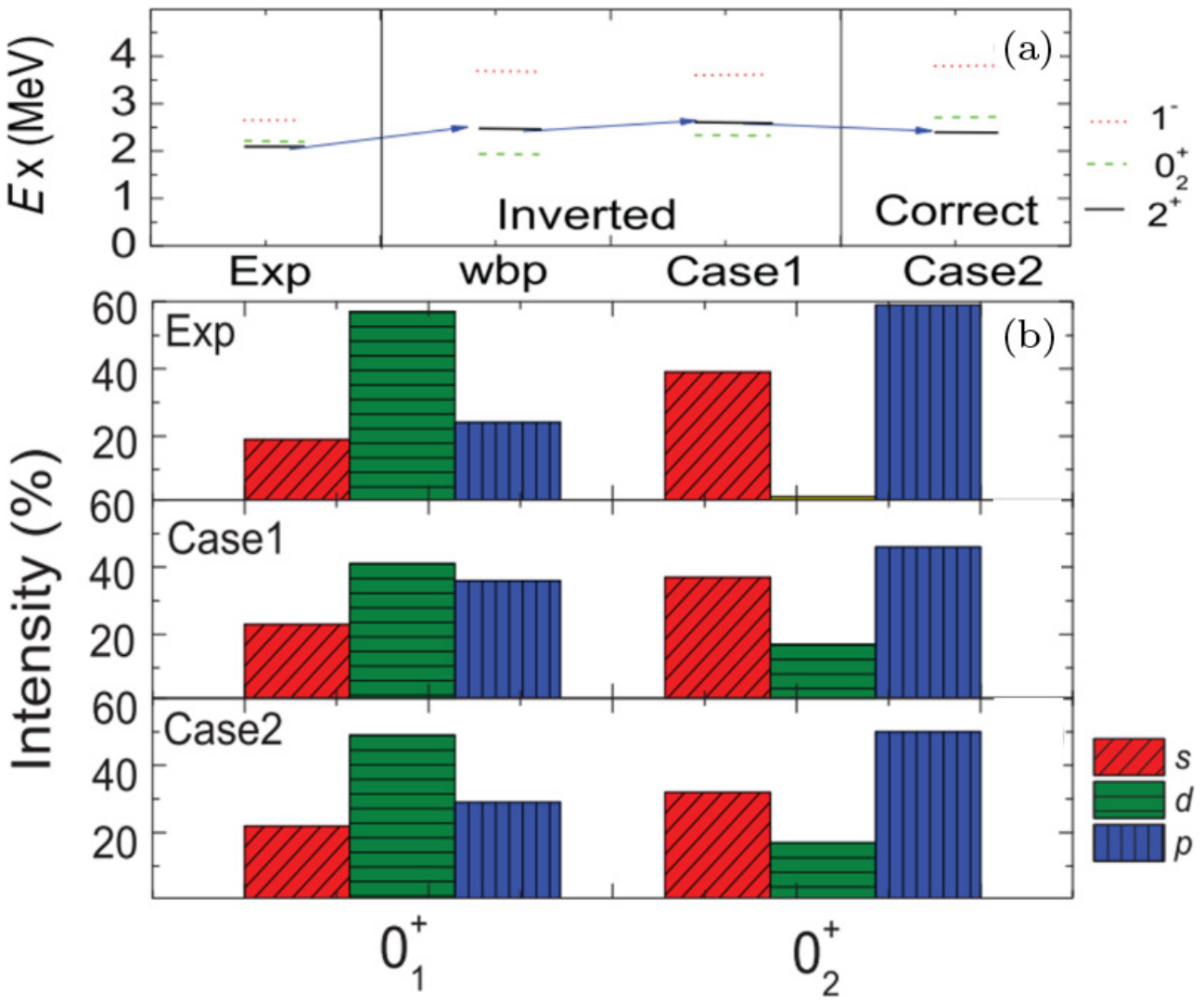

Fig. 3. From Ref. 105. Experimental and theoretical (a) excitation energies and (b) wave intensities of low-lying states in $^{12}$Be. Experimental data are newly reported in Refs. 104 and 105. "wbp" represents the CISM results using the WBP interaction and is reported in Ref. 106. "Case1" ("Case 2") refers to results calculated by the YSOX interaction without (with) reducing the SPE of $d$-orbits. It can be seen that the abnormal decrease of the $d$-orbits results in the dominance of $d$-wave functions in the ground state of $^{12}$Be. Moreover, both the lowest $0^+$ states are significantly influenced by cross-shell excitation.

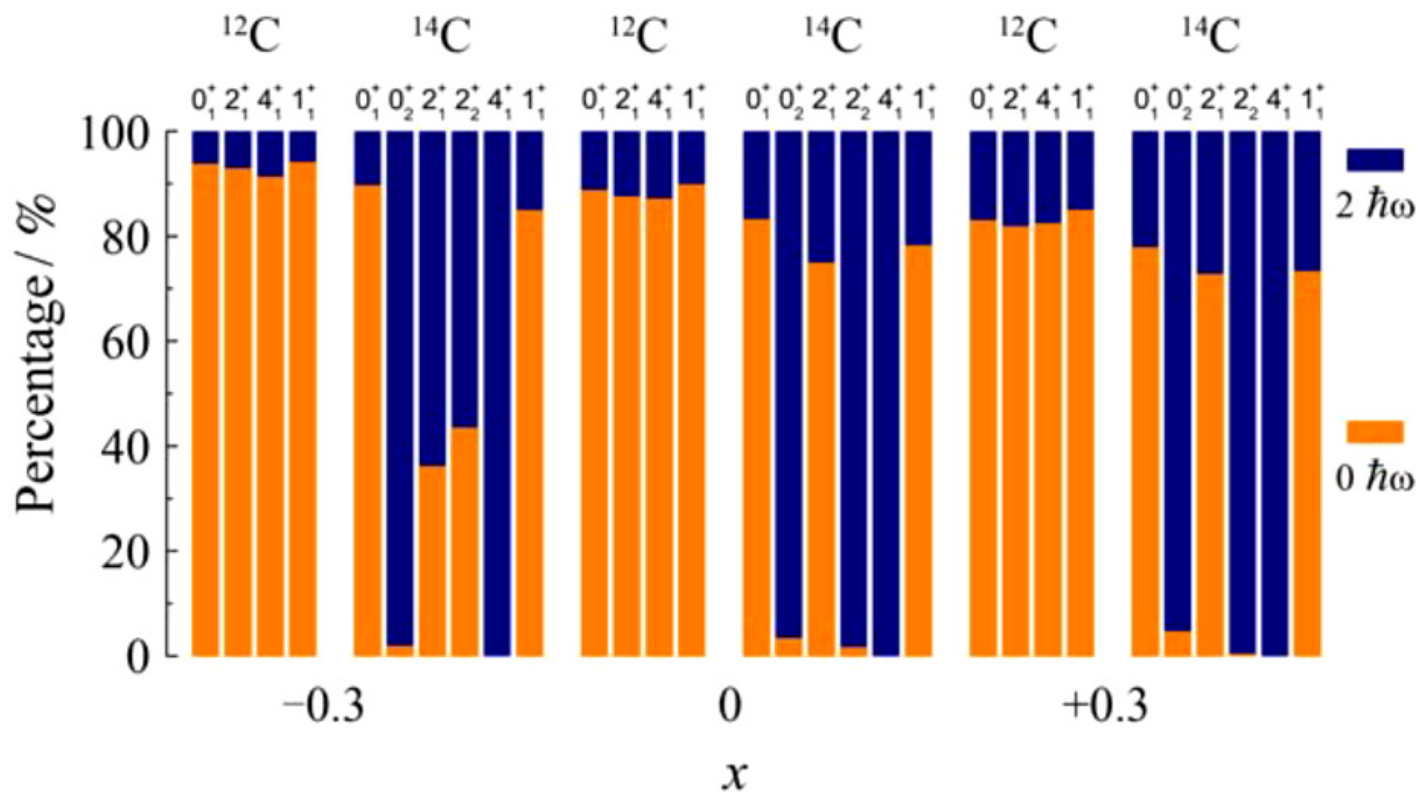

Fig. 4. From Ref. 108. Percentage of 0 and $2\hbar\omega$ configurations as a function of the $\langle pp|V|sdsd\rangle$ strength in $^{12}$C and $^{14}$C. The results are obtained by YSOX in the (0–2)$\hbar\omega$ model space.

#### 3.1.1.3. Long-lifetime problem of $^{14}$C and YSOX considering up to $6\hbar\omega$ cross-shell excitation

It is worth noting that the nucleus close to the $\beta$-stability line, $^{14}$C, had also been a long-standing challenge for the theoretical interpretation. Actually, it is a well-known and long-lived nucleus, close to the doubly magic nucleus $^{16}$O, and supposed to be semi-magic. However, almost all the theoretical models failed to reproduce its low-lying excitation spectrum and transition rates, even with the $2\hbar\omega$ excitation included. A recent study[107] has found that the inclusion of $4\hbar\omega$ configurations is necessary for $^{14}$C, which has been numerically achieved thanks to the development of parallel computation on the supercomputer. Meanwhile, its excitation spectrum and decay properties significantly depend on the non-diagonal cross-shell interactions.

CISM calculations with YSOX in the (0–2)$\hbar\omega$ model space showed that the $0\hbar\omega$ configurations were insufficient for the excited states in $^{14}$C.[108] Figure 4 illustrates the percentage of the 0 and $2\hbar\omega$ configurations in $^{12,14}$C as a function of the strength of the off-diagonal cross-shell interaction $\langle pp|V|sdsd\rangle$. It can be also found that the occupancy of $2\hbar\omega$ configurations was sensitive to the $\langle pp|V|sdsd\rangle$ strength. If the mentioned interaction elements are weakened by 30%, the $2_1^+$ ($2_2^+$) state would be more (less) dominated by the $2\hbar\omega$ configuration. Therefore, the $\langle pp|V|sdsd\rangle$ interaction, which had been rarely investigated, would impact the low-lying levels in $^{14}$C.

To better understand $^{14}$C, the cross-shell excitation up to $6\hbar\omega$ was investigated based on the Hamiltonians YSOX and WBP in Ref. 107. The necessity of including $4\hbar\omega$ configurations for $^{14}$C has been first confirmed. Figure 5 gives the experimental and theoretical levels and transition rates of positive states in $^{14}$C. It can be found that the description of $^{14}$C levels and transitions depend strongly on the $\langle pp|V|sdsd\rangle$ strength and require the inclusion of $4\hbar\omega$ configuration. On the one hand, the inclusion of $4\hbar\omega$ configurations and the reduction of the $\langle pp|V|sdsd\rangle$ strength can both result in better description of $^{14}$C levels. On the other hand, only the YSOX Hamiltonian in (0–4)$\hbar\omega$ and larger model space succeeded in reproducing the electromagnetic transitions in $^{14}$C.

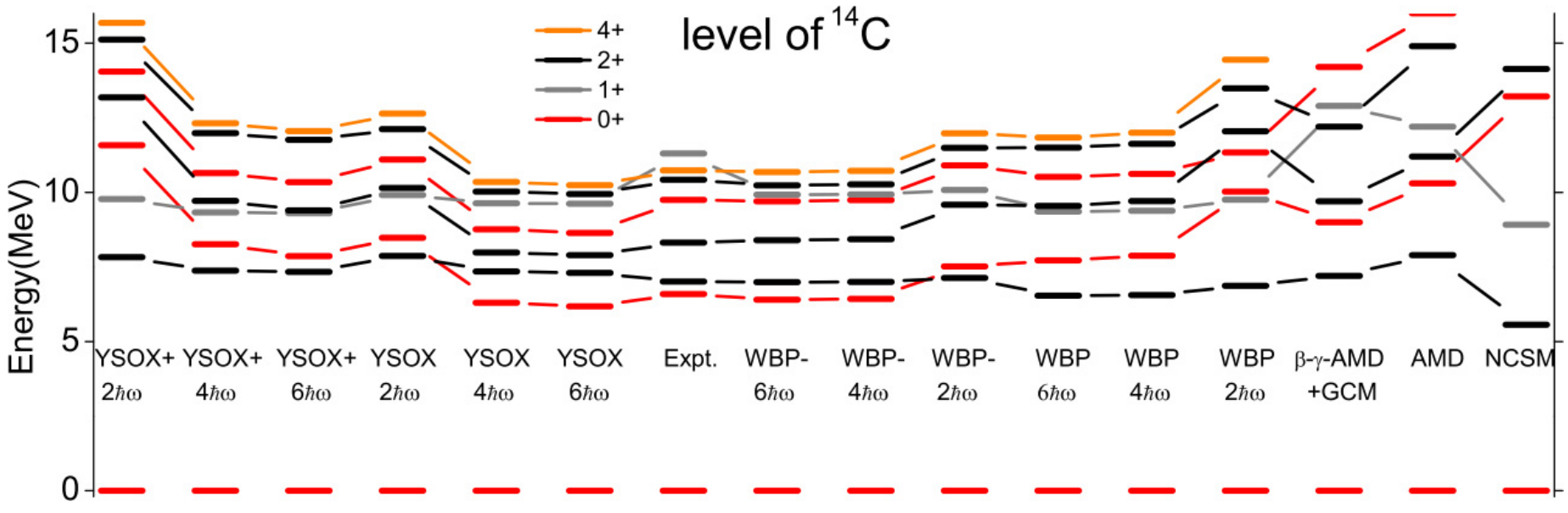

(a)

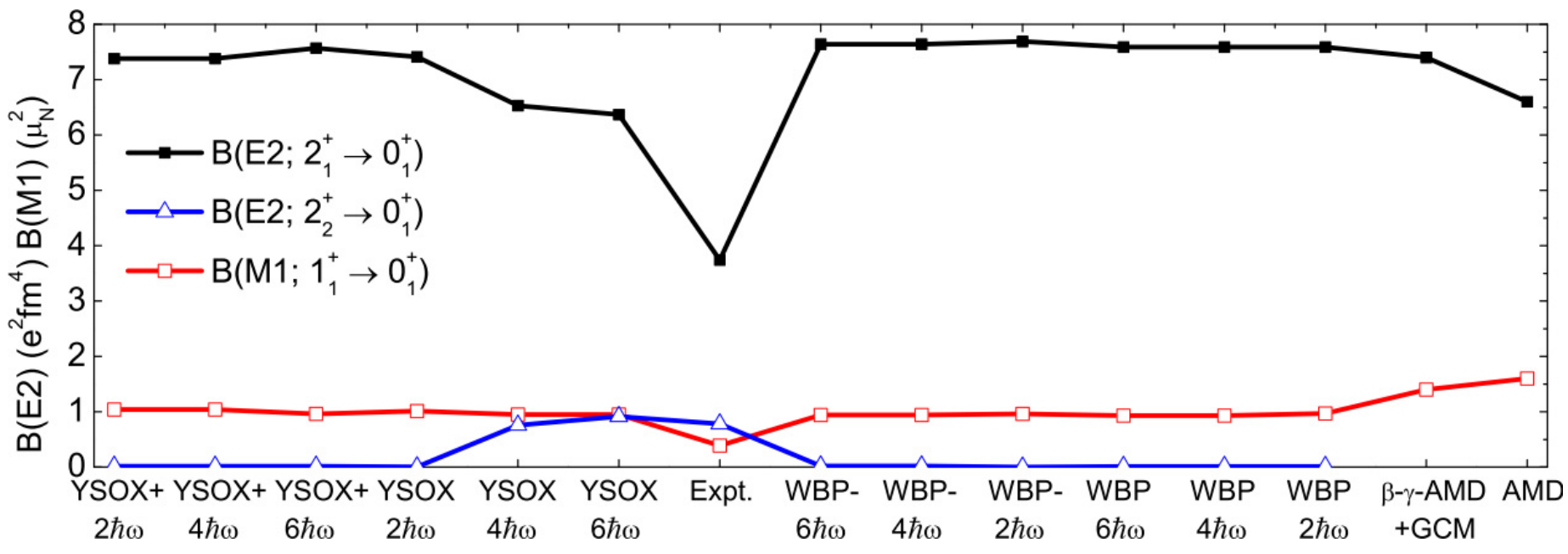

(b)

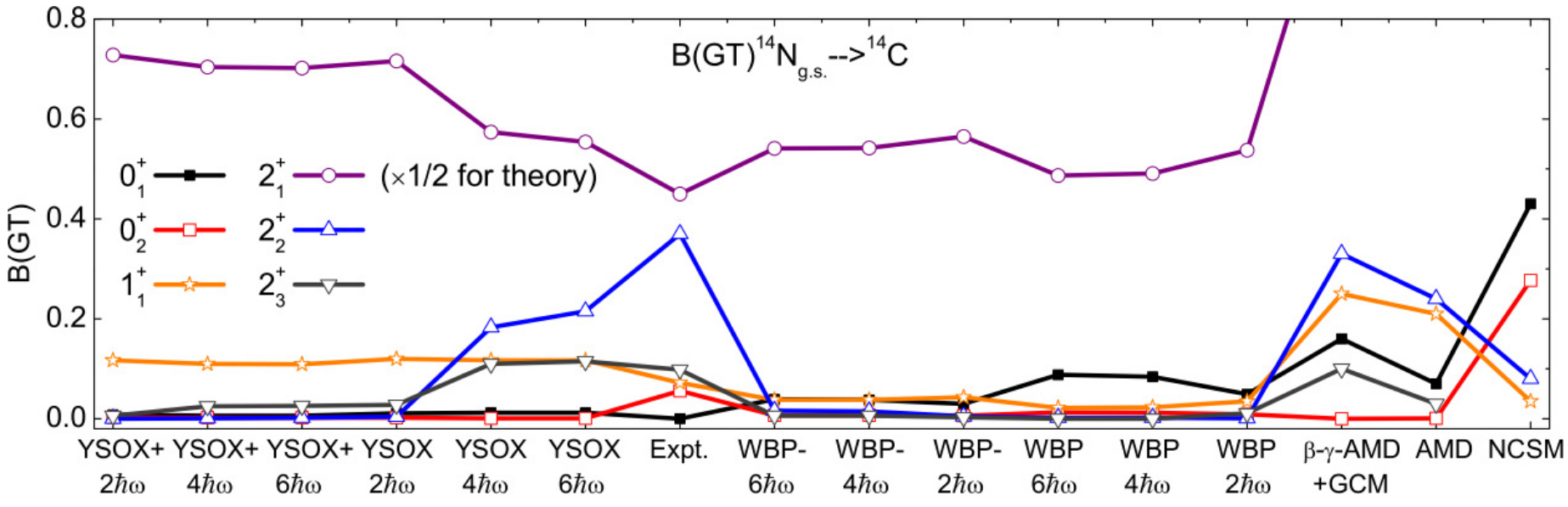

(c)

Fig. 5. From Ref. 107. (a) Levels, (b) $B$(EM) values, and (c) $B$(GT) values of states in $^{14}$C derived with different models. YSOX+ (WBP−) is the YSOX (WBP) Hamiltonian with enhanced (reduced) $\langle pp|V|sdsd\rangle$ part. Anti-symmetrized molecular dynamics (AMD), No-core shell model (NCSM) and experimental results are taken from Refs. 109, 110 and 88, 111–113, respectively. It can be seen that the models AMD, NCSM, as well as CISM in the (0–2)$\hbar\omega$ model space failed in describing the well-known, long-lived isotopes $^{14}$C.

#### 3.1.1.4. "Island of inversion" and cross-shell excitation around the $N = 20$ shell

Now, it is of great interest to discuss recent progress in the cross-shell excitation around the $N = 20$ shell. For this theme, "island of inversion"[8] is an inevitable term. It has been found that a series of nuclei with $Z \approx 11$ and $N \approx 20$ have unnatural-parity ground states, which is exactly the consequence of cross-shell excitation. Recently, experimental studies have focused on identifying the boundary of the "island of inversion", and CISM calculations well reproduced the newly measured structure and decay properties thanks to the sufficient consideration of the particle-hole ($p$-$h$) configurations.

Experiments and CISM with SDPF-U-MIX near the northern boundary have identified the inside location of $^{34}$Al.[114] A systematic study on the half-lives of nuclei within and around the "island of inversion" confirmed such a conclusion.[26] Whereas, as evidenced by the newly observed $\beta$-decay experiments of $^{34}$Al,[26,115] the $\beta$-decay daughter nucleus $^{34}$Si locates outside the island. CISM calculations with the Gogny D1S and SDPF-M interactions confirmed that the $0p$-$0h$ configuration was dominant in the ground state and also indicated that the excited states, $0^+_2$, $2^+_{1,2}$ were dominated by the $2p$-$2h$ configurations.[26,115] A recent study with SDPF-M[116] assigned the spin-parity up to $6^+$ to the observed states of $^{34}$Si and further indicated their dominant configurations in detail, as shown in Fig. 6. In short, $^{34}$Si is a doubly magic nucleus outside the "island of inversion", but cross-shell excitation plays a significant role in its low-lying excited states.

On the southern shore, detailed spectroscopy of $^{28}$F was first experimentally deduced.[27] Considering cross-shell excitation, the CISM using the SDPF-U-MIX and SDPF-MU interactions reproduced the experimental energy spectrum and explicated that $^{28}$F was located inside the "island of inversion".[27] Within the "island of inversion", the major role of cross-shell excitation on the complicated structure of $^{33}$Mg was emphasized in Ref. 117 after comparing the observed energy levels and partial cross-sections with CISM calculations using various effective Hamiltonians.

The effects of the neutron excitation across the $N = 20$ shell on neutron-rich fluorine and neon isotopes have been recently investigated in Ref. 48. The neutron-proton ratio can be even higher than 2 in such a nuclear region. Therefore, the conventional shell closure can hardly remain. After investigating CISM results derived from three Hamiltonians within (0–1) $\hbar\omega$, (0–3) $\hbar\omega$ and (0–5) $\hbar\omega$ truncated model space, it has been found that cross-shell excitation plays a significant role in the ground-state properties and low-energy levels of fluorine and neon nuclei around $N = 20$. Moreover, for the two nuclei with $N = 20$, i.e., $^{29}$F and $^{30}$Ne, the (0–5) $\hbar\omega$ cross-shell excitation was suggested to be included. Besides, it was indicated that the cross-shell parts of SDPF-M, SDPF-MU and SDPF-U should be improved in the fluorine and neon nuclei with $N \sim 20$. In other words, a more suitable cross-shell interaction for such nuclei is required.

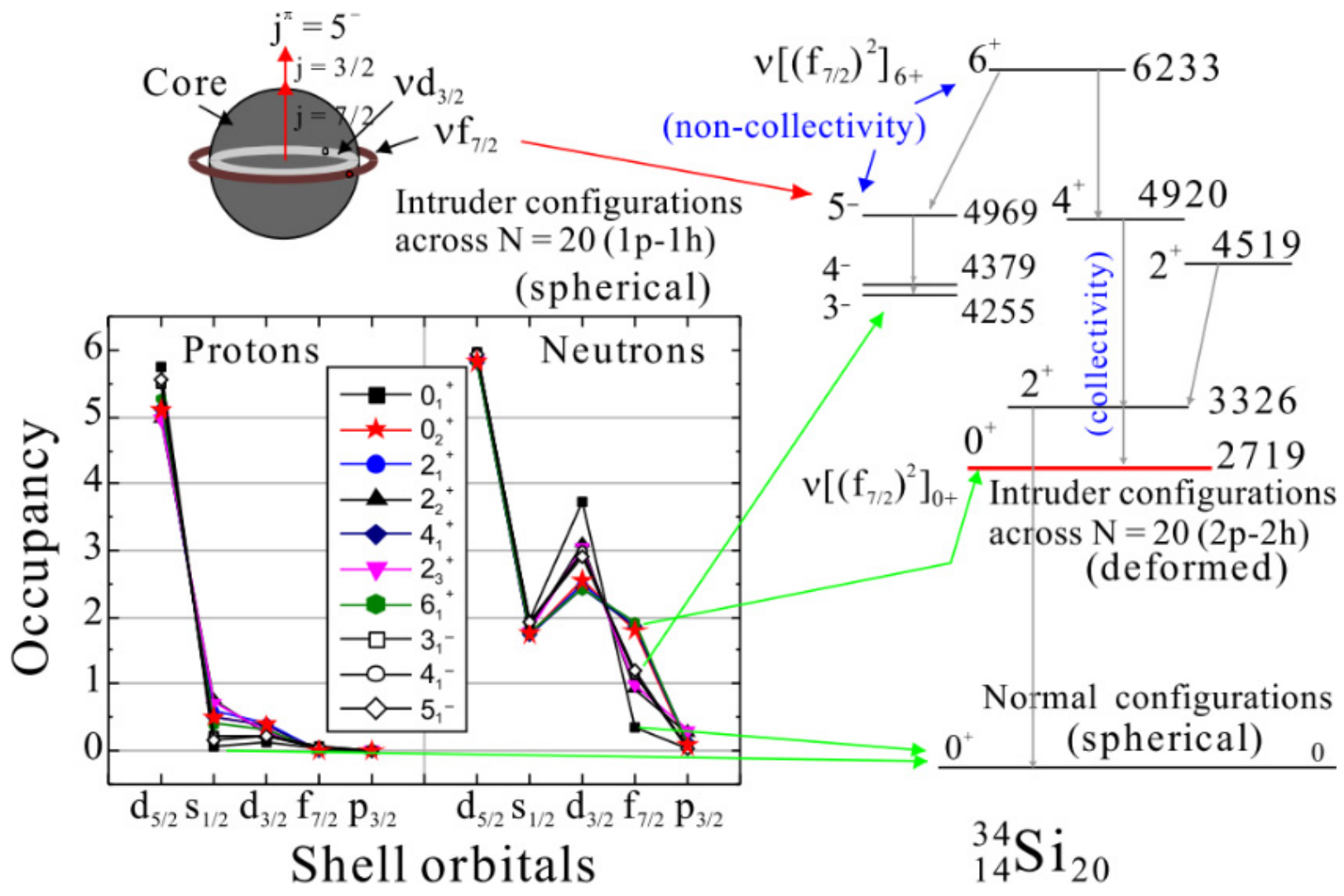

Fig. 6. From Ref. 116. Occupancy distribution of the low-lying states in $^{34}$Si. Some distinctive structures (spherical, deformed, collective and non-collective) are indicated in the level scheme, according to the CISM calculations. See details in Ref. 116.

It should be noticed that cross-shell excitation appears as well in the low-lying excited states in nuclei close to the $\beta$-stability line. Recently, low-lying states with negative parity of $^{39}$Cl have been experimentally determined, indicating the $5/2^-$ state at 1697 keV the lowest negative parity state. However, CISM calculations with ZBM2M, SDPF-MU, and SDPF-M all gave $7/2^-$ as the lowest negative parity state.[118] Once again, the cross-shell excitation for the *sdpf* region should be further investigated.

Actually, the unification of the *sdpf*-space interaction was expected in the shell-model review more than 20 years ago.[20] However, compared to the *psd* region, the *sdpf* region involves more nuclei and a larger model space. Thus, the relevant unified interaction is tougher to be constructed. Consequently, though the interaction has been improved, such an aim is still on the way.

#### 3.1.1.5. Cross-shell excitation for heavier nuclei

For heavier nuclei, the significance of the cross-shell excitation to high-lying states has also been shown. To reproduce the energy spectrum of $^{92}$Nb,[119] even though the pure configurations were sufficient for the low-excited states, mixed configurations with cross-shell excitation through CISM were employed for higher-excited states. For the nucleus $^{98}$Cd, the proton excitation across $Z = 38$ subshell was shown to be necessary in order to reproduce the excitation energy of the newly-observed $5_1^-$ state.[120] For nuclei around the doubly magic nucleus $^{132}$Sn, CISM calculations, including cross-shell excitation, agreed well with the experimental results.[121] Theoretical results showed the high-energy yrast states caused by neutron core excitations in $^{130}$In, $^{130}$Sn and $^{130}$Cd. Proton and neutron cross-shell excitations led to states in $^{130}$In of level ranging in 4.5–6.5 MeV and 2.0–4.1 MeV, respectively.

It can be concluded that cross-shell excitation is also significant for heavier nuclei. As the model space consists of more orbits, part of which are close to each other, the orbit inversion may appear more frequently. Besides, the shell structure in a nucleus may also vary with the $p$-$h$ excitation. But the limited computation ability challenges the CISM calculation in a model space consisting of plural whole major shells. Furthermore, insufficient experimental data restrict the fitting of phenomenological interactions. Therefore, only particular nuclei have been focused on. Truncation methods, which will be introduced later, are helpful in extending the understanding of cross-shell excitation.

In brief, cross-shell excitation plays an important role in plenty of light nuclei and some excitation states of medium and heavy nuclei. With the construction of the YSOX interaction considering cross-shell excitation, CISM calculations have recently succeeded in reproducing a number of neutron-rich nuclei in the *psd* region. Meanwhile, more nuclei lying within the “island of inversion” have been discovered, well interpreted by CISM. However, the cross-shell interaction is expected to be improved to describe specific nuclei, such as extremely neutron-rich F and Ne isotopes and $^{39}$Cl.

### 3.1.2. *Mirror properties in nuclei around $A = 20$*

As the nuclear force is almost charge-independent, mirror nuclei normally have very similar structures. Typically, as shown in Table 2, the energy differences between isobaric analog states of $^{51}$Fe and $^{51}$Mn are normally within 0.1 MeV.[88] Nevertheless, the Coulomb interaction and the isospin asymmetry can result in non-negligible structure differences between light mirror partners. For instance, the level of the first $1/2^+$ state in $^{13}$C is 0.72 MeV higher than that in $^{13}$N.[88] As the $\beta$-decay daughter nuclei of a mirror pair are also mirror asymmetry, the difference between the $\beta$-decay spectroscopies is another focus of the mirror pair. Without isospin asymmetry, the transition strengths of a mirror pair should be close. However, the isospin symmetry can be broken in nuclei far from the stability line. For example, a large difference in the Gamow–Teller strengths was observed in $^{24}$Si and $^{24}$Ne.[122] Meanwhile, exotic decay models have been found in the proton-rich nuclei, such as the $\beta$-delayed one-proton ($\beta p$) decay, $\beta$-delayed two protons ($\beta 2p$)[348] and $\beta$-delayed $\alpha$ ($\beta\alpha$) decay at the proton dripline[123] and $\beta$–$\gamma$–$\alpha$ decay in $^{20}$Na.[124] Furthermore, the emission of particles from the isobaric analog state (IAS) reveals the isospin mixing. Thus, the isospin symmetry and its breaking in mirror nuclei are of great interest. In the case of isospin symmetry, the knowledge of neutron-rich nuclei can guide the determination of its proton-rich mirror partner. In the other case, understanding the difference between mirror states can be crucial in gaining insights into the nuclear force.

Table 2. Comparison of energy levels of mirror states in $^{51}$Mn and $^{51}$Fe. Data are from Ref. 88.

| Spin | $5/2^-$ | $7/2^-$ | $9/2^-$ | $11/2^-$ | $13/2^-$ | $15/2^-$ | $17/2^-$ | $19/2^-$ | $21/2^-$ | $23/2^-$ |
|---|---|---|---|---|---|---|---|---|---|---|
| $^{51}$Mn (keV) | 0 | 237.3 | 1139.8 | 1488.5 | 2957.3 | 3250.8 | 3680.6 | 4139.7 | 5639.8 | 6492 |
| $^{51}$Fe (keV) | 0 | 253.5 | 1146.5 | 1516.5 | 2953.4 | 3275.7 | 3589.7 | 4097.9 | 5608 | 5471.5 |
| $\Delta$ (keV) | 0 | 16.2 | 6.7 | 28 | 3.9 | 24.9 | 90.9 | 41.8 | 31.8 | 20.5 |

The *sd* region is special because the MED is relatively significant there. The $1s_{1/2}$ orbit contributes much in the *sd* region and is less bound in the proton-rich nuclei than the neutron-rich ones. Therefore, the loosely bound effect of the $\pi 1s_{1/2}$ orbit leads to breaking the isospin symmetry. The shell model with the USD family is rather successful in the *sd* region but is mainly fitted from neutron-rich nuclei. USD, USDA, and USDB are isospin-conserving, while USDC includes the isospin-non-conserving effects. In a recent application of the USDB and USDC to $^{22}$Mg, the experimental levels and the corresponding branching ratios derived from the $\beta$ decay of $^{22}$Al have been satisfactorily reproduced.[125] The loosely bound effect should be extra considered for their application to more proton-rich nuclei.

As a recent improvement, the shift of single-particle energies and the reduction of the two-body matrix element (TBME) were proposed for describing proton-rich nuclei based on the USD family Hamiltonians (USD, USDA and USDB).[15] Considering the weakly bound effect of the $\pi 1s_{1/2}$ orbit, the reduction factors for the TBME involving the $1s_{1/2}$ orbit were evaluated with $V_{MU}$. The modified USD Hamiltonians (including wb-USD, wb-USDA, and wb-USDB, and we will not distinguish them hereafter) reproduced well the large energy differences between the nuclei $^{18}$Ne, $^{19}$Na and $^{23}$Al and their mirror partners. As an illustration, Fig. 7 shows the energy spectra of $^{18}$O and $^{18}$Ne calculated with USD Hamiltonians. It can be seen that the modified Hamiltonians give a nice description of proton-rich nucleus $^{18}$Ne.

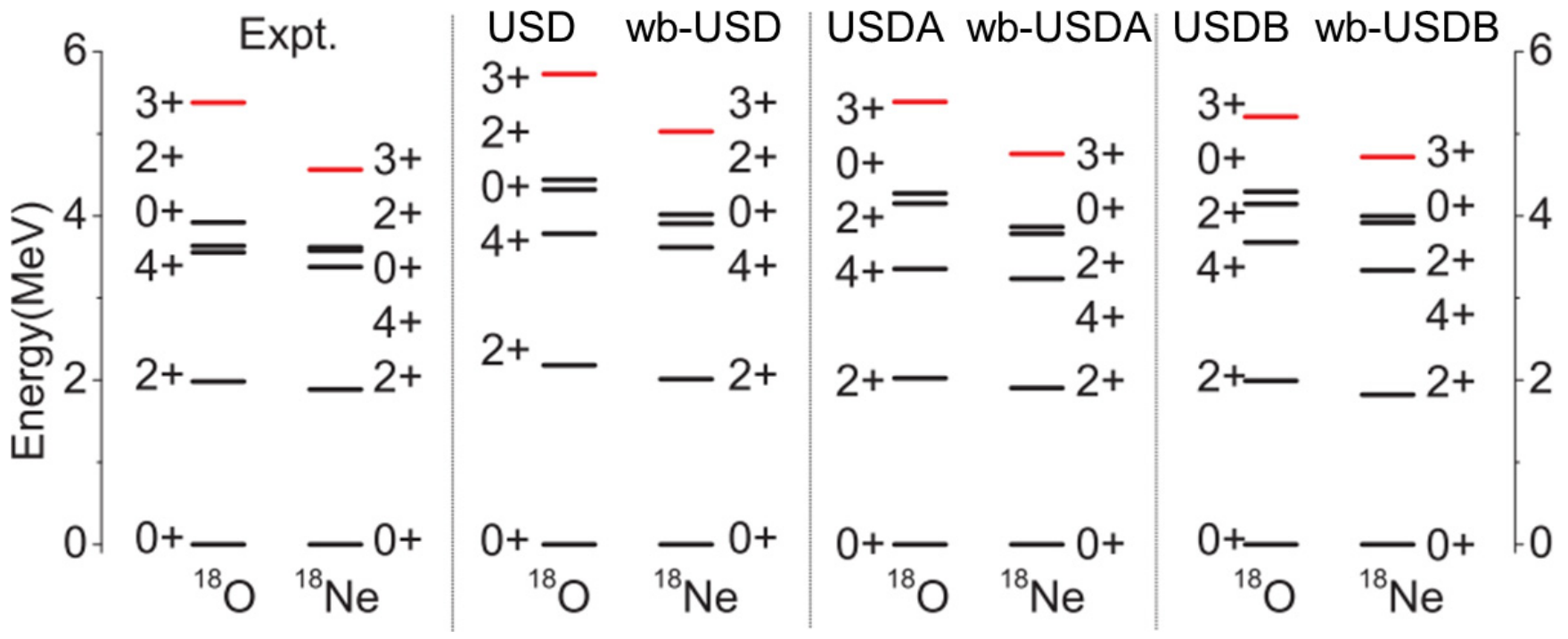

Fig. 7. From Ref. 15. Energy spectra of mirror nuclei $^{18}$O and $^{18}$Ne derived with different Hamiltonians. Experimental data are from Ref. 88.

The wb-USD type Hamiltonians also well reproduced the noticeable difference between the Gamow-Teller strength of $\beta^+$ decay ($B(\mathrm{GT}^+)$) from $^{24}$Si and that of $\beta^-$ decay ($B(\mathrm{GT}^-)$) from $^{24}$Ne, with the $B(\mathrm{GT}^+)/B(\mathrm{GT}^-)$ ratio of the $1_1^+$ state in $^{24}$Si and $^{24}$Ne measured to be 0.78 (11).[122]

Besides, the CISM with wb-USD predicted the spins of the ground states of $^{21}$Al and $^{22}$Al to be $5/2^+$ and $4^+$, respectively. The ground-state spin value of $^{22}$Al has been long debated because some $\beta$ decay experiments[126–128] of $^{22}$Al favored $4^+$ while others[123,129] supported $3^+$. As deeper investigation on $^{22}$Al, $4^+$ has been more adopted,[88,125] many other important but veiled experimental data exist in the *sd*-shell proton-rich nuclei. A series of experimental data[125,130–137] focusing on such nuclei and their $\beta$-decay spectroscopy have been reported by the RIBLL collaboration in the last years. CISM plays an important role in interpreting the isospin symmetry and the breaking evidenced by experimental discoveries. In the following of this subsection, we shed light on the latest progress made by the RIBLL[138] collaboration, especially those revealing the weakly-bound effect.

Globally, the RIBLL collaboration has focused on the $\beta^+$ decay experiments of $^{22}\mathrm{Si} \to {}^{22}\mathrm{Al}$,[130,132] $^{26}\mathrm{P} \to {}^{26}\mathrm{Si}$[133,135,137] and $^{27}\mathrm{S} \to {}^{27}\mathrm{P}$.[131,134,136] Both the $\beta p$[131–133] and $\beta 2p$[130,136,137] emissions have been observed in all three transitions. The (partial) $\beta$-decay schemes derived with the RIBLL experiments are shown in Figs. 8–10, providing fruitful information about mirror asymmetry. The CISM considering the weakly-bound effect has well interpreted the newly-observed and more detailed decay information globally. As illustrated in Fig. 10, the level scheme of $^{27}$P populated in $^{27}$S decay was generally reproduced by CISM calculations based on wb-USD. Both the levels below 2 MeV and the IAS above 12 MeV of $^{27}$P were well identified by CISM.

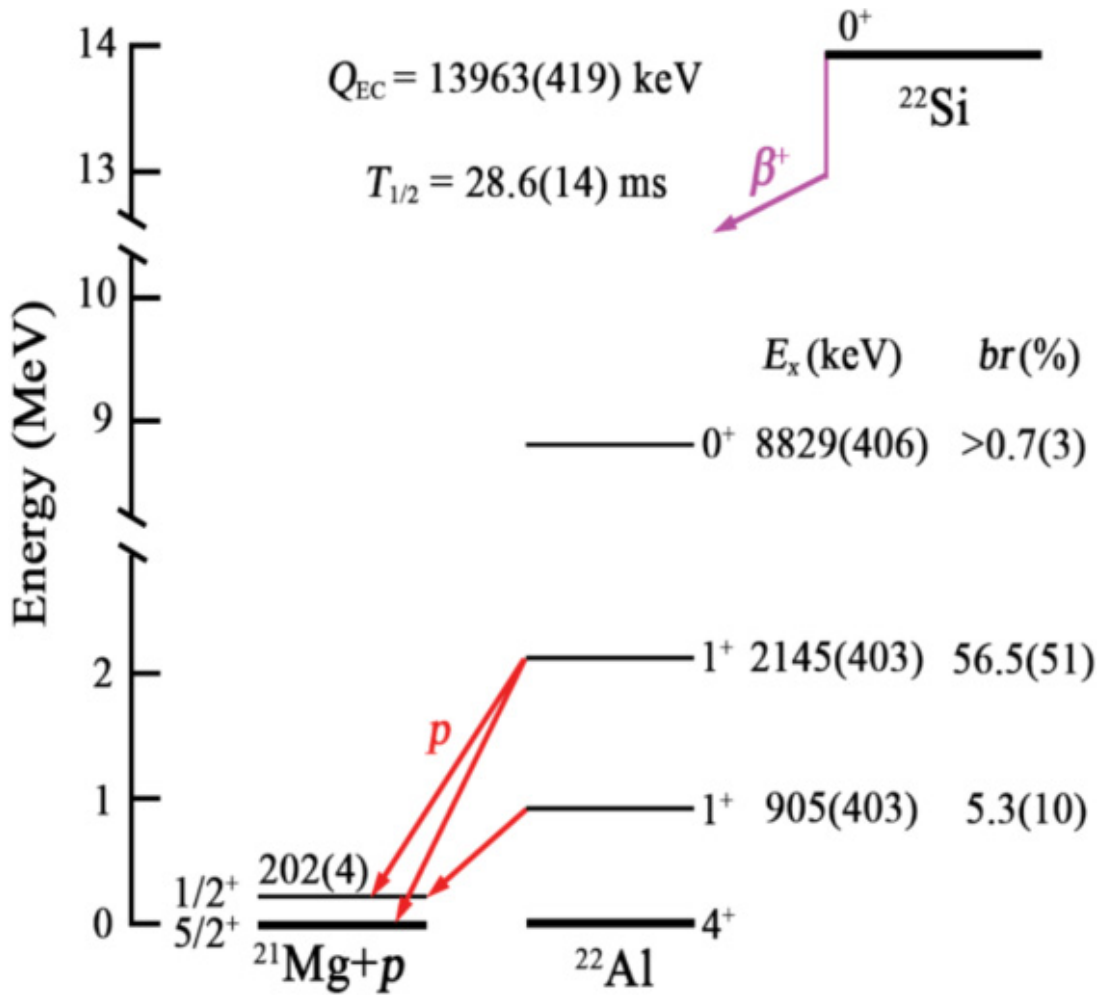

Fig. 8. From Ref. 132. Partial decay scheme of $^{22}$Si. The transitions to the two $1^+$ states are followed by one-proton emissions, while the transition to the $0^+$ state is followed by a two-proton emission.

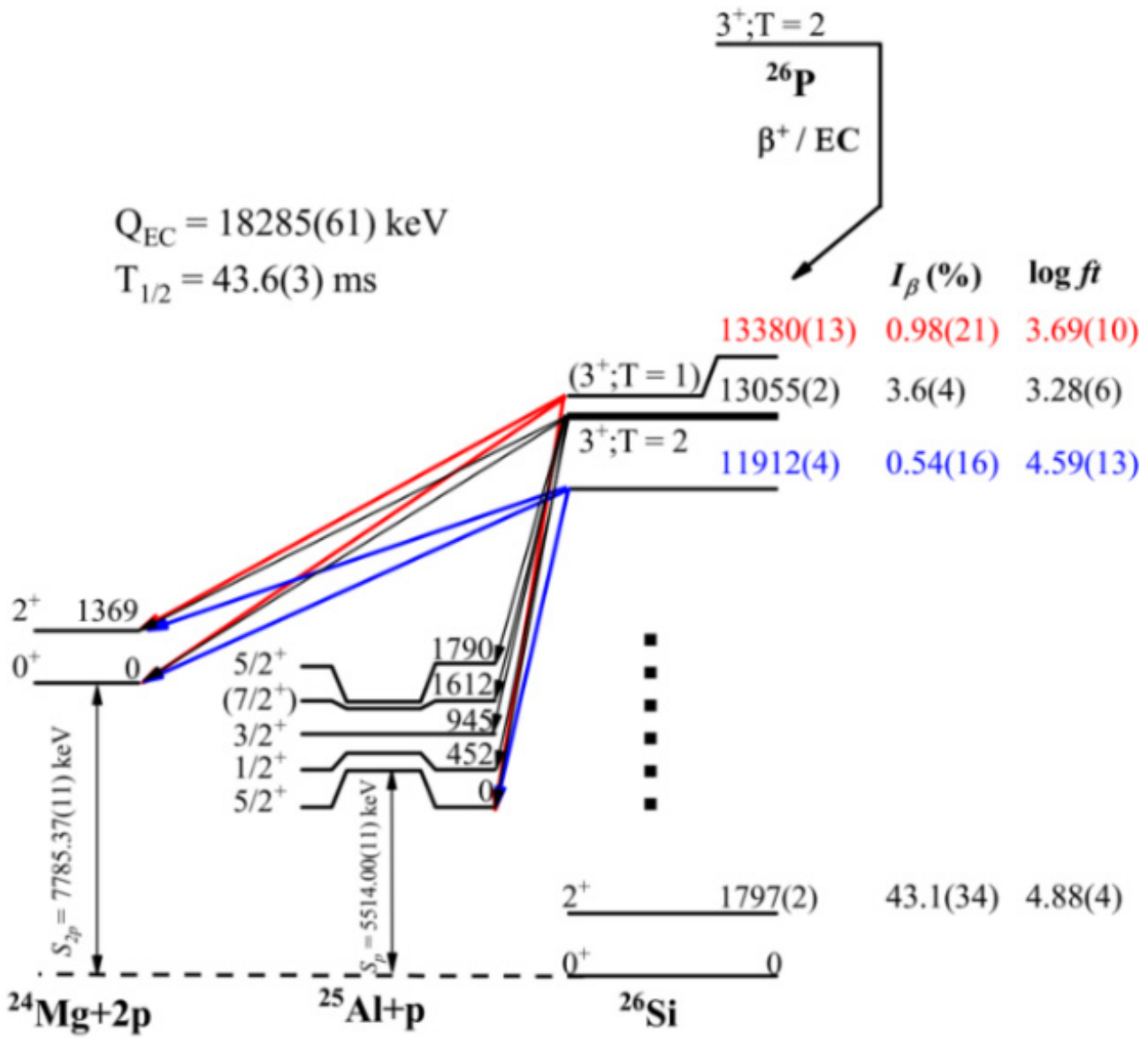

Fig. 9. From Ref. 137. Partial decay scheme of $^{26}$P. The separation energies are adopted from AME2020.[139] The branching ratios and log $ft$ values of $2_1^+$ in $^{26}$Si are from Ref. 135. Other data are experimental results of Ref. 137.

Importantly, the two-proton emission via the IAS has been identified in all three cases,[130,136,137] with the ones for $^{22}\text{Si} \to {}^{22}\text{Al}$ and $^{26}\text{P} \to {}^{26}\text{Si}$ shown in the partial decay schemes. In other words, the isospin mixing has been evidenced in these cases. CISM has successfully reproduced the excitation energy and isospin mixing of the IAS. For $^{22}$Al, the theoretical excitation energy of the IAS state ($0^+$) with wb-USD was 9020 keV,[130] in nice agreement with the measured 8829 (406) keV. For $^{26}$Si,[137] CISM calculations generally confirmed the observed transition properties. Especially, the doublet consisting of a high-lying $T = 2$, $J^\pi = 3^+$ IAS and its neighboring $3^+$ state in the daughter nucleus $^{26}$P were highlighted because of stronger mixing, higher excitation energy and larger level spacing than all the previous doublets observed in $\beta$-decay experiments. The CISM with wb-USD nicely reproduced such a strong isospin-mixing, interpreting the significant contribution of the weakly bound effect to the isospin-symmetry breaking.

A more direct way to investigate the isospin symmetry breaking is by comparing the mirror processes. In Table 3, we summarized the mirror transitions where the mirror asymmetry parameter $\delta$ ($\equiv ft^+/ft^- - 1$, where $ft^\pm$ represents the reduced transition probability) is significant. The weakly-bound effect of the $\pi 1s_{1/2}$ orbit contributes significantly to the mirror asymmetry. For the $A = 22$ isobars,[132] the $\delta$ value in the transition to the $1_1^+$ state was found to be 209(96)%, which was the largest measured in the low-lying states at the publication time. CISM confirmed this large mirror asymmetry value and further predicted the proton-halo structure of $^{22}$Al because of its high occupancy of the $\pi 1s_{1/2}$ orbit. For $A = 26$,[135] the $\delta$ value for the $2_1^+$ state was nicely reproduced only after considering the weakly-bound effect in both the Hamiltonian and the wave-function overlap. For $A = 27$,[131] eight transitions where the $\delta$ value is absolutely larger than 35% were identified with the help of CISM. The comparison also showed that the MEDs between $^{27}$Mg and $^{27}$P were up to 0.3 MeV.

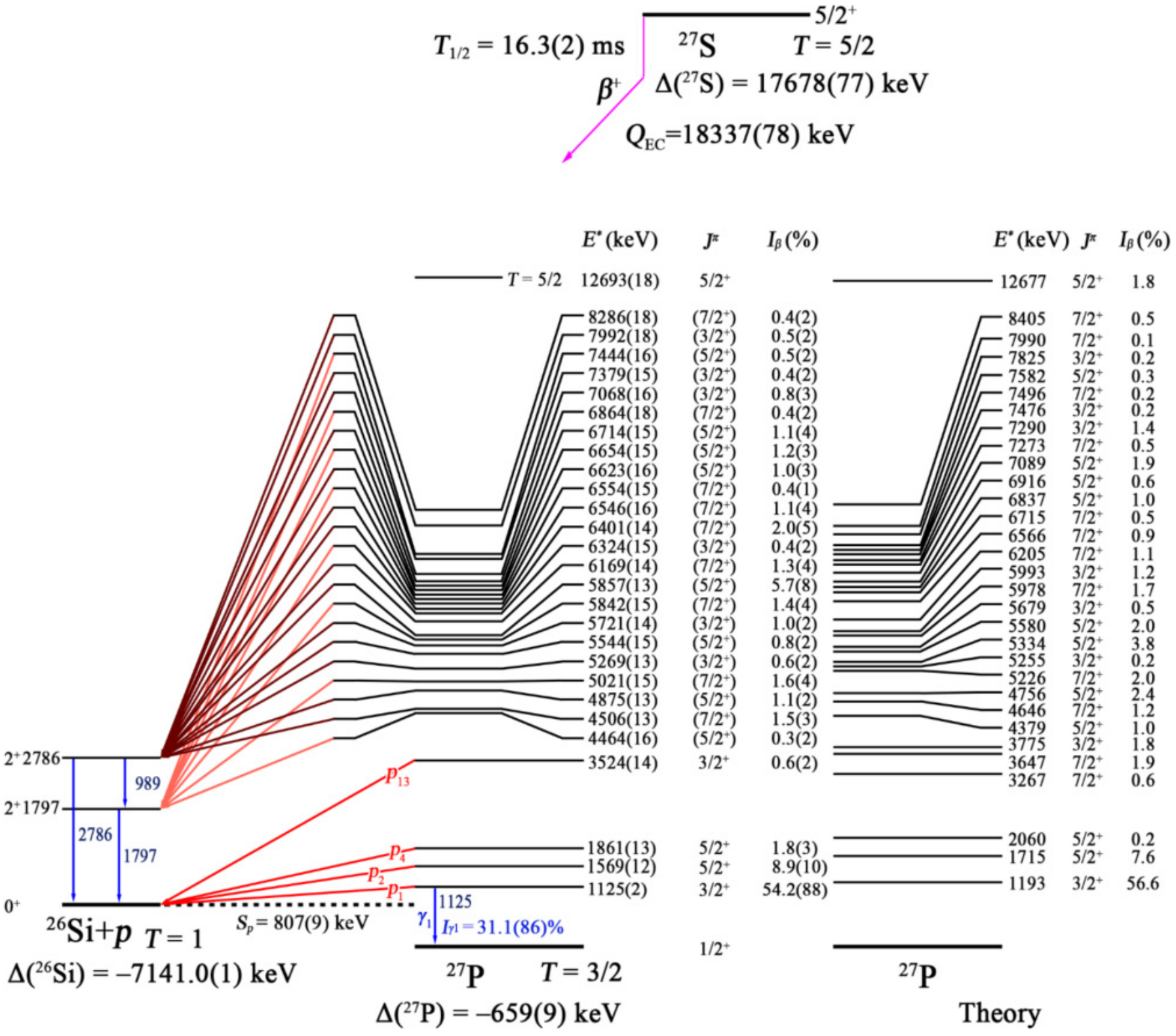

Fig. 10. From Ref. 131. Experimental decay scheme of $^{27}$S. The experimental levels of $^{27}$P populated in $^{27}$S $\beta$ decay are compared with the theoretical results derived from the CISM using wb-USD. The mass excess of $^{26}$Si is from AME2016.[140]

Also, one should notice that the combination of experimental information of the mirror nuclei and the CISM prediction is useful to the spin-assignment. As shown in the decay scheme of $^{27}$S, the populated levels are of high density and have a negligible $\beta$-decay branching ratio ($I_\beta$). With well-benchmarked Hamiltonians, the theoretical prediction of spin may be reliable. It is worth noting that the spin assignment of some states is crucial to nuclear astrophysics. One of the recent studies by the RIBLL collaboration focused on the $\beta$-decay of $^{26}$P to $^{26}$Si. It investigated the spin-assignment of the important state in $^{26}$Si at around 5.8 MeV, which was very controversial but crucial to the total $^{25}$Al $(p, \gamma)$ $^{26}$Si reaction rate in explosive hydrogen burning environments.[133] In this experimental study, $\beta$-delayed proton and $\gamma$ emission of $^{26}$P have been simultaneously measured to make the derived data more reliable.

Table 3. Comparison of excitation energies and $ft$ values of mirror pairs. The excitation energy is listed for the daughter nucleus. Data for the proton-rich nuclei are derived from recent experiments[131,132,135] by the RIBLL Collaboration, while those for the neutron-rich side are from published nuclear data sheets.[141–143] For the $A = 27$ nuclei, the underlined spins are not certainly determined[143] but favored by Ref. 131. Theoretical $\delta$ values are calculated with USD interaction but considering the isospin-non conserving force ($A = 22$, reported in Ref. 132) or the weakly bound effect ($A = 26$, reported in Ref. 135).

| | Ex (MeV) | $\log(ft^+)$ | Ex(MeV) | $\log(ft^-)$ | $\delta_{\mathrm{exp}}(\%)$ | $\delta_{\mathrm{th}}(\%)$ |
|---|---|---|---|---|---|---|
| | $^{22}$Si $\to$ $^{22}$Al[132] | | $^{22}$O $\to$ $^{22}$F[141] | | | |
| $1^+_1$ | 0.905(403) | 5.09(9) | 1.625 | 4.6(1) | 209(96) | 212 |
| $1^+_2$ | 2.145(403) | 3.83(5) | 2.572 | 3.8(1) | 7(28) | −3.4 |
| | $^{26}$P $\to$ $^{26}$Si[135] | | $^{26}$Na $\to$ $^{26}$Mg[142] | | | |
| $2^+_1$ | 1796.1(2) | 4.88(4) | 1808.81(16) | 4.7148(12) | 46(13) | 55 |
| $3^+_1$ | 3756.5(3) | 5.78(10) | 3941.48(17) | 5.870(14) | −19(19) | |
| $2^+_3$ | 4138.4(9) | 5.46(9) | 4332.02(17) | 5.62(1) | −31(15) | |
| $3^+_2$ | 4185.4(11) | 5.6(2) | 4350.02(17) | 5.33(1) | 86(86) | |
| | $^{27}$S $\to$ $^{27}$P[131] | | $^{27}$Na $\to$ $^{27}$Mg[143] | | | |
| $3/2^+$ | 1125(2) | 4.44(8) | 984.69(8) | 4.300 | 38(26) | |
| $5/2^+$ | 1569(12) | 5.16(5) | 1698.06(10) | 4.99 | 48(18) | |
| $5/2^+$ | 1861(13) | 5.82(8) | 1940.06(9) | 6.3 | −67(7) | |
| $3/2^+$ | 3524(14) | 6.04(14) | 3490.9(4) | 5.76 | 91(62) | |
| $5/2^+$ | 4464(16) | 6.24(21) | 4150.0(5) | 6.81 | −73(14) | |
| $5/2^+$ | 4875(13) | 5.59(8) | 4553.0(6) | 5.82 | −41(11) | |
| $7/2^+$ | 4506(13) | 5.50(7) | 4776.3(7) | 5.75 | −44(10) | |
| $5/2^+$ | 5021(15) | 5.39(9) | 4992.6(9) | 5.59 | −37(14) | |

The CISM with the wb-USD Hamiltonians well reproduced the determined levels and suggested the observed 5945.9(40) keV state to be $4^+$, which agreed with mirror analysis. Finally, the total $^{25}$Al $(p, \gamma)$ $^{26}$Si reaction rate, calculated with the newly-observed data and wb-USD, was consistent with previous studies.

The supplemented experimental data in neutron-rich nuclei may be helpful in investigating the rapid proton capture process. In another RIBLL report, the mass excess of $^{27}$P ($^{27}$S) has been determined (constrained) through $\beta$ decay, indicating that $^{27}$S is not a significant waiting point in the rapid proton capture process.[134]

In summary, CISM calculations have succeeded in describing the light proton-rich nuclei and their difference from the mirror pair. Considering the weakly-bound effect of the $1s_{1/2}$ orbit, modified USD Hamiltonians have been successfully applied to interpret the latest experimental progress achieved by the RIBLL collaboration, mainly concerning the $^{22}$Si $\to$ $^{22}$Al, $^{26}$P $\to$ $^{26}$Si and $^{27}$S $\to$ $^{27}$P decays.

### 3.1.3. *New isomers and new nuclides in medium and heavy mass region*

In the medium and heavy mass regions, both the isotopic and the isotonic chains are much longer than those in the light mass region. The systematics is shown in the nuclei with the same proton (neutron) number. Figure 11 illustrates a typical example of the systematic variation of the excitation energy of the $2_1^+$ states ($E(2_1^+)$) in the even–even nuclei in the long isotopic chain near $Z = 50$. A recent $\beta$-decay experiment of $^{140}$Te[144] indicated that its $E(2_1^+)$ and $E(4_1^+)$ follow the systematic trend shown by the lighter isotopes. The $\beta$-decay branching ratio to the low-lying $1^+$ states for odd-odd nuclei gradually decreases from $^{136}$I to $^{140}$I.[145] Besides, the $\alpha$-decay half-life is believed to be a simple function of a few variables, such as the proton and neutron number.[146–148]

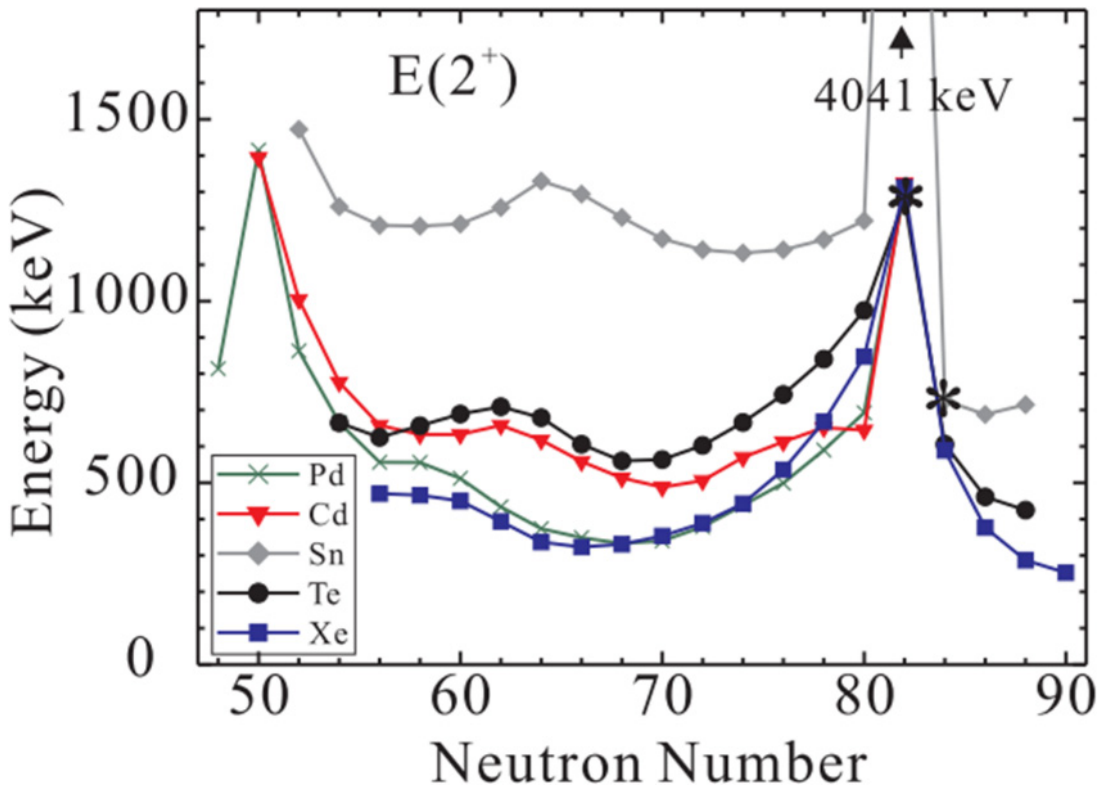

Fig. 11. From Ref. 144. The excitation energy of the $2_1^+$ state in Pd, Cd, Sn, Te and Xe isotopes. Experimental data are from Refs. 88, 144, 149–151.

The isomerism has been systematically found near the doubly magic nuclei of medium and heavy mass. The appearance of an isomeric state may indicate the narrow gap between two adjacent orbits or the different coupling of nucleons inside one orbit. Properties of the isomeric states of isotopes or isotones are usually similar. As the shell gradually evolves, they follow notable trends. For instance, the $1/2^-$ isomer systematically emerges in the odd-$A$ nuclei of the southwest vicinity of $^{100}$Sn. The excitation energy of the $1/2^-$ isomer increases in the isotopic (isotonic) chain when the neutron (proton) number approaches 50.[152] Moreover, the $13/2^+$ isomeric state can be found in $N = 123$ odd-$A$ isotones, including $^{207}$Po, $^{209}$Rn, $^{211}$Ra and $^{213}$Th.[153] The half-life of such $13/2^+$ isomer varies systematically with mass number. Shell evolutions are revealed by systematic trends.

Numerous isomers have been newly observed,[152–158] and the systematics study on isomerism has been more complete. A well-developed Hamiltonian in sufficiently-large model space is crucial in the CISM framework to accurately interpret a new isomer. This work will shed light on the effective Hamiltonians based on the $V_{MU}$ force plus the M3Y spin-orbit force ($V_{MU} + LS$), the CD-Bonn potential renormalized by the G-matrix approach, and the Kuo–Herling interactions,[78] applied in medium and heavy nuclei around $^{132}$Sn and $^{208}$Pb.

On the one hand, a series of effective Hamiltonians based on the interaction $V_{MU} + LS$ have been developed in the last decade. In practice, one derives the TBMEs with the $V_{MU} + LS$ interaction and fixes the SPEs with well-known experimental data such as the excitation energy. Such Hamiltonians have been rather promising in interpreting the systematic isomerism and the shell evolutions in various nuclear regions, including the *jj*44 region around $^{100}$Sn,[152] the *jj*45 region around $^{100}$Sn[154] and $^{132}$Sn.[155,156]

For the *jj*44 region around $^{100}$Sn, the excitation energy of the low-lying $1/2^-$ isomers in $^{87}$Mo, $^{91}$Ru and $^{95}$Pd has been first measured, as reported in the latest paper.[152] The new experimental data complete the systematic study of the $1/2^-$ isomers in odd-$A$ nuclei of the southwest vicinity of $^{100}$Sn. Combining the existing observed results for the $1/2^-$ isomers in $N = 47$ and 49 isotones, it was found that the excitation energy increases with the proton number from $Z = 40$–46 and that the $1/2^-$ isomers lie higher in the isotopes with two more neutrons. Calculations with $V_{MU} + LS$ indicated that such a systematic trend originates from the tensor force. When a proton (neutron) pair is added to the $\pi(\nu)1g_{9/2}$ orbit, the repulsive tensor force between the $\pi 1g_{9/2}$ protons and the $\nu 1g_{9/2}$ neutrons will be enhanced. As a result, the gap between $\nu 1p_{1/2}$ and $\nu 1g_{9/2}$ will be enlarged, and the $1/2^-$ isomer has higher excitation energy.

For the *jj*45 region around $^{100}$Sn, the first precision mass measurements were performed for the $1/2^-$ isomeric and $9/2^+$ ground states in $^{101}$In.[154] It was found that the excitation energies of the $1/2^-$ isomer in odd $^{101-113}$In were similar, which illustrated the stable subshell gap of $Z = 40$. CISM calculations performing in the *jj*45 model space with the $V_{MU} + LS$ interaction reproduced such results well. The strong configuration mixing between the $\nu 2d_{5/2}$ and $\nu 1g_{7/2}$ orbits in the indium isotope can be seen in Fig. 12(e). The proton-neutron interactions between these two neutron orbits and two proton orbits, $\pi 2p_{1/2}$ and $\pi 1g_{9/2}$, keep a relatively stable $Z = 40$ subshell gap. Vice versa, the same proton-neutron interactions can induce configuration-dependent shell evolution (also called type-II shell evolution)[159] of the $\nu 2d_{5/2}$ and $\nu 1g_{7/2}$ orbits from the ground state to the isomeric states in these indium isotopes. Indium isotopes are ideal systems to see such an effect with one proton hole below the $Z = 50$ shell. As shown in Fig. 12(d), calculation results indicate that the gap between two neutron orbits varied from ground to isomeric states in the same indium isotope.

For the *jj*45 region near $^{132}$Sn, the first observations of the $1/2^-$ isomers in $^{123,125}$Ag[155] and a long-lived isomer in $^{127}$Ag[156] were reported. The Hamiltonian derived similarly as in Ref. 154 reproduced well the excitation spectra of $^{123,125}$Ag. The $Z = 40$ subshell gap in Ag was pointed out to be reduced as approaching $N = 82$, as illustrated in Fig. 12(c). According to CISM calculations, the tensor force was suggested to be one crucial cause of this shell evolution. Figure 12(a) illustrates that removing the tensor force will significantly shift the proton ESPEs in odd-$A$ silver. In addition, the increasing occupancy in the $\nu 1h_{11/2}$ orbit also contributes to the ordering inversions of the single-particle levels near $^{123}$Ag. More clearly, the rise of the neutron number in the $\nu 1h_{11/2}$ orbit lifted the level of the $\pi 1g_{9/2}$ orbit, which is shown in Fig. 12(b). Besides, a high-spin and long-lived isomer in $^{127}$Ag was first reported in Ref. 156. Similarly, CISM calculations achieved agreement with the observed energy levels of $^{127}$Ag. More importantly, the Gamow–Teller strengths of the reported isomer to states in $^{127}$Cd were well reproduced by the CISM, for which the configuration-dependent shell evolution was also found to account.

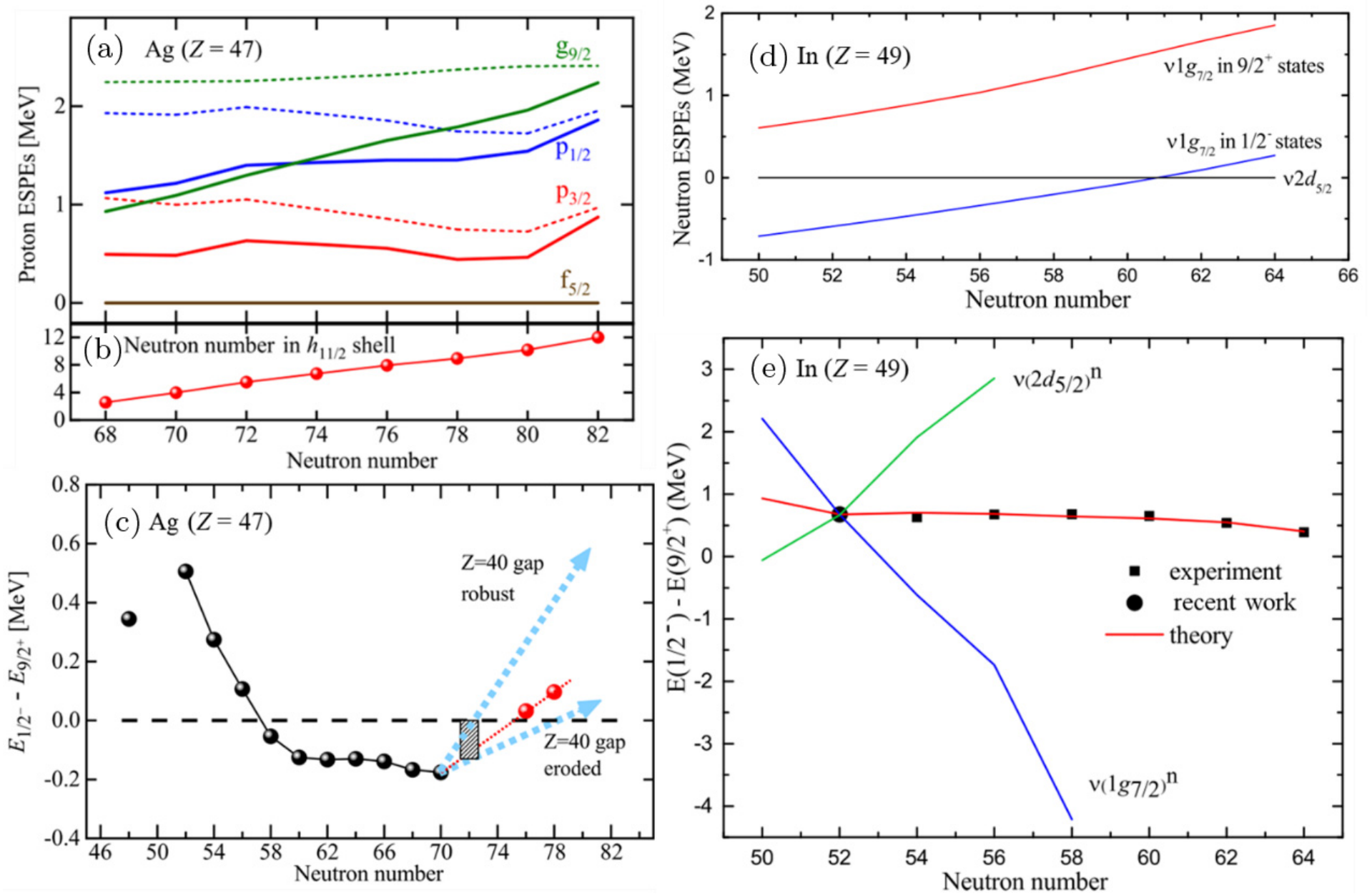

Fig. 12. (Color online) From Refs. 154 and 155. (a) Proton effective single-particle energies, (b) Neutron occupation number in the $\nu 1h_{11/2}$ orbit and (c) Energy difference between the $1/2^-_1$ and $9/2^+_1$ states in odd-$A$ silver isotopes. (d) Neutron effective single-particle energies of $\nu 1g_{7/2}$ versus $\nu 2d_{5/2}$ for $9/2^+$ ground states and $1/2^-$ isomers and (e) Excitation energies of the $1/2^-$ isomer in odd-$A$ indium isotopes. Experimental data are taken from Refs. 88, 154, 155 and 160. All the theoretical results are derived in the *jj*46 model space with $V_{MU}$ + LS, except that in (a), the dashed lines are obtained after removing the tensor force, and that in (e), the green (blue) line is obtained when neutrons occupy only the $\nu 2d_{5/2}$ ($\nu 1g_{7/2}$) orbit.

On the other hand, the $V_{MU}$ + LS has been applied as the cross-shell part of new Hamiltonians in the medium and heavy nuclear regions, similar to its application in the *psd* and *sdpf* regions.

In the southeastern vicinity of $^{132}$Sn,[161] a Hamiltonian in the *jj*46 model space was first constructed with the $V_{MU}$ + LS for the proton-neutron part and G-matrix CD-Bonn potential for the others. It should be noticed that the central force strength of $V_{MU}$ was set to be stronger in the *jj*46 model space than in the *psd* and *sdpf* regions. CISM calculations succeeded in reproducing one-neutron separation energies, $B(E2)$ values, and excitation energies of the isomeric states in observed isomers $^{134,136,138}$Sn, $^{130}$Cd and $^{128}$Pd, and predicted other candidate isomers in nuclei, including $^{135}$Sn, $^{131}$Cd, $^{129}$Pd, $^{132,134}$In and $^{130}$Ag. For illustration, Fig. 13 gives the energy levels in Sn isotopes, and Table 4 presents the prediction details of the possible isomers. One should notice that a $19/2^-$ state in $^{129}$Pd was predicted to be isomeric and a candidate for neutron radioactivity. Besides, a $5^-$ isomeric state was predicted in $^{134}$In, which was subsequently observed.[162]

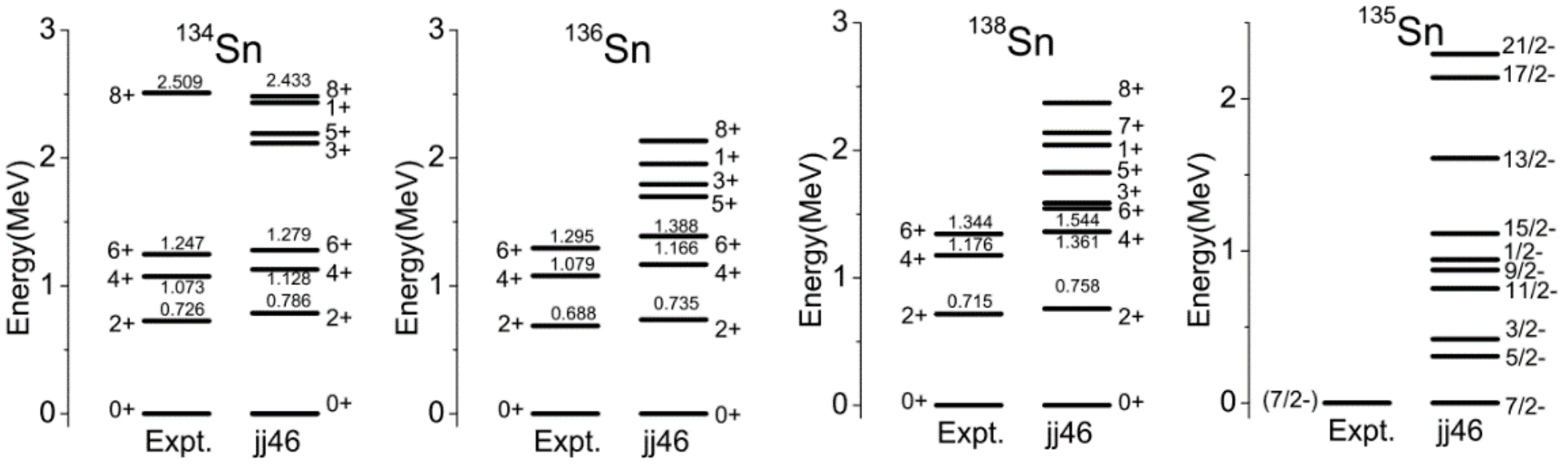

Fig. 13. From Ref. 162. Comparison of energy levels calculated with CISM and experimental results in Sn isotopes.

In the region "south" of $^{208}$Pb,[34] a Hamiltonian based on KHHE,[78] KHPE[79] and $V_{MU}$ + LS was constructed in the model space consisting of five proton orbits within $50 < Z \leq 82$ and 13 neutron orbits within $82 < N \leq 184$. The proton-proton interaction starting from KHHE was improved to describe well the binding energy of Pb, Tl, Hg, Au, Pt and Ir isotopes around $N = 126$. The constructed Hamiltonian nicely described the excitation energies and electromagnetic properties of nuclei in the south vicinity of $^{208}$Pb. Possible isomers have been proposed in such nuclear regions through the CISM calculation for nearly 100 nuclei, approaching the present limit of computational ability. The possible isomers with relatively small $E2$ transition probability are shown in Table 4.

Table 4. Prediction of possible isomers in the southeastern vicinity of $^{132}$Sn and the region "south" of $^{208}$Pb. $\delta E$, $B(E2)$ and $\tau_{1/2}$ are, respectively, the theoretical energy difference, reduced strength, and half-life of the $E2$ transition from the initial state $J_i^\pi$ to the final state $J_f^\pi$. Data are from Refs. 34 and 161. It should be noted that the $5^-$ isomer has been observed.[162]

| Nuclei | $J_i^\pi$ | $J_f^\pi$ | $\delta E$ (MeV) | $B(E2)$ ($e^2 fm^4$) | $\tau_{1/2}$ ($\mu s$) |
|---|---|---|---|---|---|
| $^{135}$Sn | $21/2^-$ | $17/2^-$ | 0.147 | 93.96 | 0.09 |
| $^{132}$In | $5^-$ | $7^-$ | 0.067 | 1.75 | 239.62 |
| $^{133}$In | $17/2^+$ | $13/2^+$ | 0.257 | 48.36 | 0.01 |
| $^{134}$In | $5^-$ | $7^-$ | 0.074 | 27.69 | 9.20 |
| $^{131}$Cd | $19/2^-$ | $15/2^-$ | 0.14 | 100.03 | 0.11 |
| $^{130}$Ag | $5^-$ | $7^-$ | 0.072 | 18.72 | 15.62 |
| $^{129}$Pd | $19/2^-$ | $15/2^-$ | 0.146 | 12.88 | 0.66 |
| $^{130}$Pd | $6^+$ | $4^+$ | 0.213 | 185.41 | 0.01 |
| $^{215}$Pb | $17/2^+$ | $13/2^+$ | 0.331 | 28.08 | 0.0047 |
| $^{215}$Pb | $21/2^+$ | $17/2^+$ | 0.119 | 31.44 | 0.18 |
| $^{213}$Pb | $17/2^+$ | $13/2^+$ | 0.294 | 33.46 | 0.0069 |
| $^{213}$Pb | $21/2^+$ | $17/2^+$ | 0.123 | 0.45 | 12.13 |
| $^{213}$Tl | $13/2^+$ | $9/2^+$ | 0.091 | 0.01 | 1812.75 |
| $^{213}$Tl | $17/2^+$ | $13/2^+$ | 0.047 | 0.03 | 444.55 |
| $^{212}$Tl | $11^+$ | $9^+$ | 0.114 | 2.27 | 2.93 |
| $^{211}$Tl | $13/2^+$ | $9/2^+$ | 0.088 | 66.27 | 0.14 |
| $^{211}$Tl | $17/2^+$ | $13/2^+$ | 0.049 | 18.66 | 0.65 |
| $^{210}$Tl | $11^+$ | $9^+$ | 0.106 | 140.76 | 0.053 |
| $^{210}$Hg | $6^+$ | $4^+$ | 0.100 | 69.26 | 0.12 |
| $^{210}$Hg | $8^+$ | $6^+$ | 0.061 | 28.03 | 0.45 |
| $^{209}$Hg | $17/2^+$ | $13/2^+$ | 0.380 | 433.65 | 0.00016 |
| $^{209}$Hg | $21/2^+$ | $17/2^+$ | 0.121 | 213.71 | 0.029 |

For the region "north" of $^{208}$Pb, new isomers and new nuclides have been discovered and should be emphasized here. New isomers in the nuclei such as $^{213}$Th[153] and $^{218}$Pa[33] and dozens of new nuclides, including $^{204,205}$Ac,[163,164] $^{207}$Th,[165] $^{214–216}$U,[166–168] $^{220,223,224}$Np,[169–171] etc., have been discovered in the last decade. The discovery of new nuclides and new elements has been a long-time major scientific problem. While the synthesis of new nuclides challenges the experimental setup and techniques, each extension to the known nuclear chart is a benchmark of the theoretical models. Nevertheless, there are few available Hamiltonians for those heavy nuclei. For nuclei with $82 < Z \leq 126$ and $126 < N \leq 184$, KHPE has been a commonly used one. For those lying on the "northwest" of $^{208}$Pb, the combination of the KHPE (proton-proton part), KHHE (neutron-neutron part), and $\mathrm{V_{MU}} + \mathrm{LS}$ (proton-neutron part) interactions is very helpful.

In the nuclear region with $Z > 82$ and $N \geq 126$, the CISM calculation with KHPE was helpful in understanding the newly-discovered states with shell evolutions and proposing the spin assignment. With the report of new isomeric states of $^{212}$Rn in Ref. 172, the excitation spectra theoretically established by the CISM agreed globally well with the new experimental level scheme of the semi-magic nucleus $^{212}$Rn. Particularly, the new state measured at 2121 keV was suggested to be a $3^{(-)}$ collective state, suggesting the core excitation from $^{208}$Pb. An $\alpha$-decaying isomer lying only 83(6) keV above the ground state was newly found in $^{218}$Pa.[33] CISM calculations suggested that the spin and parity of this isomer and the ground state should be $1^-$ and $8^-$, respectively. For lighter even mass $N = 127$ isotones, the first $1^-$ state tended to be the ground state. In fact, the excitation energy of the $8^-$ isomer decreases from $^{210}$Bi to $^{216}$Ac. Thus, the evolution of isomeric states in $N = 127$ isotones was suggested because the proton-neutron interaction changed from particle-particle to particle-hole type.[28] Moreover, the ground state of the short-lived isotope $^{223}$Np[170] was proposed to be a $9/2^-$ state considering the observed $\alpha$-decay width and corresponding CISM calculations. While the single-proton separation energy of $^{223}$Np was shown to follow the systematical trend of its lighter isotones, the $Z = 92$ shell closure was disapproved near $N = 126$.

In the nuclear region with $82 < Z$, $N \leq 126$, the Hamiltonian based on KHPE, KHHE and $\mathrm{V_{MU}} + \mathrm{LS}$ was used to interpret the intriguing $\alpha$-decay properties revealed in the region around the new nuclides. The latest experiments showed that the $\alpha$-decay reduced width $\delta^2$ was abnormally enhanced in $^{214,216}$U. The strong proton-neutron interaction between $\pi 2f_{7/2}$ and $\nu 2f_{5/2}$ orbits was suggested to explain the more substantial $\alpha$ formation probability in light uranium isotopes,[168] illustrated in Fig. 14(a). Along with the discovery of $^{207}$Th, a systematic study shows the regular and distinct odd–even staggering of $\alpha$-decay energy $Q_\alpha$ in the northwest of $^{208}$Pb, which can be seen in Figs. 14(b) and 14(c). CISM reproduced such odd–even staggering by considering the configuration mixing of nucleons lying on higher orbitals beyond the fermi surface.[165]

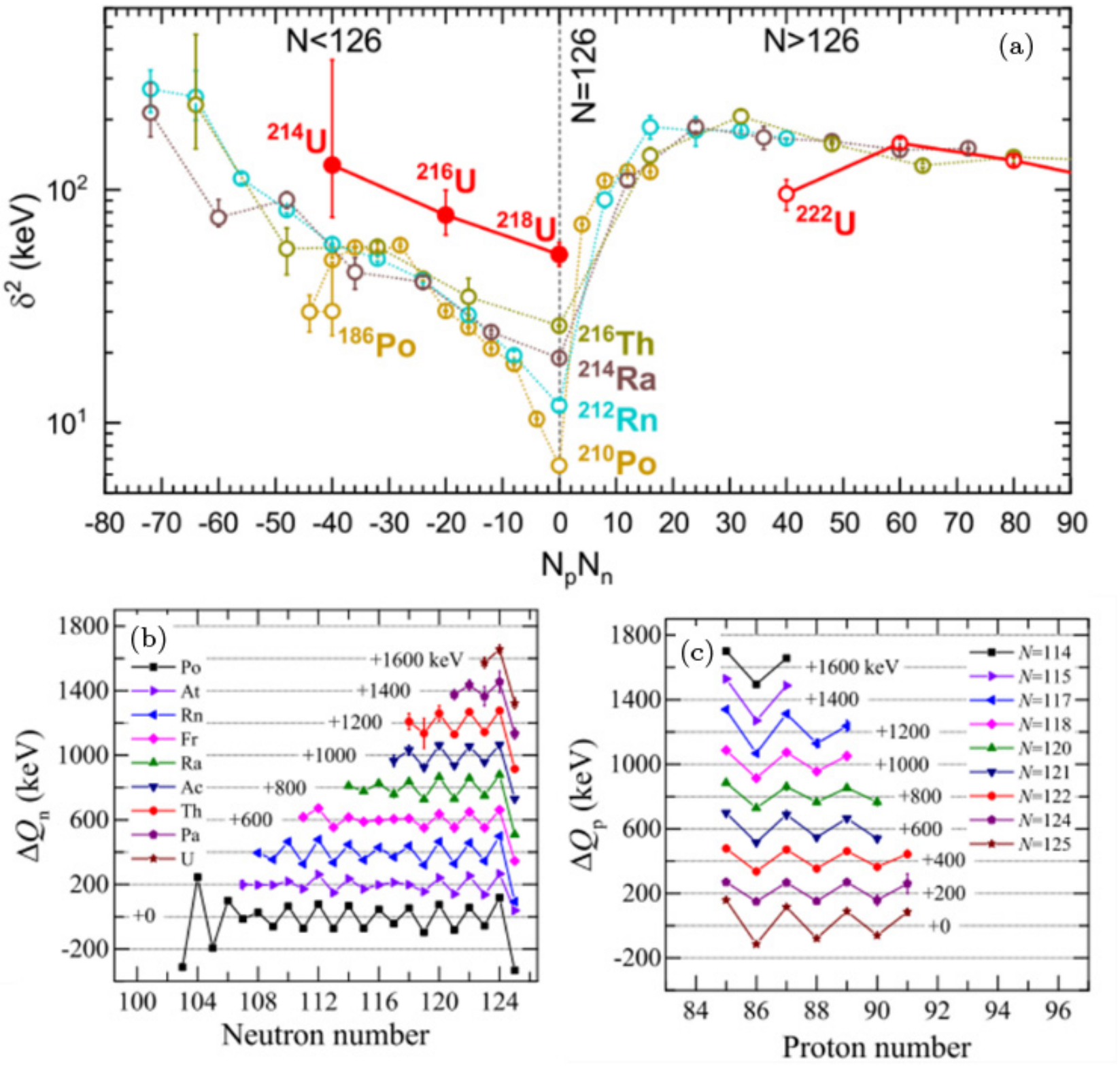

Fig. 14. From Refs. 165 and 168. Systematics of the $\alpha$-decay properties in (a) $82 < Z$, $N \leq 126$ nuclei, and (b) and (c) $Z > 82$, $N \leq 126$ nuclei. For details, $N_p = Z - 82$, $N_n = N - 126$, $\Delta Q_n(Z) = 1/2$ $(2Q_\alpha(Z) - Q_\alpha(Z+1) - Q_\alpha(Z-1))$, $\Delta Q_p(N) = 1/2(2Q_\alpha(N) - Q_\alpha(N+1) - Q_\alpha(N-1))$. $\delta^2$ and $Q_\alpha$ represent, respectively, the reduced width and energy of $\alpha$-decay.

In conclusion, over the last few years, CISM has generally well reproduced the isomerism in the medium and heavy nuclear region, for which the configuration-dependent shell evolution has been highlighted.[28,154,156] It has been found that the proton-neutron interaction and configuration mixing are crucial to interpret the shell evolution reflected by the isomers.[28]

### 3.1.4. *Towards a unified nuclear interaction for medium and heavy mass region*

Nuclear structure data are deeply required both on nucleosynthesis[173] and nuclear engineering.[174–176,349,350] As one of the most important theoretical nuclear structure models, CISM is not frequently utilized in nuclear data studies, mainly for three reasons.[32] First, the computation consumption increases significantly with the valence nucleons in the model space, making it difficult to calculate medium and heavy nuclei far from the doubly magic nuclei. Second, the hypothesis of spherical symmetry in a limited model space constrains the ability of CISM to describe large deformations. Third, the requirement of high-precision nuclear data for the application is challenging. Above all, with the increasing computation ability, the investigation of CISM for precisely deriving various nuclear structure properties in medium and heavy nuclei is significant. For this purpose, it is expected that an effective nuclear force can be built to describe the nuclei in medium and heavy mass regions in a unified way based on CISM.[32]

At present, a systematic shell model study remains to be realized to provide detailed spectroscopic data for supporting the experimental study and promoting the understanding of nuclear force.[177] It seems to be explorable since CISM calculations have been carried out to study the spectroscopic properties of nuclei from light to heavy and from the stability line to the driplines.[178] On the one hand, CISM reproduced well the masses, excitation spectra, decay properties and many other structure data. On the other hand, analyzing the nuclear force provided a reasonable interpretation of exotic phenomena such as shell evolution. Considering the present computational ability, CISM is capable of calculating more than 1000 nuclei.

Therefore, we have recently employed an effective nuclear force $\mathrm{V_{MU} + LS}$ including the central, spin-orbit, and tensor forces to describe medium and heavy nuclei near $^{132}$Sn and $^{208}$Pb. In detail, $\mathrm{V_{MU} + LS}$ comprises the M3Y spin-orbit force and $\mathrm{V_{MU}}$, which consists of the Gaussian central force and the bare $\pi + \rho$ tensor force. As discussed in previous sections, $\mathrm{V_{MU} + LS}$ has been combined with other commonly used phenomenological interactions for constructing new and improved Hamiltonian in the *psd*,[25] *sdpf*,[71] *jj*46,[162] *jj*56 + *jj*57[34] and *jj*66[168] model spaces. It is a promising effective nuclear force for the unified construction of the Hamiltonians.

The primary investigation of the unified nuclear force starts from the central force, which is known to be typically stronger than the spin-orbit and tensor forces. According to the value of isospin and spin of two nucleons, the central force can be divided into four terms: C00 (isospin $T = 0$, spin $S = 0$), C01 (isospin $T = 0$, spin $S = 1$), C10 (isospin $T = 1$, spin $S = 0$) and C11 (isospin $T = 1$, spin $S = 1$). Previous studies in semi-magic nuclei gave strength intervals for proton-proton and neutron-neutron central forces.[32] In detail, Fig. 15 illustrates the root mean square (RMS) of the error of energy levels varying with C10 forces. According to Fig. 15, the strength of the proton-proton (neutron-neutron) central force is suggested to be 110–115% (100–105%) of the original strength proposed in Ref. 69.

Based on these studies, calculations were firstly performed for 220 states of semi-magic nuclei and 605 states of non-magic nuclei using proton-proton (neutron-neutron) interactions with 115% (105%) relative to the original strength. The overall calculations were in nice agreement with experimental observations, with the root mean square error (RMSE) in energy levels valuing approximately 0.2 MeV.

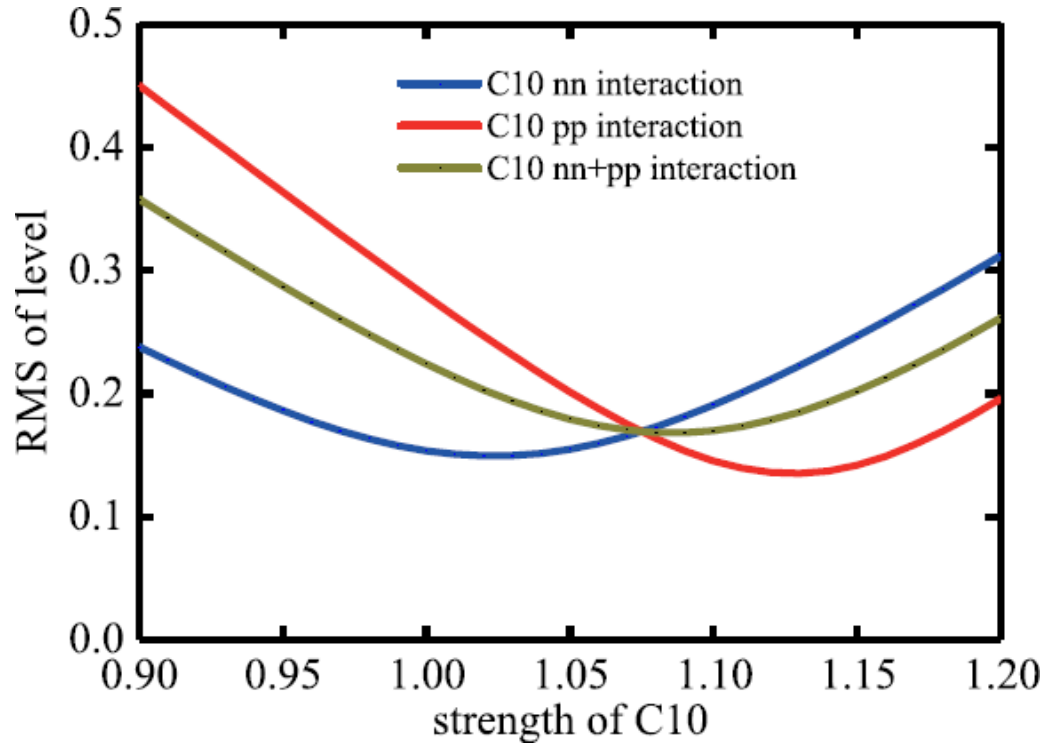

Fig. 15. From Ref. 32. RMSE of nearly 200 levels of semi-magic nuclei around $^{132}$Sn and $^{208}$Pb varying with the strength of C10.

However, for the energy levels of non-magic nuclei, the theoretical values for each region were always higher than the experimental values on average. One possible solution to further improve the calculation accuracy is weakening the proton-neutron interactions (referred to as C(pn)).

In a recent work, C(pn) was fully or partially attenuated. Figure 16 shows the RMSE of energy levels of 605 states in 116 non-magic nuclei varying as a function of C(pn) strength. The central force strengths of C01(pn) and C11(pn) channels need to be weakened to reduce the overall RMSE, while C10(pn) needs to be enhanced. The remaining C00(pn) affects little — the energy spectrum. The weakening of the central force for isospin 0 (1) is dominated by C01(pn) (C11(pn)). Lower RMSEs

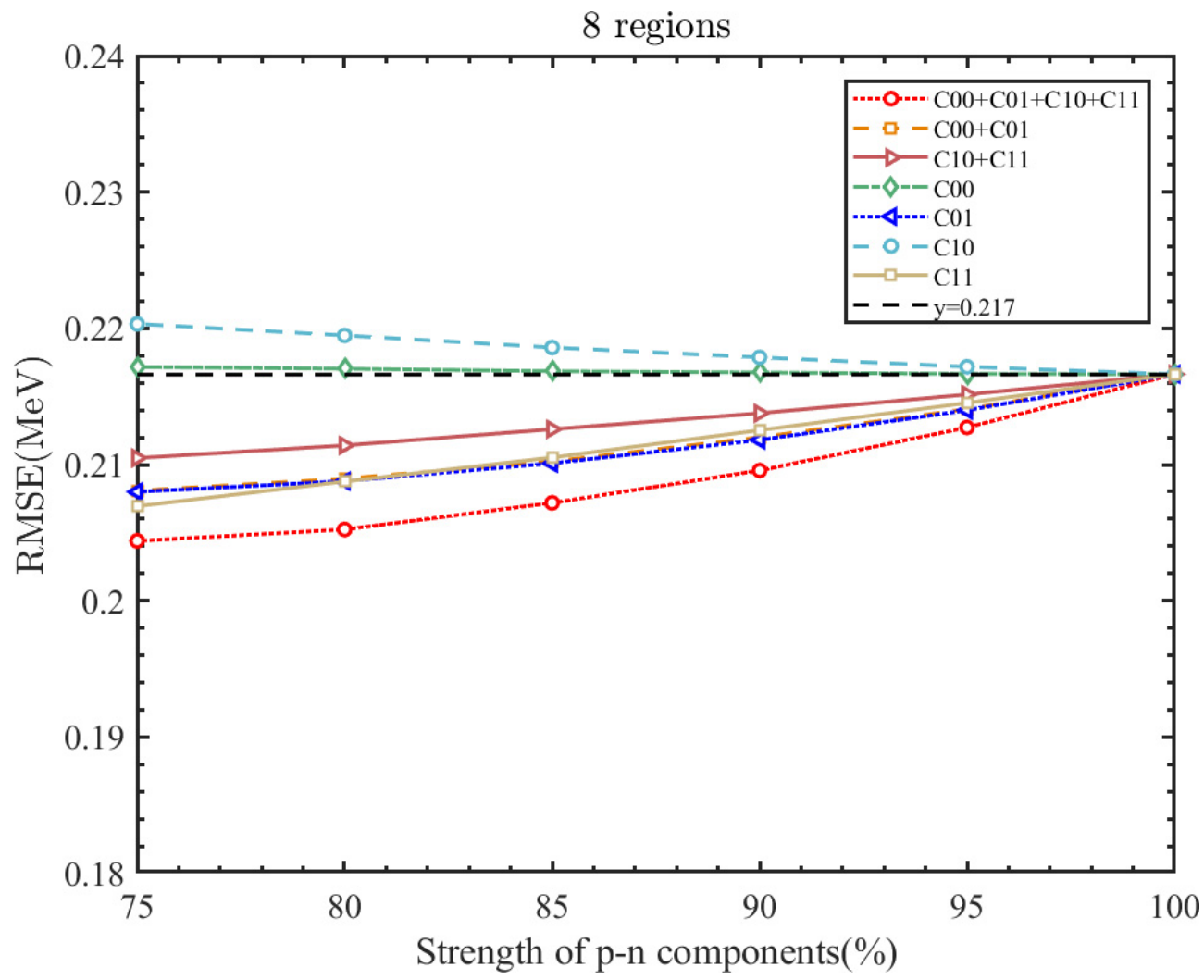

Fig. 16. RMSE of energy levels of 605 states in 116 non-magic nuclei varying as a function of C(pn) strength.

Table 5. Deviation from theoretical to experimental energy levels in nuclei near $^{132}$Sn and $^{208}$Pb, with the original proton-proton (neutron-neutron) central force enhanced by 15% (5%), and proton-neutron central force reduced by 25%. Nuclei in the region *jj*45-132 are calculated in the *jj*45 model space and located near $^{132}$Sn, while nuclei in the region *jj*67-208 are calculated in the *jj*67 model space and located near $^{208}$Pb, and so on.

| Region | Mean ($E_{\rm th}$-$E_{\rm ex}$) (MeV) | RMSE (MeV) | N_nuclei | N_states |
|---|---|---|---|---|
| *jj*45-132 | 0.022 | 0.205 | 14 | 70 |
| *jj*46-132 | 0.049 | 0.183 | 3 | 12 |
| *jj*55-132 | 0.093 | 0.186 | 27 | 155 |
| *jj*56-132 | 0.153 | 0.202 | 13 | 69 |
| *jj*56-208 | 0.240 | 0.359 | 18 | 77 |
| *jj*57-208 | 0.020 | 0.086 | 5 | 20 |
| *jj*66-208 | 0.025 | 0.101 | 21 | 120 |
| *jj*67-208 | 0.074 | 0.171 | 15 | 82 |
| West of $^{132}$Sn | −0.003 | 0.046 | 2 | 10 |
| East of $^{132}$Sn | 0.006 | 0.165 | 7 | 41 |
| South of $^{132}$Sn | −0.017 | 0.184 | 7 | 52 |
| North of $^{132}$Sn | 0.173 | 0.210 | 3 | 13 |
| West of $^{208}$Pb | 0.159 | 0.191 | 2 | 9 |
| East of $^{208}$Pb | 0.001 | 0.098 | 6 | 37 |
| South of $^{208}$Pb | −0.007 | 0.185 | 7 | 36 |
| North of $^{208}$Pb | 0.024 | 0.063 | 4 | 22 |
| Total | 0.071 | 0.193 | 154 | 825 |

could be obtained by weakening all C(pn), and the RMSE decreased by about 6% for a 25% weakening.

However, the reduction mentioned above is not considered significantly throughout weakening the proton-neutron central force from 100% to 75%. Due to the involvement of two kinds of nucleons, the proton-neutron interactions are more complex. When only one or several types of C(pn) with the same spin and isospin are weakened, it is difficult to significantly improve the description accuracy for numerous nuclei near $^{132}$Sn and $^{208}$Pb with the effective nuclear force in the $V_{\rm MUC}$ form. Table 5 provides the detailed deviation of 825 theoretical energy levels, obtained with reduced C(pn), compared with experimental data in Ref. 88. Deeper studies about each type of force are thus planned.

Finally, the uncertainty of the central force is of great interest, as the central force makes the most contributions. The uncertainty of theoretical calculations was investigated in many nuclear theory models, including the liquid drop model,[179] the energy density functional theory,[35] and the Weizsäcker–Skyrme mass model.[180] This kind of study was not performed until very recent years.[147] The uncertainty study of CISM is of great challenge and significance. Uncertainty analysis of the shell model in the medium-mass and proton-rich region has been done in Ref. 181. The authors gave recommended formulas to calculate the binding energy and other properties from the shell model energies for $Z, N = 30$–$50$ nuclei. They estimated the uncertainty within the bootstrap framework, as they did for existing $\alpha$ decay formulas in Refs. 147 and 351.

### 3.1.5. *Seniority conservation in the medium and heavy mass region*

Seniority (noted "$\nu$" in this work), regarded as the number of the particles that are not paired to zero spin ($J = 0$),[182] is a helpful quantum number to understand the nuclear structure and electromagnetic properties. The seniority symmetry has many interesting consequences.[183] For instance, the excitation energy of the $2_1^+$ state in nuclei of the same $j$-shell is constant. Besides, $B(E2)$ value with $\Delta\nu = 0$ in mid-shell nuclei vanishes, while the $B(E2; 0_1^+ \to 2_1^+)$ value presents a parabolic behavior, reaching a maximum at mid-shell. However, seniority is not always a good quantum number. It can be conserved for interactions with certain conditions, such as the pure pairing interaction. For arbitrary interactions, the seniority for a single-$j$ shell system is naturally conserved when $j \leq 7/2$. In practice, the generalized seniority[184] considering several $j$-orbits for one state is of interest. Progress has been made in the study of seniority and its generalization in recent years.

In Ref. 185, analytical energy expressions for the systems with 3 and 4 fermions in a single-$j$ shell were derived based on the decomposition of the angular momentum, making it easier to understand the isomers in single-$j$ shell nuclei. Meanwhile, a condition for seniority conservation in such systems was presented.

In recent years, partial seniority conservation has been better known for single-$j$ systems with $j \geq 9/2$. It has been found that, for instance, $\nu = 4$, $J = 4$ and $J = 6$ states exist and do not mix with other states in the $(9/2)^4$ system, i.e., the system with four particles (holes) in the $j = 9/2$ shell, no matter what are the interactions.[186] These special states are often referred to as "solvable" states. The partial seniority conservation in the $j = 9/2$ shell has been analytically proven in previous works.[187,188] In Ref. 189, the authors focused on the solvable states of single-$j$ systems with $j$ up to $15/2$ and derived analytic expressions for the eigenvalues of those solvable states.

In practice, the special solvable $\nu = 4$, $J = 4$ and $J = 6$ states in the $j = 9/2$ shell have been helpful in understanding the structure and electromagnetic properties of semi-magic nuclei. In a $(9/2)^4$ system, these special $\nu = 4$, $J = 4$, 6 states are potentially as low-lying as $\nu = 2$, $J = 2$, 4, 6, 8 states. In addition, the $E2$ transitions between $\nu = 2$ states are expected to be relatively weak, while those between the special $\nu = 4$ states and $\nu = 2$ states are much stronger, as illustrated in Fig. 17. Following such rules, the dominant seniority configurations of the observed low-lying states can be illustrated with the experimental $B(E2)$ values. Therefore, the $B(E2)$ transitions of low-lying states in Ni isotopes and $N = 50$ isotones below $^{100}$Sn were analyzed in Refs. 190–193.

Firstly, the $8_1^+$ isomer in $^{94}$Ru[88] was well interpreted in Ref. 190. The $8_1^+$ state and its neighboring $6_1^+$ state are both calculated to be $\nu = 2$ states, which means a very weak $E2$ transition possibility between the two. In addition, the small $B(E2)$ strength for $6^+ \to 4^+$ transition and unexpected large $B(E2)$ strength for $4^+ \to 2^+$ transition in $^{94}$Ru were experimentally derived.[191,192] Such results were ascribed to the interference between the wavefunctions of the special $\nu = 4$ states and $\nu = 2$ states, either destructively or constructively,[192] as shown in Fig. 18. Calculations indicated that the dominant seniority of $4^+$ states varied abruptly for a certain narrow interval of the non-diagonal cross-orbital TBMEs. In that narrow interval, the special $\nu = 4$ states and $\nu = 2$ states mix strongly, which also could cause the sharp transition in the $N = 50$ isotones $^{96}$Pd and $^{94}$Ru. The experimental $B(E2)$ values for $4_1^+ \to 2_1^+$ and $2_1^+ \to 0_1^+$ transitions in $^{92}$Mo and $^{94}$Ru and for $6_1^+ \to 4_1^+$ and $4_1^+ \to 2_1^+$ transitions in $^{90}$Zr were newly reported in Ref. 193, where the new experimental data and CISM calculations both supported the (approximated) seniority conservation in $N = 50$ nuclei within the $g_{9/2}$ shell.

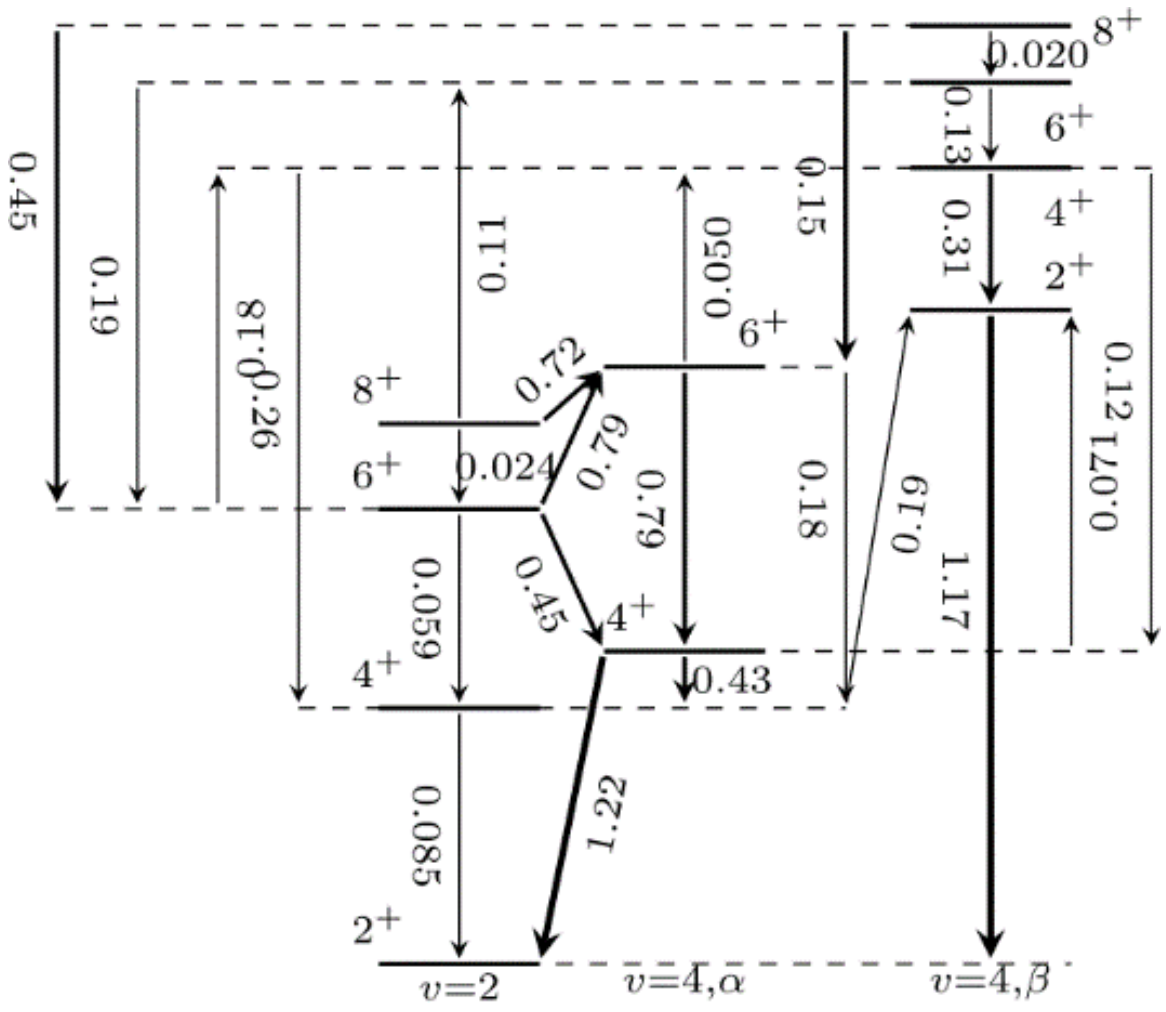

Fig. 17. From Ref. 190. The $E2$ transition strengths relative to the $B(E2; 2_1+ \to 0_1+)$ for the $(9/2)^4$ system calculated using a seniority-conserving interaction. The $\nu = 4$, $\alpha$ states are the two special states which exhibit partial dynamic symmetry, while the $\nu = 4$, $\beta$ states are the other non-solvable $\nu = 4$ states.

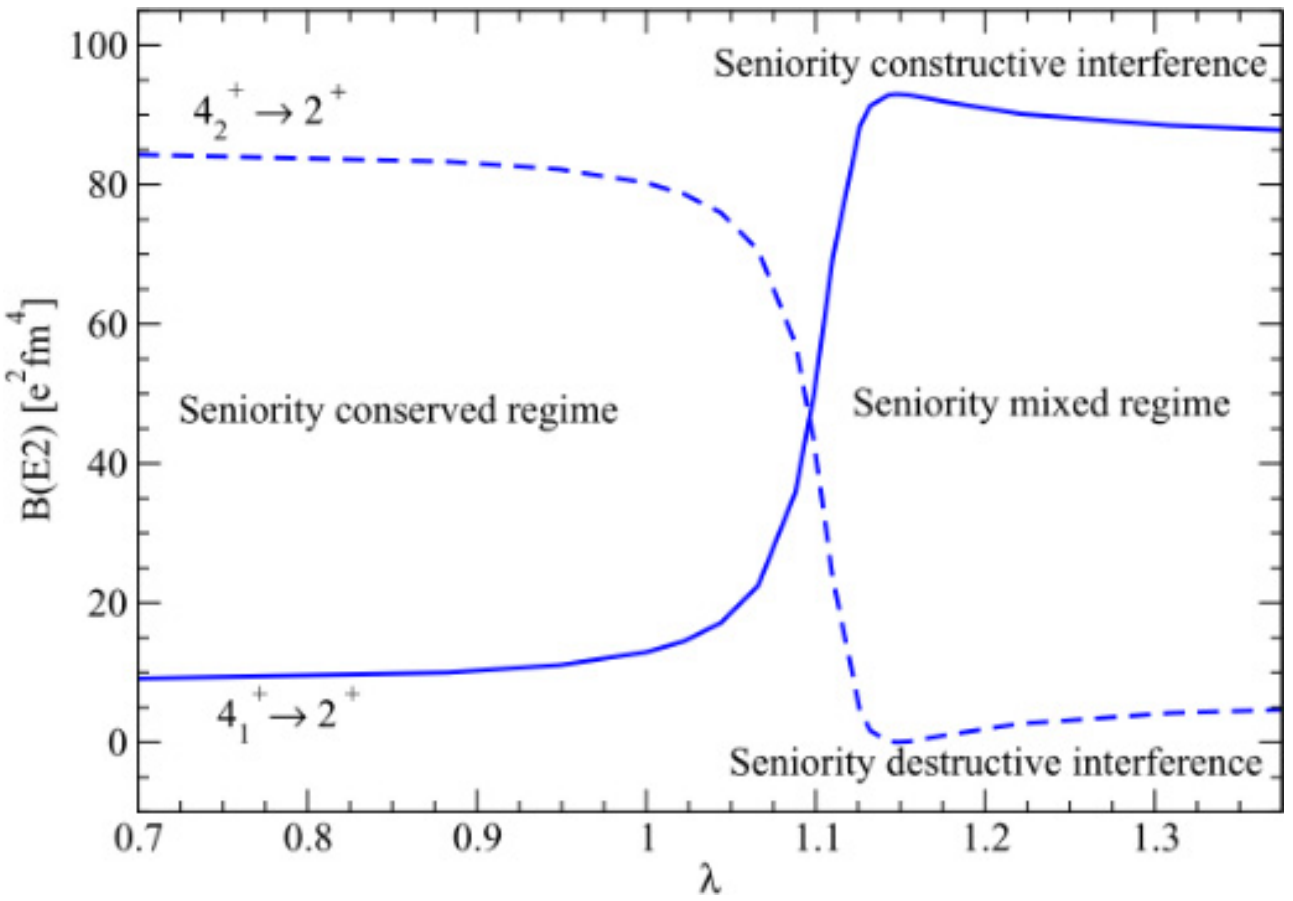

Fig. 18. From Ref. 192. The calculated $B(E2)$ strengths for $4^+ \to 2^+$ transitions in $^{94}$Ru varying with the strength parameter of non-diagonal TBMEs. The calculations were reached with the jun45 interaction in the *jj*45 model space in the CISM framework.

It should be noticed that Refs. 191 and 192 gave different theoretical explanations on the origin of seniority mixing, while the measured lifetimes of the $4_1^+$ states in $^{94}$Ru in Refs. 192 and 193 did not agree with each other. Therefore, investigations on the electromagnetic properties of the isotopes $^{56-78}$Ni and isotones $^{78}$Ni–$^{100}$Sn are anticipated to further study the seniority conservation in $(9/2)^n$ systems.

The Sn isotopes are another type of nuclei to investigate seniority conservation. As mentioned above, in the exact seniority conservation scheme, the excitation energy of the yrast state is constant, which is almost the truth for the $2_1^+$ state along the whole Sn isotopic chain. Moreover, the $B(E2; 0_1^+ \rightarrow 2_1^+)$ value in Sn isotopes is expected to vary parabolically when the neutron number increases from 52 to 80 and reaches the maximum in $N = 66$. However, the shallow minimum of $B(E2; 0_1^+ \rightarrow 2_1^+)$ was observed for $^{116}$Sn by measuring the lifetime of even-even $^{112-124}$Sn.[194] Disagreement can be seen in Refs. 194 and 195 on the question of whether such an anomaly is related to the reduced collectivity around $^{116}$Sn.[194] Reference 195 ascribed the reduced $B(E2; 0_1^+ \rightarrow 2_1^+)$ at mid-shell to the different rates of filling of orbits between light and heavy isotopes but not to the reduced collectivity, and suggested the conservation of the generalized seniority in Sn isotopes. The agreement between experimental results and the calculations within a generalized seniority scheme is shown in Fig. 19. Results also showed that the inversion of the neutron $d_{5/2}$ and $g_{7/2}$ orbits and the neutron excitation from the $g_{9/2}$ orbit together cause a small enhancement of the value in light Sn isotopes.[196]

A monopole-optimized effective interaction was proposed for light Sn isotopes in Ref. 197. In the model space involving the neutron orbits $g_{7/2}$, $d_{5/2}$, $d_{3/2}$, $s_{1/2}$ and $h_{11/2}$, unknown SPEs and the $T = 1$ monopole interactions starting from the CD-Bonn potential, were optimized by fitting the available binding energies of 157 states in $^{102-132}$Sn. Based on the optimized interaction, the $B(E2; 0^+_1 \to 2^+_1)$ strength of even $^{102-130}$Sn was systematically calculated in Ref. 196. As shown in Fig. 20, the calculations globally well reproduced the experimental results after considering the isospin dependence of the effective charge proposed by Bohr and Mottelson.[198]

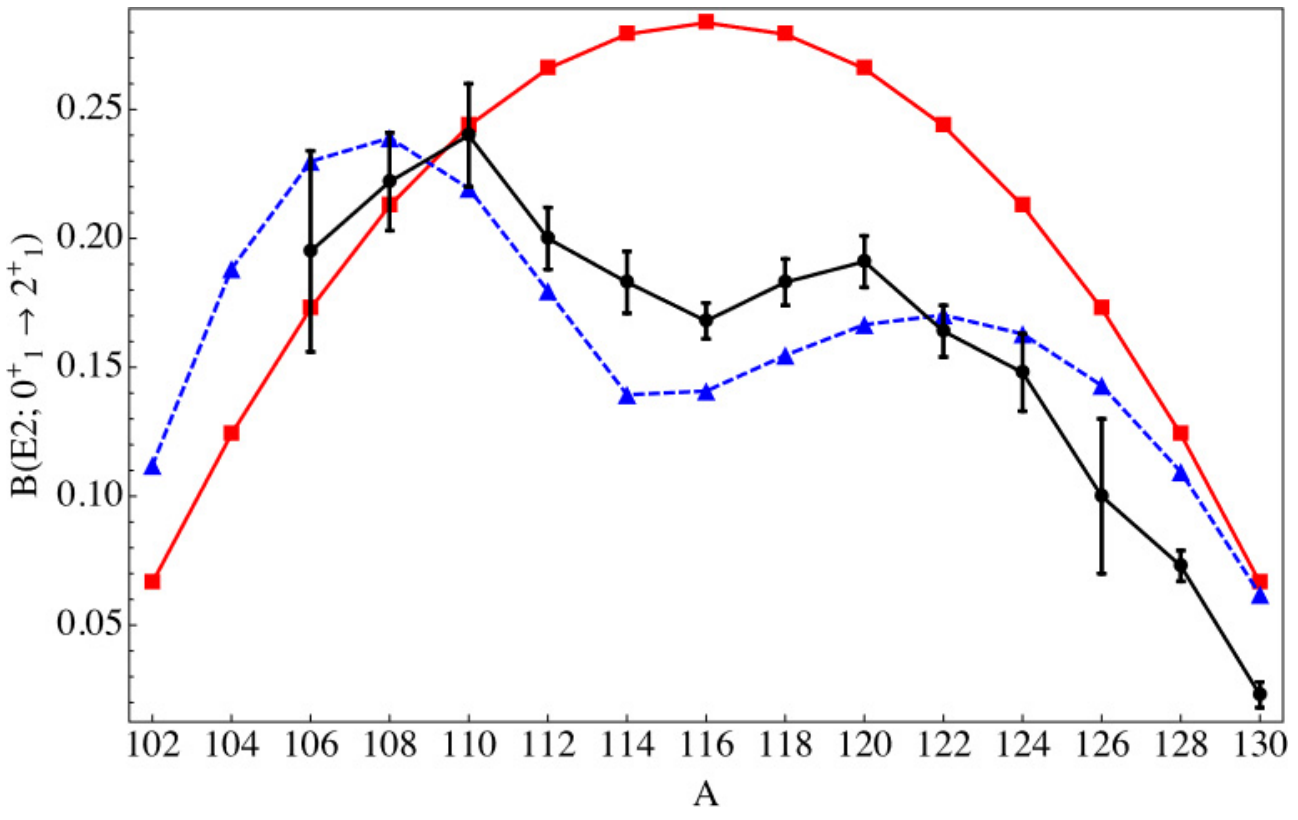

Fig. 19. (Color online) From Ref. 195. Experimental and calculated $B(E2)$ values for Sn isotopes. Experimental data (black circles, solid lines) are from Refs. 194, 199–201. The red solid line with squares (blue dashed line with triangles) is the result of exact seniority (generalized seniority) calculated in Ref. 195.

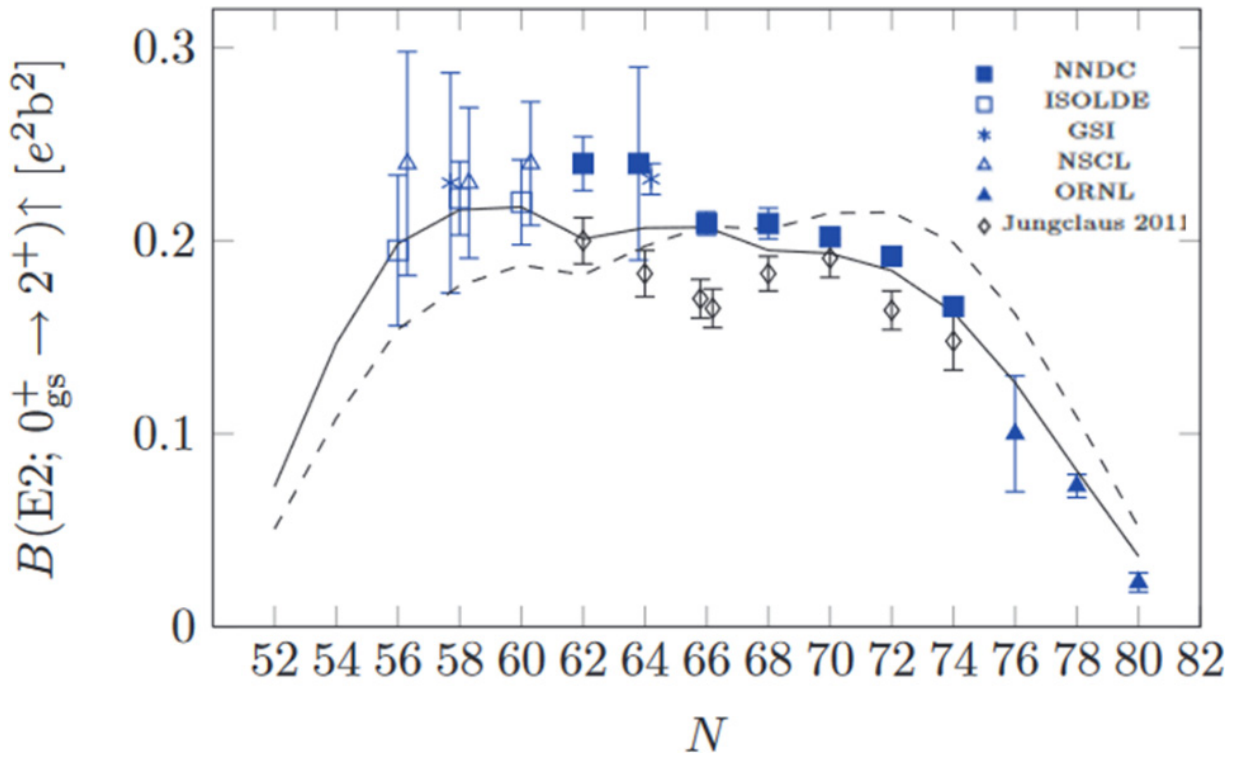

Fig. 20. From Ref. 196. Experimental and calculated $B(E2)$ values for Sn isotopes. Experimental data points are from Refs. 88, 194, 199–202. The dashed (solid) curves are from CISM calculations using the effective interaction proposed in Ref. 197 with constant (isospin-dependent) effective charge.

In the Te isotopic chain, the proton-neutron correlation competes with the seniority coupling, leading to complicated patterns. CISM calculations using the same interaction mentioned for Sn isotopes were performed to systematically study the excitation energies and transition rates of Te isotopes.[203] On the one hand, the proton-neutron correlation is strong in Te isotopes near $N = 50$. Thus, the yrast spectrum is roughly equidistant, and the $B(E2)$ value is nearly constant along the yrast line, presenting a collective pattern. On the other hand, when approaching $N = 82$, the proton-neutron correlation becomes less competitive, and the $B(E2)$ value decreases rapidly from $J = 4$–$8$, providing the possible existence of the $8^+_1$ isomer and satisfying the seniority coupling pattern. Globally, the available experimental $B(E2; 0^+_1 \to 2^+_1)$ agrees reasonably with the calculated values, which show a parabolic behavior from $N = 52$ to 82 as indicated in the seniority scheme. To illustrate the above phenomena, Fig. 21 shows the experimental and calculated excitation energies of $2^+_1$ and $4^+_1$ and $B(E2; 0^+_1 \to 2^+_1)$ in even $^{104-134}$Te. It should also be noted that experimental data for $B(E2; 0^+_1 \to 2^+_1)$ in $^{108,112}$Te have not been reported until the last decade,[204,205] and data in Te isotopes are still required to validate the robustness of the shell closure and the seniority scheme.

In the extreme case, the neutron-proton correlation is rather enhanced in $N = Z$ nuclei so that the spin-aligned neutron-proton (paired to the maximal angular momentum) coupling scheme rather than the usual seniority coupling scheme (nucleons are paired to the zero spins) dominates in the low-lying states, which is fairly unusual yet still observed in the heaviest $N = Z$ nucleus $^{92}$Pd.[206] As illustrated in Fig. 22, its yrast states are equidistant, distinctly different from the semi-magic nuclei $^{96}$Pd known as a $(9/2)^4$ system. CISM calculations indicated that the $T = 0$ $(T = 1)$ neutron-proton interactions contribute the most to the $^{92}$Pd ($^{96}$Pd) spectrum, favoring the dominating role of isoscalar pairs. The nearly constant excitation energy between the consecutive states can be attributed to the relative motion of the spin-aligned neutron-proton pairs.

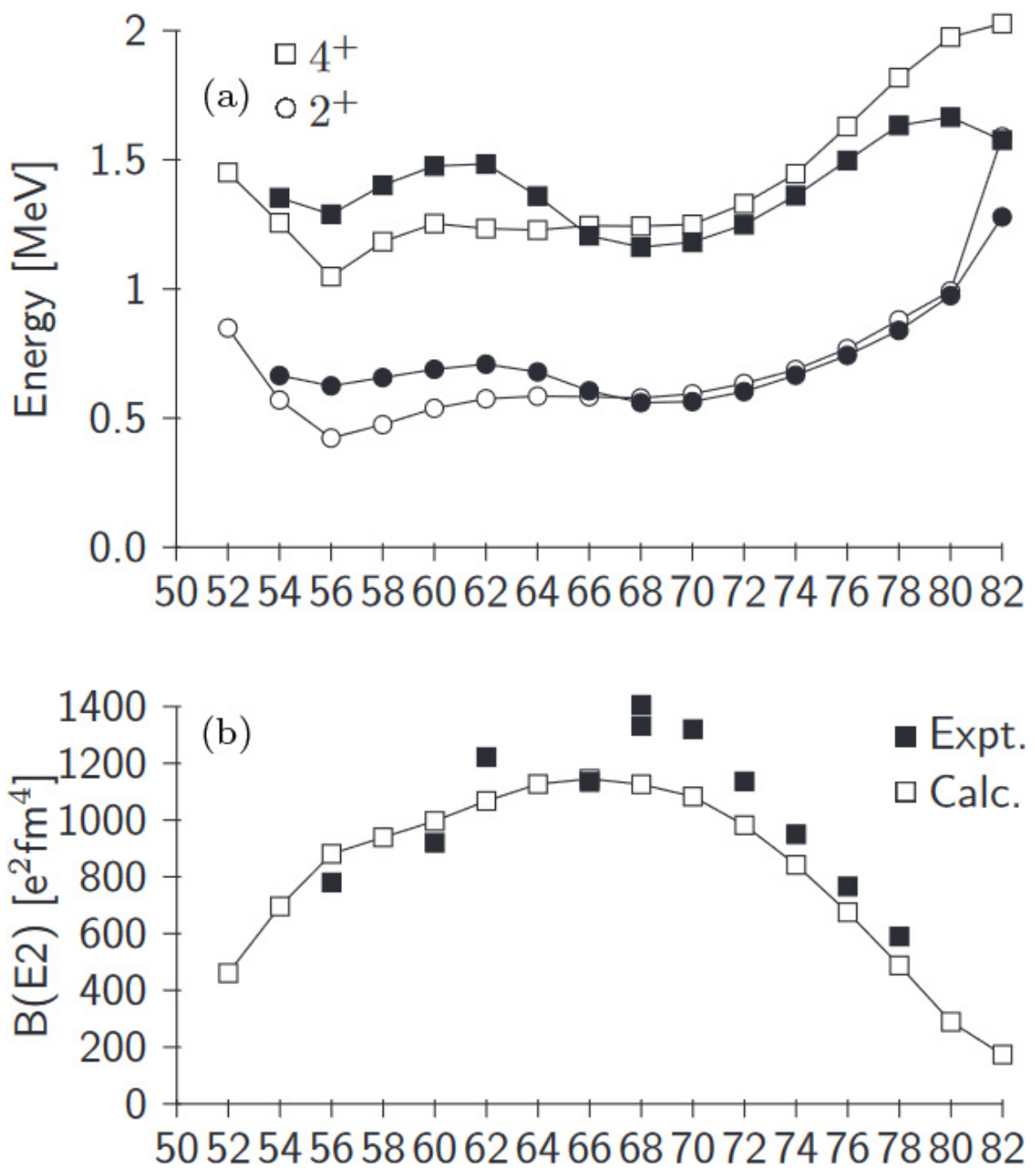

Fig. 21. From Ref. 204. The experimental and calculated: (a) excitation energies of $2_1^+$ and $4_1^+$ and (b) $B(E2; 0_1^+ \rightarrow 2_1^+)$ in even $^{104-134}$Te. The experimental data are from Refs. 88, 204 and 205, while the theoretical results are from CISM calculations[203] using the interaction proposed Ref. 197.

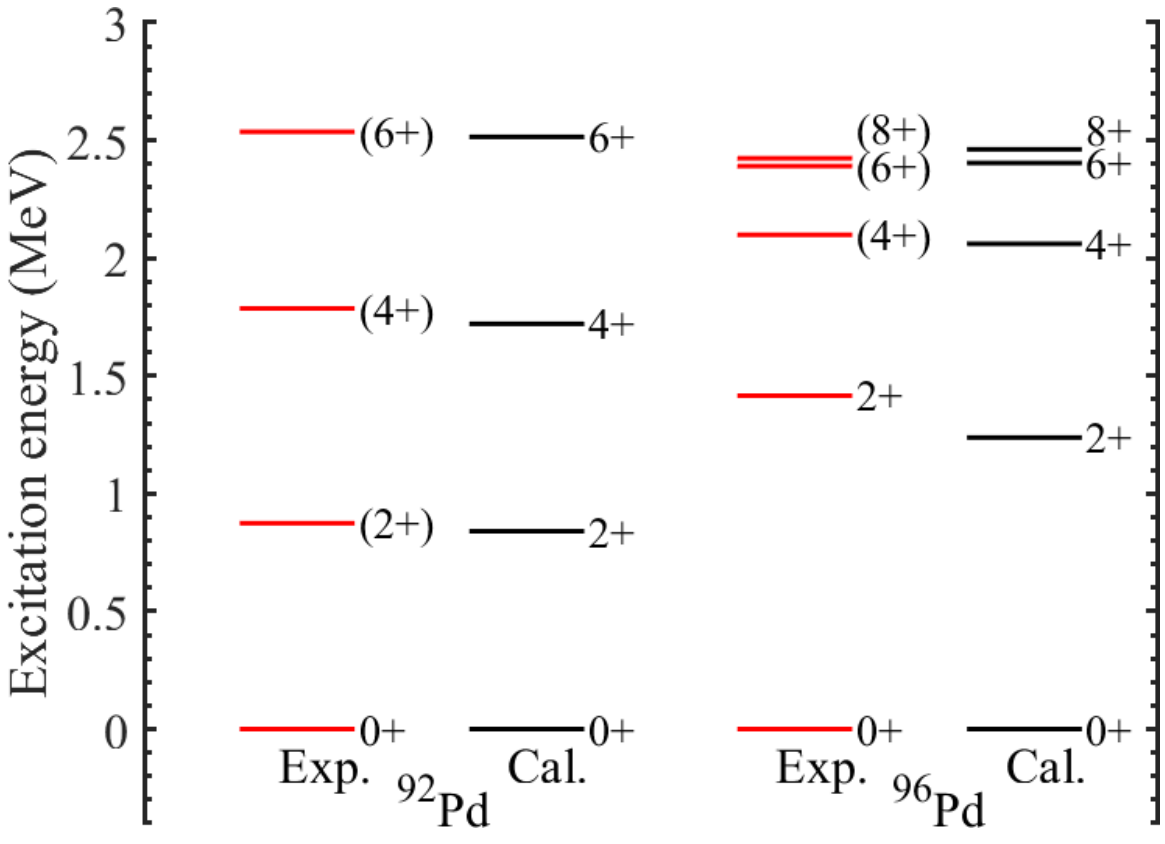

Fig. 22. Experimental and calculated excitation spectra of $^{92,96}$Pd. Experimental data are from NNDC[88] while the theoretical results are calculated with JUN45 interaction.

Calculations[207] also predicted that the $B(E2)$ value in the spin-aligned neutron-proton pair phase is nearly constant as a function of spin and selective. In other words, nuclei dominated by the spin-aligned neutron-proton pairs have a vibrational-like low-lying spectrum and rotational-like relation between $B(E2)$ and spin. Furthermore, $^{96}$Cd was predicted to be dominated by the same coupling scheme as $^{92}$Pd. The unusual structure of the two even–even nuclei in the $g_{9/2}$ shell was understood with the rather strong TBME $\langle \pi g_{9/2} \nu g_{9/2} | \pi g_{9/2} \nu g_{9/2} \rangle_{J=9}$.

Overall, the competition between the seniority coupling and neutron-proton correlation plays an important role in the nuclear structure. Nuclei, including $^{96}$Pd and $^{94}$Ru, were well reproduced in the seniority scheme, with the help of the special solvable $\nu = 4$, $J = 4$ and 6 states in the $(9/2)^4$ system. The transition rates in such nuclei were well understood with seniority mixing. The generalized seniority scheme was suggested to remain in Sn isotopes, though reduced $B(E2)$ was experimentally discovered at around mid-shell. The neutron-proton correlation was shown to affect the structure of light Te isotopes, and spin-aligned neutron-proton pairs were suggested to illustrate the low-lying states in $N = Z$ $^{92}$Pd due to the strong neutron-proton correlation. Nevertheless, both experimental and theoretical investigations are anticipated further to validate the above findings in the last two decades. Especially, the $B(E2)$ data in $^{92}$Pd and $^{96}$Cd are urgently in need to benchmark the predicted unusual dominance of the $J = 9$, $T = 0$ neutron-proton pairs in those nuclei.

In conclusion, significant progress has been achieved in nuclear structure studies using CISM in recent years. On the one hand, the feasibility of the shell model has been tested. Advancements in computational capacity have enabled CISM to consider cross-shell excitations in light nuclei and interpret neutron-rich nuclei, such as $^{14}$C, providing insights into configuration mixing and evolution. Moreover, recent progress in nuclear physics experiments has allowed for systematic studies on isomerism in medium and heavy nuclei, providing valuable insights into the evolution of shell gaps. Besides, CISM has been used to interpret the different patterns (seniority-like, vibrational-like, rotational-like and so on) shown in semi-magic nuclei and the nearby nuclei. On the other hand, there have been significant advancements in understanding nuclear forces. The comparison between mirror nuclei in the light-mass nuclear region has been focused on, which is meaningful to understand the isospin-non-conserving force. The unified description of nuclei near $^{132}$Sn and $^{208}$Pb has been preliminarily attempted in the medium and heavy nuclear regions. Efforts have been made to develop a unified form of effective nuclear force, with preliminary studies on the strength of the central force. The competition between different pairing interactions has been partly investigated, with the first evidence for the spin-aligned neutron-proton paired phase discovered. These studies are crucial for advancing our understanding of the nuclear force, and further in-depth investigations are warranted. Experimental data are especially expected to testify the new models proposed in recent decades.

## 3.2. *Related models (Truncation methods and other methods)*

### 3.2.1. *Truncation methods*

Conventional CISM calculations are limited by the large computing resources in the full configuration space. Restricting the number of particle-hole excitations is the simplest and the most usual way to truncate the model space to a level within the present computing limit. There are many successful examples using such a truncation method to describe nuclei with a large full configuration space. However, for heavily deformed nuclei with strong collective correlation, the configuration mixing is so complex that the particle-hole excitation should be largely maintained. Alternative truncation methods have been proposed and largely developed, including the NPA approach and various importance-truncation methods.

#### 3.2.1.1. Nuclear shell model with the nucleon-pair approximation

The NPA is a practical approach to reducing the gigantic configuration spaces of the nuclear shell model and also leads to simpler pictures in physics.[208] Due to the residual pairing interaction between nucleons, a pair of valence nucleons has lower energy when the spin of the pair is zero ($S$ pair). Moreover, the excitation energy augments with the break of $S$ pairs. In the interacting bosons model (IBM) proposed by Arima and Iachello,[36] the building blocks are spin-zero and spin-two bosons ($sd$ bosons), which explain the ground states with spin zero for all even-even nuclei and the first excited states with spin two for most even-even nuclei.[209]

In the late 20th century, Chen used collective nucleon pairs, including SD pairs, as the building blocks of the model space for even systems, gave the analytical formulas for the matrix elements of the Hamiltonian, and finally proposed a formalism for the nuclear shell model, named as the nucleon-pair shell model (NPSM).[209] Zhao refined the formalism of the NPSM and called the method as NPA to the shell model.[210] Recently, He *et al.* developed an $M$-scheme NPA code, which could run significantly faster and allows one to reach much larger NPA model spaces.[211] Lei *et al.* revisited the M-scheme NPA with uncoupled representation, which significantly improves the computational efficiency.[345] Furthermore, a state-of-the art version of the NPA approach has been very recently suggested.[346] The new version takes the J-scheme basis sates and uses the M-scheme commutators. Consequently, it reserves the elegant simplicity of traditional J-scheme version and further increase the computational efficiency by 1-2 orders of magnitude on the basis of Ref. 345. Various numerical calculations of the nuclear shell model using NPA were applied in interpreting and predicting spherical and transitional nuclei from light to heavy, such as $^{42-48}$Ca,[212–215] $^{72-80}$Zn,[215] $^{102-130}$Sn,[216,217] $^{126-142}$Te,[218] and $^{208-214}$Ra.[219] Good descriptions of their low-lying states show the validity and efficiency of NPA in simplification. In recent works, besides SD pairs, higher-spin pairs have been included. For short, nucleon pairs with spin $J = 0$–6 are normally denoted by $S$, $P$, $D$, $F$, $G$, $H$ and $I$, respectively.

Recently, the NPA shell model was applied to investigate $^{132}$Ba in the works of Lei *et al.* In Refs. 220 and 221, $H$ pairs representing two neutron holes in the $h_{11/2}$ orbit and the negative-parity pairs of lowest energy were as well building blocks of neutron valence space besides the SD pairs. The full consideration of these three parts resulted in good agreements on experimental energy spectra, $g$ factors, $B(E2)$ values, and reasonable explanations of back-bending phenomena and shape coexistence in $^{132}$Ba. It was shown that configurations of negative-parity pairs were essential to reproduce its experimental $B(E2, 10_1^+ \to 8_1^+)$ value. In addition, the $(\nu h_{11/2})^{-2}$ alignment, leading to the back-bending phenomena, was reasonably explained by the negative-parity pairs and better by spin-ten $H$ pairs.

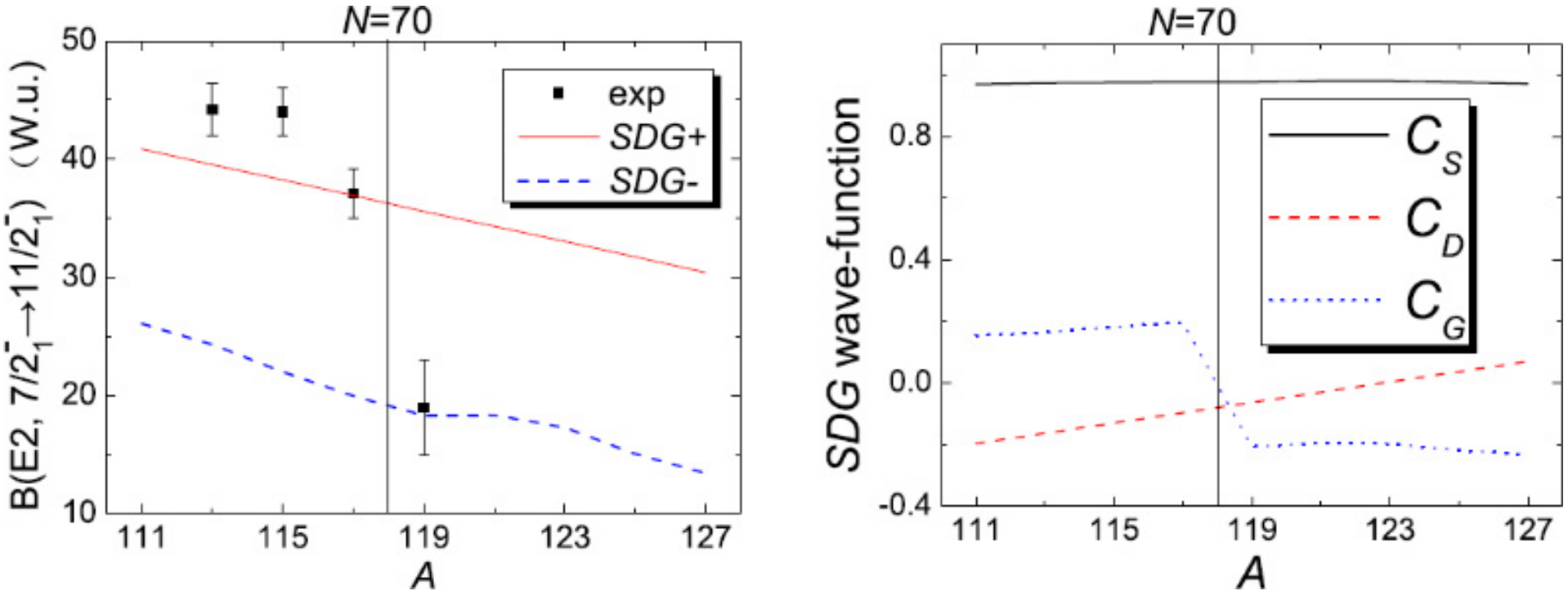

Fig. 23. From Ref. 224. Phase change at $N = 70$ indicated by the electromagnetic features and wave-functions. SDG+ is from the SDG wave functions with $C_G > 0$, and SDG$^-$ with $C_G < 0$.

Meanwhile, the quadrupole moment $Q$ and $B(E2, 7/2_1^- \to 11/2_1^-)$ values in odd-mass $^{111-127}$Cd isotopes[222] were well reproduced by the NPA shell model. As shown in Fig. 23, the sudden changes of these electromagnetic features at $N = 70$ confirmed the phase changes of the extracted wave functions of the $11/2^-$ isomers, for which $C_G$ was suggested to be the deciding factor. Furthermore, robust correlations between $Q(2_1^+)$ and $Q(2_2^+)$ were studied in random-interaction ensembles[223] and a variational approach was proposed for pair optimization to improve the accuracy of the NPA.[224]

The isospin symmetry was included in the NPA shell model by Fu *et al.* in Ref. 225. This work derived analytic expressions for the matrix elements of one- and two-body interactions and overlaps between states in the collective nucleon-pair basis with conserved spin and isospin. This generalized model was expected to describe the correlation properties, especially the proton-neutron pairing and the quartet condensation, of nuclei in the $N$–$Z$ regime. The specific application of this proposed NPA with isospin symmetry can be seen in Refs. 226–231. It was applied to the shell model based on usual nuclear interactions for $N = Z$ nuclei. Studies on overlaps between wave functions obtained in both the full shell-model space and various truncated subspaces supported the validity of NPA and indicated the contributions of different nucleon pairs.

As discussed in Sec. 3.1.5, the dominance of the spin-aligned neutron-proton pair in $^{92}$Pd was first experimentally evidenced.[206] With the generalized NPA method and the JUN45 interaction,[226] the important role of isoscalar (IS) spin-aligned pairs was confirmed again in low-lying states of $^{96}$Cd, $^{94}$Ag and $^{92}$Pd. It was suggested that IS pairs were rather important for $0_1^+$ and $2_1^+$ states of $^{94}$Ag. Meanwhile, the isovector (IV) SD pairs contributed more to $0^+$ and $2^+$ states of $^{96}$Cd and $^{92}$Pd than IS pairs. Besides, explanations were given for "level-inversion" isomerism appearing in $^{96}$Cd and $^{94}$Ag. As a complement, Ref. 228 showed the essential position of quartet correlation in ground states of $^{92}$Pd.

A more detailed study on low-lying $T = 0$ states $^{96}$Cd was done in Ref. 231, where the lowest seniority scheme (LS), the spin-aligned (SA) pairs, the $J_{\max}$ pairs, the spin-one (SO) pairs, the IV pairs and the IS pairs were all considered separately. Results showed that LS and SO pair approximations were over-approximated for those states. The SA and $J_{\max}$ pair approximation described relatively well some yrast states while unsatisfactorily reproducing others. Both the IV and IS pair approximations demonstrated close agreement with wave functions, energy spectra and electromagnetic properties obtained in the full shell-model space since the two approximations contained more nucleon pairs. For illustration, Fig. 24 presents the overlaps between wave functions obtained in the full shell-model space and those in the NPA subspaces for the positive parity $T = 0$ yrast states. Among the six types of pair approximations mentioned, the IV and IS pair approximations showed the best agreement with the case in the full shell-model space.

Reference 227 had already shown that both IV and IS pair approximation described well the ground band states of $^{20}$Ne and $^{24}$Mg while SA did not, using the USDB interaction and a schematic pairing plus quadrupole–quadrupole interaction. Furthermore, LS + SA + SO pairs-constructed subspace was proved to describe the low-lying $T = 0$ states very well.

For a more global view, similar studies for ground states of 10 even–even $N = Z$ nuclei and $T = 0$ yrast states of six odd–odd $N = Z$ nuclei were also shown in Refs. 229 and 230, respectively. Competitions between IS and IV pairs appeared, and the sensitivity of the validity of NPA was more clearly seen. For instance, IV S pairs performed better than IS P (SO) pairs in describing ground states of even–even semi-magic nuclei. And SO pairs became more relevant in ground states when the spin-orbit coupling strengthened.

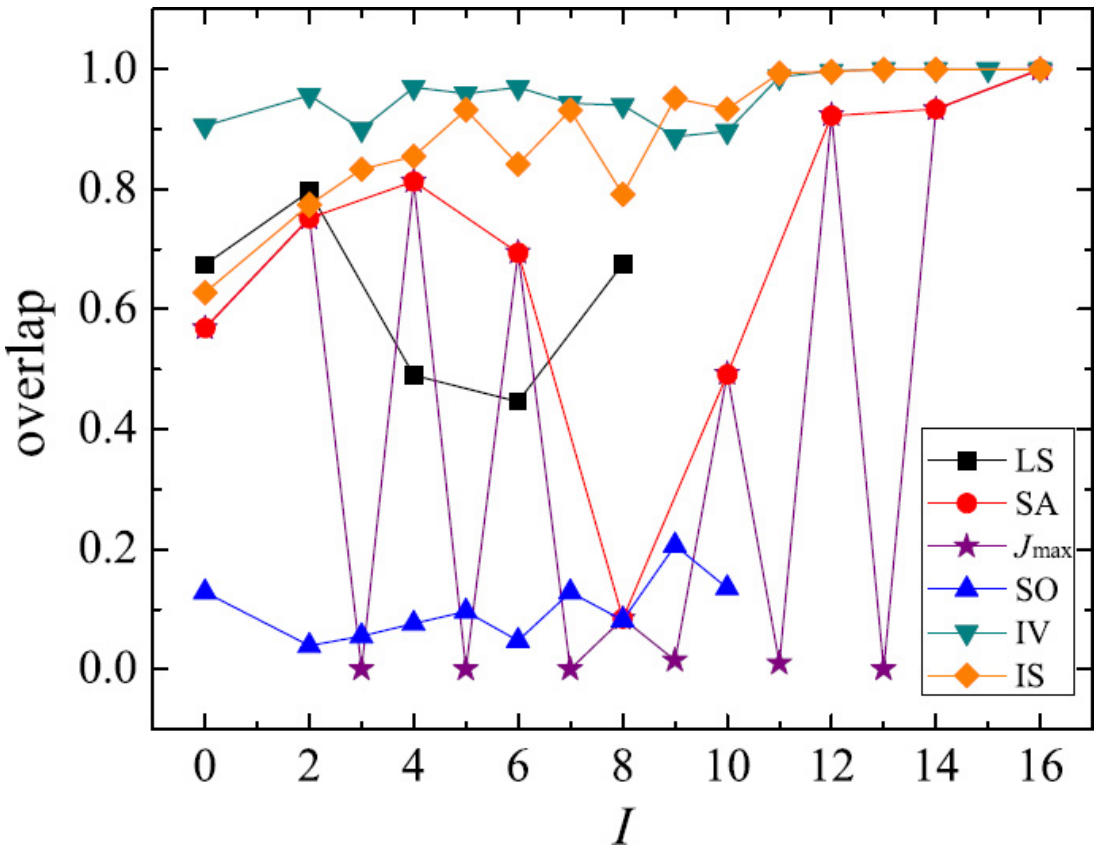

Fig. 24. From Ref. 231. Overlaps between wave functions obtained in the full shell model space and those in the NPA subspaces for the yrast states of $^{96}$Cd with positive parity and isospin zero.

Cheng *et al.* developed the NPA with particle-hole excitations to perform calculations in multi-major shells.[220] In this approach, basis states are constructed by valence-particle pairs in upper major shells, valence-hole pairs in lower shells, and pairs consisting of one valence particle in upper shells and one hole in lower shells (i.e., phonons). With such an approach, $^{100}$Sn was calculated by considering both proton and neutron particle-hole excitations, up to $4p$–$4h$, across the 28–50 and 50–82 shells with the CD-Bonn potential. The NPA with particle-hole excitations is also very useful in the microscopic study of low-lying states of exotic nuclei and nuclear shape coexistence.

The NPA has been proved to be a compact truncation to full configuration–interaction diagonalization in shell-model spaces, but selecting good pairs was a long-standing problem. Recent NPA applications chose pairs by using the generalized seniority scheme and showed that seniority-based pairs worked well for collective states in semi-magic nuclei and vibrational nuclei, but they did not work well for rotational bands in deformed nuclei. In Refs. 232 and 233, Fu *et al.* proposed two alternate approaches to generate collective pairs for NPA calculations for deformed nuclei: the approach based on the Hartree–Fock calculation and the conjugate gradient minimization approach. They showed that the SU(3) symmetry could be realized in a pair-truncated subspace. For example, a six-proton-six-neutron system in the $pf$ ($sdg$) shell was reproduced exactly with SDG (SDGI) pairs; using just SD pairs cannot describe the full SU(3) collectivity. A mapping from Elliott's fermionic SU(3) model to systems with $s$, $d$, $g$, ... bosons was also presented. It was also shown that SDG pairs obtained by the new approaches gave good descriptions for low-lying states of the rotational bands in $^{52}$Fe and $^{108}$Xe and the phenomenon of shape coexistence in $^{68}$Se and $^{68}$Ge.

On the whole, the NPA is an important approach to reducing the dimension of the shell-model Hamiltonian matrix. In the last decade, researchers have improved the NPA by including the higher-spin pairs beyond the SD pairs, considering the isospin-symmetry, developing the cross-shell excitation interaction, etc., so that the shell model using the NPA could be more accurate and treat more exotic nuclei. The validity of the generalized approaches and the selection of good nucleon pairs could be the future subject of the NPA.

#### 3.2.1.2. Monte Carlo shell model

In the full configuration space, a physical state may be represented with more than a thousand basis states, which is physically unnatural. Evidently, only a few of the basis states are relevant for finite low-lying states. Thus, one attempts to select the important basis states to span a tractable model space, called the importance-truncation method.

One well-developed importance-truncation scheme is the MCSM,[37–40] which applies the quantum Monte Carlo method to generate the Hilbert subspace for diagonalization. MCSM starts from the imaginary-time operator, which can turn an arbitrary state (nonorthogonal to the ground state) to the ground state when the time is long enough. In the framework of MCSM, one discretizes the time and transforms the imaginary-time operator to an integration form with the Hubbard-Stratonovich transformation. Then, the Monte Carlo sampling can be naturally applied to numerically calculate the integration. All the basis states required for describing the ground state should be generated during the process. Therefore, one can select the important basis states which are relevant to the realistic low-lying states from the states randomly generated with the Monte Carlo sampling. A candidate basis state is accepted only if it provides a sufficiently smaller eigenenergy. Finally, one diagonalizes the effective Hamiltonian in the subspace spanned by the generated basis states, where the low-lying states can be simultaneously derived.

MCSM was first proposed by Honma *et al.* in 1995[37] and improved soon with the acceleration of convergency by sampling around the Hartree–Fock solutions,[38] the restoration of symmetries by projection methods,[38,39] the further compression of the basis space by refining the candidate basis states.[234] Later, energy variance extrapolation methods were introduced to estimate the exact energy and other observables which would be obtained in the full valence space.[235] The energy variance extrapolation, together with the variational procedure[236] led to an enhanced version of MCSM suitable for massively parallel computers.[237] The advanced MCSM has been extended to *ab initio* no-core systems[238] and can now treat a system with $M$-scheme dimension up to $10^{29}$,[239] showing an advantage in calculation efficiency for large model space and mid-shell heavy nuclei.

One of the most prominent applications of MCSM has been the interpretation of the deformed structure and electromagnetic properties resulting from collective motion, which are hard (even impossible) to treat in a full valence space. In the past decade, MCSM has succeeded in interpreting the unusual shape deformation, coexistence and evolution in Ni,[239–243] Sr,[244] Zr[245] and Hg[246,247] isotopes, shedding light on configuration-dependent shell structure and type-II shell evolution.[159]

Recent experimental studies have extended the low-lying scheme and related electromagnetic properties in even $^{64-70}$Ni isotopes.[241–243,248,249] MCSM calculations in the $pf$-$g_{9/2}$-$d_{5/2}$ model space well reproduced the latest observations and revealed the triple shape coexistence in $^{68,70}$Ni.[239] As indicated in the type-II shell evolution, the particle-hole excitation varies significantly with the ESPEs and the shell structure depending on the configuration. When the particle-hole excitation reduces the shell gap, the reduction of the shell gap favors the hole excitation across the gap. In such a way, two close states have rather different configurations and nearly degenerate shapes coexist. Taking $^{68}$Ni, for example, the occupation numbers in the orbits, such as $g_{9/2}$ and $f_{5/2}$, vary significantly in the three $0^+$ states, producing spherical, oblate and prolate shapes. Figures 25 and 26 illustrate, respectively, the level scheme and the occupation distribution in $^{68}$Ni.

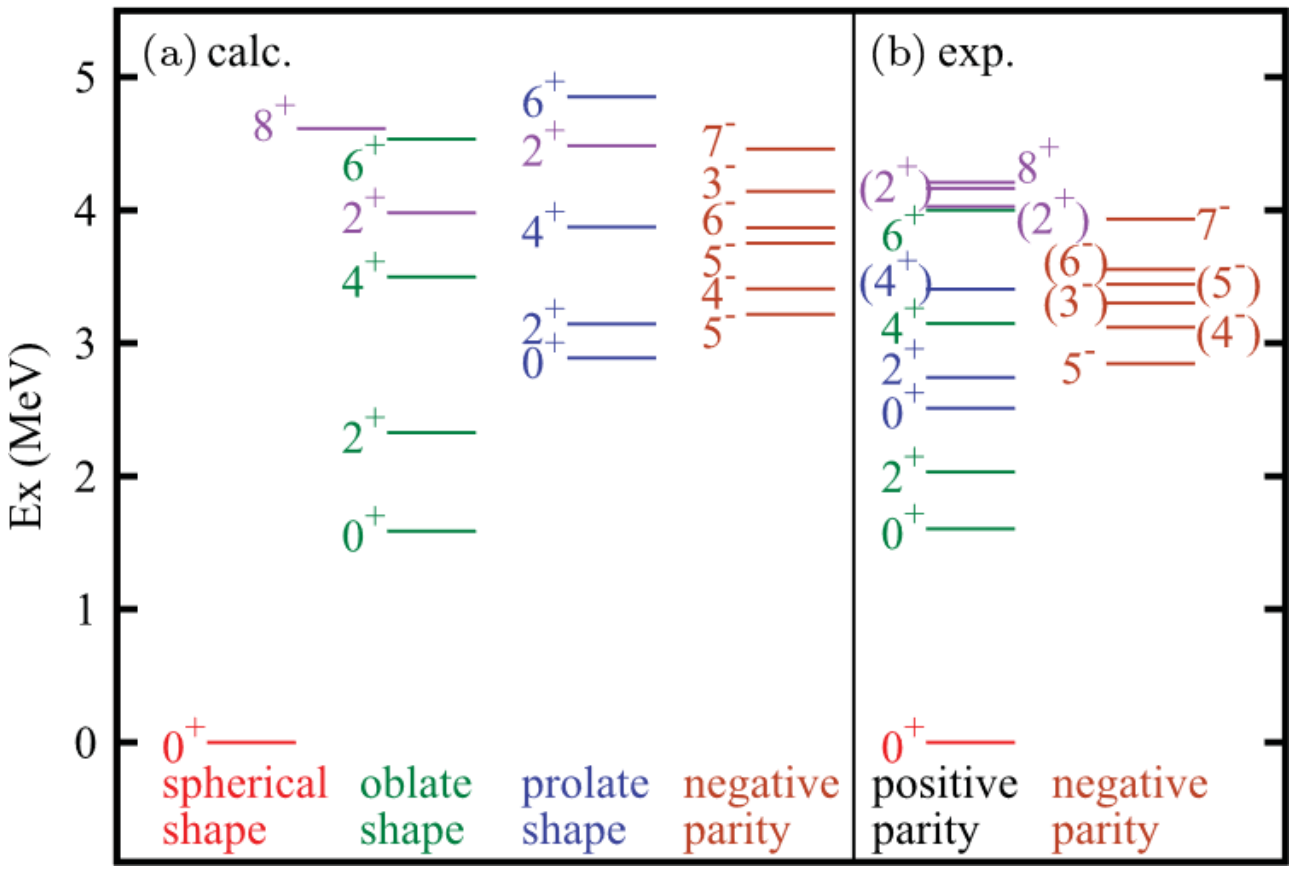

Fig. 25. From Ref. 239. Calculated (a) and experimental (b) level scheme of $^{68}$Ni. Theoretical data are from MCSM calculations in Refs. 240 and 243, while experimental results are from Refs. 241, 243, 248 and 249.

Similarly, the abrupt transition occurring in the shape of Sr[244] and Zr[245] isotopes around $N = 60$ has been recently revealed with the type-II shell evolution. The study using MCSM in the large model space, including 8 proton and 8 neutron orbits above $^{68}$Ni, succeeded in the theoretical interpretation. In such nuclei, the spherical and deformed (or the weakly deformed and strongly deformed) states coexist with a small energy gap. From the spherical state to the deformed one, some binding energies are lost due to particle-hole excitation, but others are gained thanks to the quadrupole deformation. Normally, the losses are larger than the gains. However, now that the ESPEs shift as more particles occupy in particular orbits, presenting the type-II shell evolution, additional monopole binding energy can be obtained. The total gains compete with losses along the isotopic chain, at one point in which the deformed state begins to locate even lower than the spherical one.

Moreover, the same mechanism has been applied to interpret the Hg isotopic chain, except that the odd–even staggering in pairing energy also influences the subtle balance, resulting in a unique shape staggering from $^{181}$Hg to $^{187}$Hg.[246,247] The endpoint of the shape staggering in Hg isotopes has not been determined until recently.[246] The largest-ever MCSM calculations, performed in the model space and including 11 proton and 13 neutron orbits above $^{132}$Sn, are in nice agreement with the observed charge radii, energy levels and transition rates.

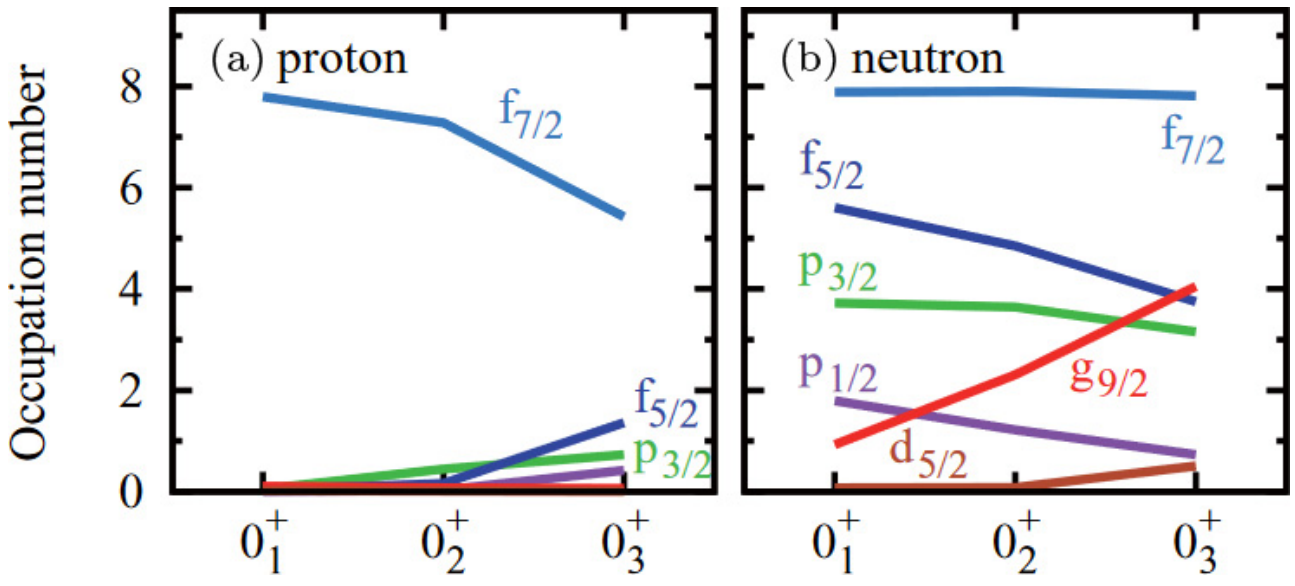

Fig. 26. From Ref. 240. Occupation number for the $0^+$ states of $^{68}$Ni in (a) proton and (b) neutron orbits.

It is worth noting that the model spaces described in this subsection are too large for direct diagonalization, while the particle-hole excitation plays a crucial role in interpreting the intriguing shapes, especially for exotic nuclei. A suitable truncation method is thus promising to break the present computing limits.

For another importance-truncation method, one generates the candidate states with particle-hole excitations from a reference state, and estimates the relevance of each candidate state via many-body perturbation theory.[250,251] In the recently developed correlated-basis method,[252] one first diagonalizes the Hamiltonian in each partition, and selects the basis vectors possessing lower unperturbed energies to span the subspace for final diagonalization. The truncation method based on generalized seniority has also been largely developed. The computing scheme for generalized seniority based on a spherical single-particle basis has been accelerated,[253] while that on a deformed basis has been proposed.[254] Besides, the simple truncation scheme based on monopole interactions[255] has also been tested.

#### 3.2.2. *Generalization with bases other than the spherical harmonic oscillator basis*

The only thing to distinguish the NPASM or the MCSM from standard CISM is that an additional selection of relevant bases is performed in the formers. One should notice that the importance-truncation shell methods, including the NPASM and the MCSM, still start from the harmonic oscillator basis. However, there are other types of shell models that use bases other than the spherical harmonic oscillator (HO) basis, such as the deformed Nilsson basis,[41] the Wood–Saxon (WS) basis, the Hartree–Fock (HF) basis and the complex Berggren basis.[256,257] HO is beneficial to practical calculations, as it is the only one with analytical solutions. Also, mature techniques have been developed in the HO bases to decompose a two-body system into the center of mass motion and the relative motion, which is more troublesome for other bases. Nevertheless, other types of bases can be attractive for understanding realistic nuclei. For example, it is much simpler to express the deformed nuclei with the deformed Nilsson basis,[258] and more accurate to study the weakly bound states with the WS basis[259] or the Berggren basis.[256,257,260,261] With the development of related calculation techniques, such kinds of shell models have gradually completed the standard CISM to understand the nuclear structure more comprehensively. In this section, we would like to shed light on the recent progress in the PSM using the deformed basis and the GSM using the Berggren basis.

##### 3.2.2.1. Projected shell model

The treatment of high-spin states and deformed nuclei is difficult for CISM.[41] In such cases, large model space and complex configuration mixing are in need. To this end, the PSM attempts to describe the excited states with simple configurations by using a deformed basis, which takes both the advantages of the deformed mean-field approach and CISM.[41,258] PSM is called projected because it starts from the intrinsic states with broken rotational symmetry and projects the intrinsic states onto good angular momenta to restore the rotational symmetry. In the framework of PSM, the highly deformed nuclei are described with few configurations. For example, the states with negative parity in $^{50,58,62}$Mn were calculated to be the two-quasiparticle configurations originating from the couplings of one proton from the *pf* shell and one neutron from the $g_{9/2}$ orbit.[262]

In the last decade, PSM calculations have been performed to investigate the collectivity properties of nuclei around the "island of inversion" in Refs. 263 and 264, where the authors applied the improved isospin-dependent Nilsson potential[265] for the deformed basis and the Lipkin–Nogami approach for the pairing correlations. As a result, the excitation spectra, $B(E2)$ values and $g$ factors of neon and magnesium isotopes were well reproduced. Furthermore, the shell erosion near $N = 20$ was indicated, through the variation of $B(E2)$ values and $g$ factors as a function of neutron number, both for neon and magnesium isotopes. For illustration, it was found that the quasi-proton configurations dominated in the yrast states of $^{20,22,24}$Ne, while quasi-neutrons configurations dominated in those of $^{28,30,32}$Ne. Besides, in nuclei including $^{28,38}$Mg and $^{28,30,32}$Ne, two quasi-neutron configurations were proved to be important for yrast states at spin $I$ more than 6.

Due to the combinational complexity in practice, PSM had been confined in a configuration space including only no more than four-quasiparticle states. Thanks to the Pfaffian method, the model basis of PSM was recently extended to include up to 10-quasiparticle states.[266] Such extended PSM was first shown to reproduce the observed yrast and sidebands of $^{134}$Nd well. For illustration, the nice agreement between the energy levels calculated by the extended PSM and experimental data can be seen in Fig. 27. Moreover, the significance of the inclusion of 8- and 10-quasiparticle states can be seen in Fig. 28. Without the two kinds of multi-quasiparticle states, the yrast states with spin higher than 40 could not be well described. In addition, the PSM calculations suggested the collectivity reduction and the intrinsic-structure change at very high spins in $^{134}$Nd. The extended PSM was able to investigate the evolutional properties at very high spins in fast-rotating nuclei.

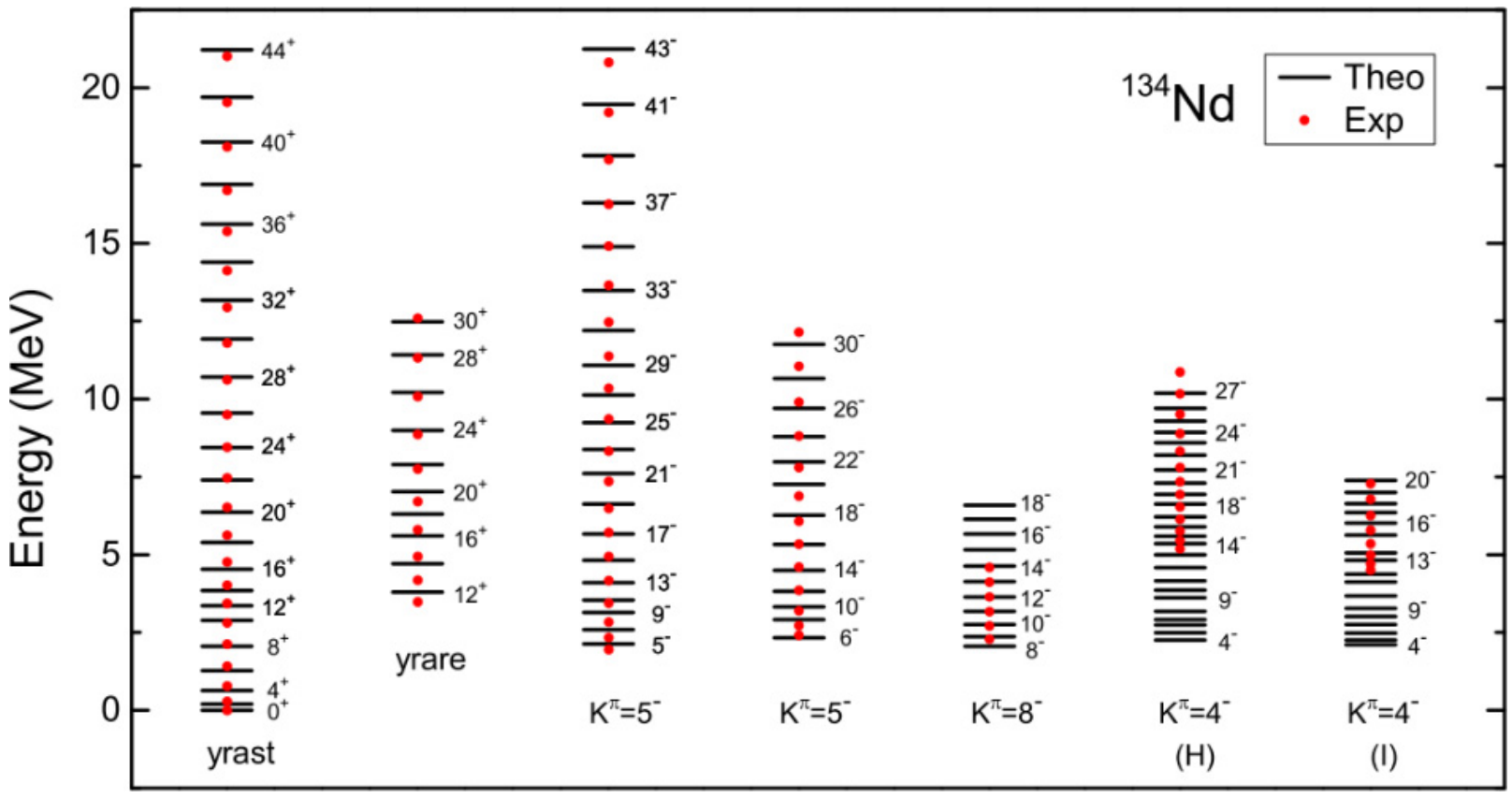

Fig. 27. From Ref. 266. Energy levels of $^{134}$Nd calculated by the extended PSM.

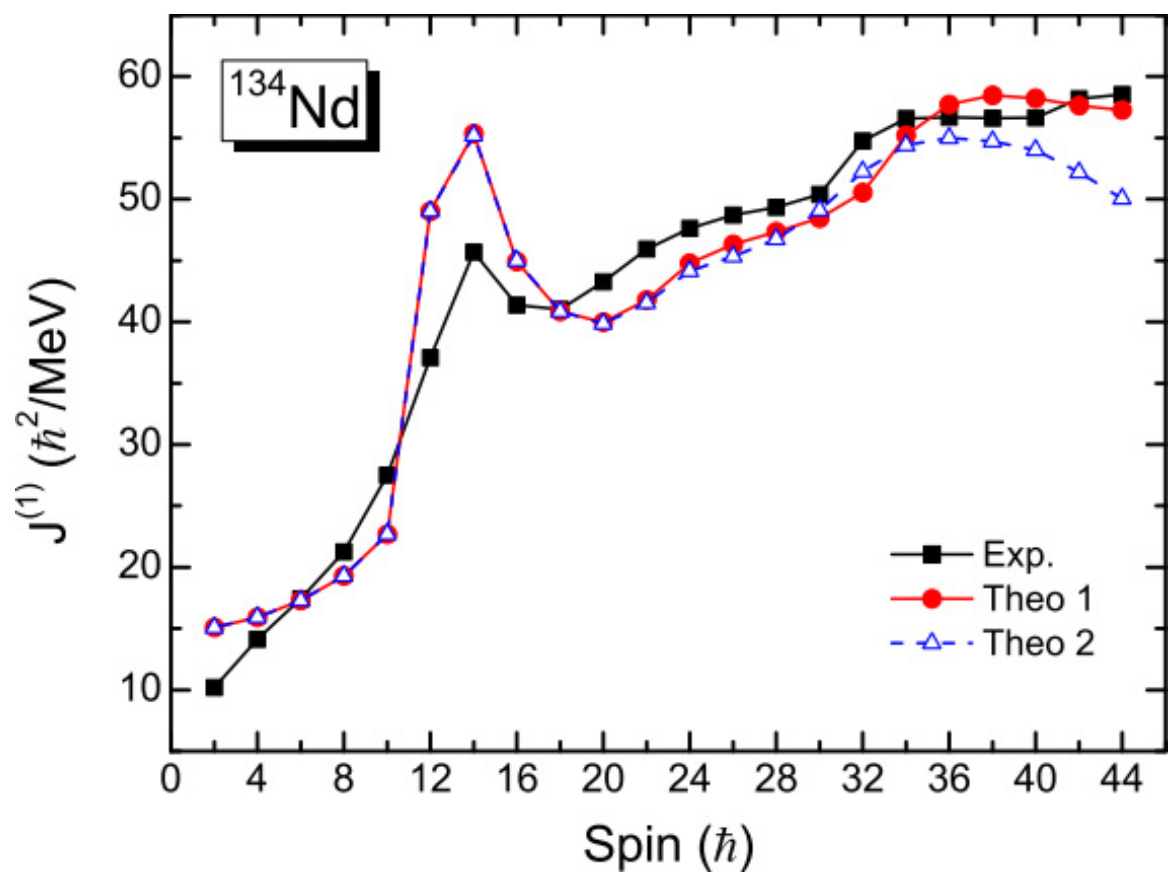

Fig. 28. From Ref. 266. Calculated and experimental moment of inertia of the yrast band for $^{134}$Nd. "Theo 1" refers to the theoretical result obtained in the full configuration space, while "Theo 2" the one without 8- and 10-qp states.

The extended PSM was then employed to reproduce high-$K$ isomers and their rotational bands in proton-rich krypton isotopes[267] and tungsten isotopes.[268] Unlike the studies in Ref. 268, the configuration space including up to eight-quasiparticle states was sufficient in Ref. 267. It was shown that the extended PSM calculations reproduced overall well the rotational bands of krypton isotopes. Meanwhile, the necessity of including six-quasiparticles and eight-quasiparticles was presented to interpret states of very-high spin. The degrees of freedom, including the triaxial and gamma-soft shapes, were suggested to be considered in order to illustrate the structure evolution of krypton isotopes. In Ref. 268, PSM calculations described the $K$-isomeric bands of even-mass neutron-rich W nuclei, predicting their dominant multi-quasiparticle structures and rotational properties.

To sum up, as a method to treat high-spin states in deformed nuclei, PSM has been recently extended to include more than four-quasiparticle states, which makes it able to reproduce high-$K$ isomeric states and their rotational bands.

#### 3.2.2.2. Gamow shell model

The GSM[256] can be considered as a generalized shell model that replaces the harmonic oscillator (HO) basis with the Berggren (Gamow) basis[269] in the complex-energy plane. The Berggren basis comprises bound, resonant and continuum states on an equal footing and is usually generated from the WS potential or HF equations. It is noteworthy that the complex-momentum space is too huge to calculate.

Meanwhile, the GSM faced various technical problems, such as selecting physical states and treating the non-resonant part of the continuum. As a result, the Berggren basis was not applied to CISM until the 21st century,[270,271] more than 30 years later than its proposition.[269] In the first works, the resonant states in $^{80}$Ni[270] and $^{18}$O[271] were calculated by considering two valence particles to benchmark the GSM calculation. Furthermore, GSM, together with couple-channel methods, has also been extended to calculate the radiative capture and scattering reaction, in which structure and reaction properties are unified in the same framework.[272]

Thanks to the consideration of the resonance and continuum effect, GSM has advantages in reproducing and predicting exotic properties of weakly bound and unbound nuclear systems. With more and more experimental investigations on the exotic nuclei near the drip lines, the advantages of GSM have been largely confirmed, showing the essential role of the continuum coupling. For example, investigations on the newly observed unbound states in $^{12}$Be demonstrated the significance of the continuum coupling, which was better reproduced by GSM than by the standard CISM.[260] Recent experiments of the Borromean nucleus $^{17}$B surprisingly gave the smallest *s*- or *p*- orbital component among the known halo nuclei, which was interpreted with the continuum effect. For the observed states in $^{16}$B derived from the quasi-free neutron knockout reaction of $^{17}$B, GSM correctly gave the ground state, while other theoretical models failed,[261] as illustrated in Fig. 29. In addition, GSM was applied to study $^{25-31}$F in the *sdpf* region, suggesting the two-halo structure of $^{31}$F resulted from the continuum coupling and nucleon–nucleon correlations.[273]

In the last decade, the GSM has been largely developed with the extension to *ab initio* theories, which will be introduced in the following section.

### 3.2.3. *Extension to the ab initio theories*

*Ab initio* theories, in other words, the first principles, start from realistic nuclear forces and use nearly exact quantum many-body methods to solve nuclear systems.

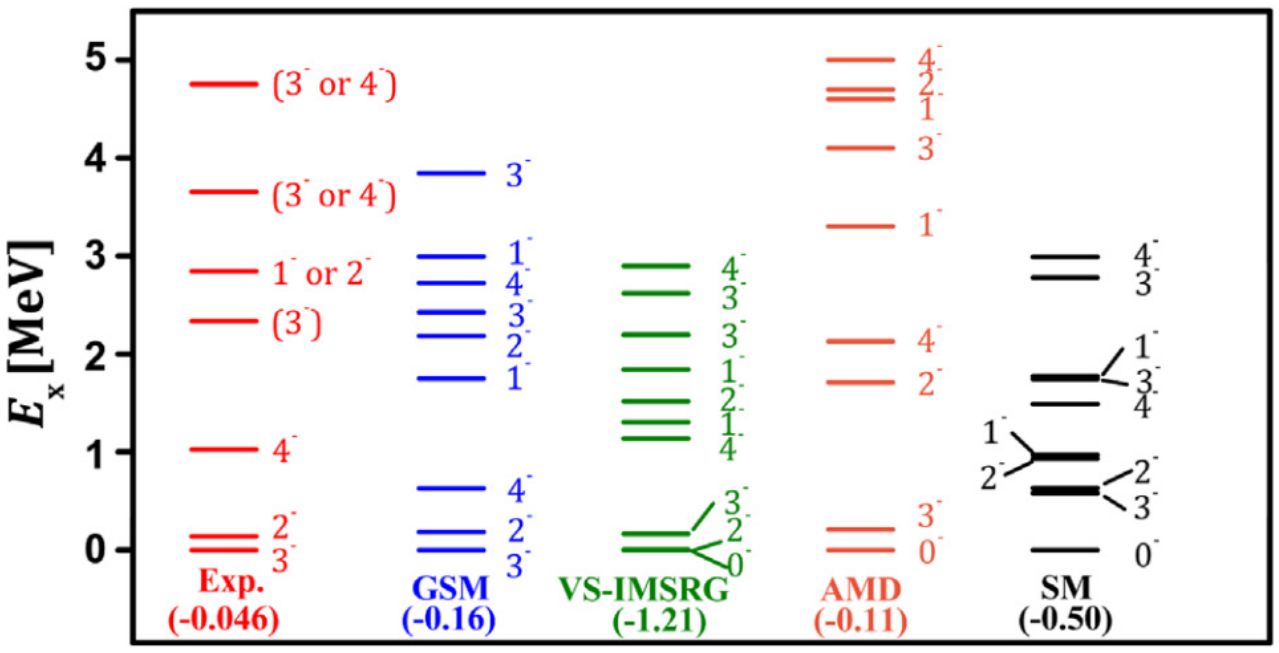

Fig. 29. From Ref. 261. Comparison of observed and theoretical low-lying levels in $^{16}$B. The data shown in parentheses are $S_n$ values in MeV. Theoretical results are calculated from the CISM, GSM, valence-space in-medium similarity renormalization group (VS-IMSRG), and anti-symmetrized molecular dynamics plus generator coordinate method (AMD). It can be seen that the GSM reasonably reproduced the observed low-lying states of $^{16}$B.

The realistic nuclear force is emphasized in the *ab initio* study because it is independent of nuclear structure data (except for the lightest nuclei, such as $^{2,3}$H) and aim at all the nuclei. Though not emphasized, effective nuclear force has been more frequently used in CISM. Moreover, most of the CISM progress summarized in the previous sections is based on the phenomenological Hamiltonian. In some cases, the TBMEs are fitted from numerous structure data. In other cases, the TBMEs are derived with an effective nuclear force whose strengths are determined by fitting structure data. The deficient construction of realistic Hamiltonians may be ascribed to various reasons. First, the realistic nuclear force had been a long time not sufficiently precise. Especially, the knowledge of the strong repulsive core is still incomplete. Meanwhile, the nuclear force derived with nucleon–nucleon scattering experiments is distinct from the one in the many-body nuclear system, where complex many-body correlations count significantly. One has to both remove the strong repulsive core of the realistic nuclear force to speed up the convergency and transform the Hamiltonian defined in an infinite Hilbert space to the effective one in a finite model space.

In the last three decades, theoretical investigations based on the realistic nuclear force have been boomingly developed. On the one hand, the high-precision realistic nuclear forces, including Argonne potential,[274] Nijmegen potentials,[275] charge-dependent (CD)-Bonn potential,[276,277] and chiral effective field potential[278–280] have been constructed. On the other hand, the renormalization methods which are more advanced than the classical $G$-matrix approach,[19] including the $V_{\text{low-}k}$ method,[281–283] the similarity renormalization group (SRG)[284] and the unitary correlation operator method (UCOM)[285] have been developed. Till now, the shell model based on the realistic nuclear force (also called RSM) has provided reasonable results for certain properties compared with the experimental results, though it still needs further improvement.

To make it clear, CISM is one of the most important many-body methods. NPASM and MCSM can be considered as the CISM with importance truncation, while PSM and GSM are generalized models of CISM. Other quantum many-body methods for the *ab initio* calculations include the No-core shell model (NCSM),[286] coupled-cluster (CC),[287,288] in-medium similarity renormalization group (IMSRG),[289] self-consistent Green's function (GF),[290,291] many-body perturbation theory (MBPT),[292–294] Bruckner–Hartree–Fock,[295–297] etc. NCSM is a special case of CISM where no frozen core is considered. Meanwhile, importance-truncation methods are usually needed for the NCSM calculations. Therefore, when using high-precision realistic nuclear force, the border between CISM (and related SM studies) and *ab initio theories* is not distinct.

Actually, different many-body methods are normally in reasonable agreement when the many-body correlations have been sufficiently considered, as is proved in Refs. 292, 293, and 298. In specific cases, the *ab initio* approaches have their own strengths and weaknesses because they all have to fold or neglect a part of correlations. It is of great interest to combine different methods to understand more nuclei more precisely. Besides NCSM, the MCSM[299,300] and the GSM[301–303] in the no-core cases have been developed. Meanwhile, both the Gamow basis[304] and the deformed basis[305] have been extended to the IMSRG methods. Such combinations help to understand the exotic nuclei more precisely.

On the other hand, only very light nuclei (around mass up to 20) can be calculated by the *ab initio* approaches where all the nucleons are active. Otherwise, semi-magic Ca isotopes and Ni isotopes have been investigated with the *ab initio* approaches, assuming a closed shell.[306–308] The CISM choosing a valence space is an efficient many-body method to treat heavier nuclei. It is worth noting that *ab initio* approaches, including NCSM,[309,311] CC[311] and IMSRG[312] have been recently used to generate the RSM effective interactions in the valence space. Performing the shell model calculations (CISM, MCSM, NPASM, etc.) on the basis of the *ab initio* (NCSM, IMSRG, CC, GF, etc.) effective interactions may be an economical solution to high precision calculations of heavier nuclei.

Following this section, recent progress in the *ab initio* CISM and GSM studies using realistic nuclear forces will be particularly introduced.

#### 3.2.3.1. CISM with realistic nuclear forces

High-precision realistic nuclear force and suitable renormalizations are the perquisites of successful *ab initio* calculations. Here, we would like to shed light on applying the CD-Bonn potential and the chiral effective field potential. Featured progress in the RSM framework has been the inclusion of the three-body force. The three-body force plays a significant role in the reproduction of nuclear structure in $A \leq 12$ nuclei,[313,314] as well as the anomalous long lifetime of $^{14}$C[315] and surprising location of the neutron drip line of oxygen isotopes.[87] The many-body correlation will be extra induced when softening the nuclear force in the RSM framework. With the development of the chiral effective potential where the three-body force can be naturally given, the significance of the three-body force has been emphasized. Meanwhile, another choice has been the phenomenological consideration of the three-body effects[317] in the other realistic potential, such as CD-Bonn.

The high-precision CD-Bonn potential proposed at the beginning of this century[276] departs from the one-boson-exchange mechanism and is defined in momentum space. Compared with the first realistic effective Hamiltonian Kuo–Brown (KB)[317] which was constructed from the $G$-matrix renormalized Hamada–Johnston potential,[318] the realistic Hamiltonian based on the CD-Bonn potential is closer to the empirical interactions in the sense of containing more correlations.[319]

The CISM study using renormalized CD-Bonn potential has been widely achieved in the last two decades. For benchmarking, it well reproduced the nuclear spectra and electromagnetic properties of nuclei around $A = 50$ in the *pf* shell.[320,321] On the one side, the collective properties shown in the odd–odd self-conjugate nuclei were reasonably described.[320] On the other side, the energy differences of the mirror nuclei there were well reproduced thanks to their charge-dependency property as exemplified in Fig. 30.[321]

There is a growing consensus that including the three-body force is important to improve the accuracy of realistic Hamiltonians. To include the three-body force, the feasibility of applying empirical adjustments to parameters such as the SPEs and monopole terms of TBMEs of the realistic Hamiltonian has been discussed.[319] The *pf*-shell Hamiltonian based on the CD-Bonn potential with $V_{\text{low-}k}$ renormalization and the pioneering realistic Hamiltonian KB was employed to describe neutron-rich Ca isotopes. It was shown that the adjustments of the SPEs were sufficient for the CD-Bonn interaction, while both SPEs and monopole terms needed to be adjusted for the KB interaction in order to obtain satisfactory results.

It is more convenient to include the three-body force with the chiral effective potential, which was constructed from chiral symmetry breaking of low-energy Quantum Chromodynamics (QCD).[280,323] The chiral effective potential was first employed in the NCSM due to its nonlocal property.[314] A series of NCSM calculations, including the three-body force of the chiral effective potential, have confirmed the significance of the three-body force in describing the ground-state energies, low-lying levels and electromagnetic transitions, particularly in light nuclei.[314,315,324] Actually, the present computation ability can only support NCSM calculations of $A \leq 20$ nuclei. Therefore, study assuming a frozen core is rather helpful. For this purpose, approaches including the classical and extended many-body perturbation theories (MBPT, also known as the folded-diagram theory)[325] and the newly-developed *ab initio* theories were employed to derive the effective Hamiltonian in the finite model space.[311,312,326] For example, a second renormalization has been proposed so that the NCSM effective Hamiltonian defined in the no-core basis space can be projected onto the valence space beyond a core.[310] One can divide the effective interactions into three types: the phenomenological interactions, the MBPT realistic interactions and the *ab initio* interactions.

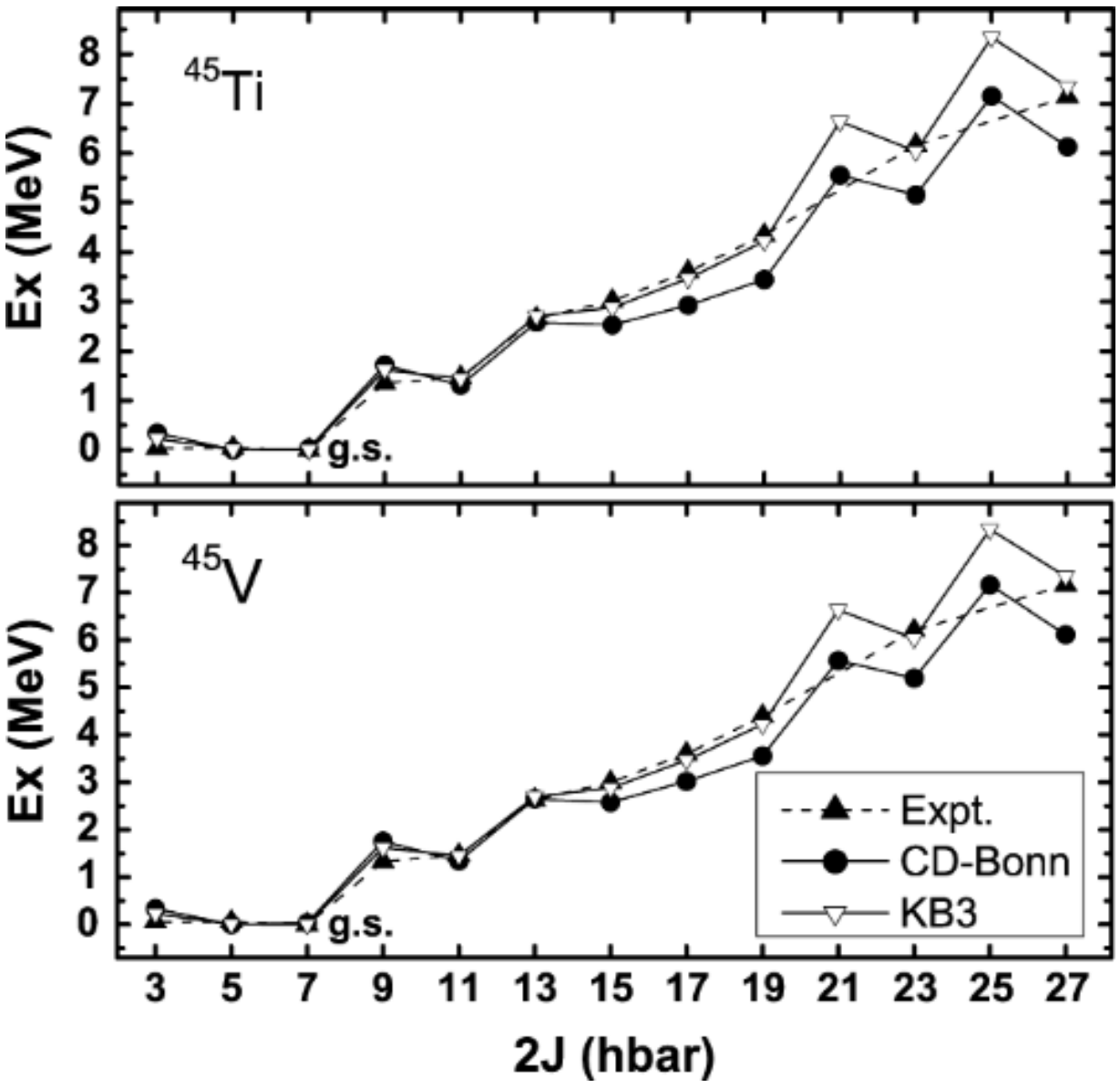

Fig. 30. From Ref. 321. Comparison of experimental and calculated yrast bands for $^{45}$Ti and $^{45}$V. Experimental data are from Ref. 322. KB3 interaction[53] is effective and isospin-conserving.

In such studies, aside from the intrinsic three-body force genuinely arising in the many-body system, the three-body force is as well induced by the two-body potential in the renormalization process. To introduce their contributions to the effective Hamiltonian, a recent study[327] included the three-body force within chiral perturbation theory at next-to-next-to-leading order ($N^2LO$)[314] and added the three-body correlations among valence nucleons due to space truncation. For the later process, the contributions from the monopole components of the second-order three-body diagrams were evaluated and then explicitly added to the preliminary calculated ground-state energies. It should also be noted that the many-body perturbation theory, named as the energy-independent linked-diagram perturbation theory, was employed to project the Hamiltonian onto the valence space. Figure 31 shows ground-state energies of light-mass $N = Z$ nuclei calculated by NCSM and RSM, where the improvement due to the two types of the three-body force could be seen separately. The NCSM calculations, including the three-body potential, are closer to the experimental data than those without them. In the NCSM and RSM calculations, the inclusion of the three-body force corrects the ground-state spin information of $^{10}B$. Compared with NCSM, the RSM explicitly considering the induced three-body correlations obtained satisfactory results of the ground-state energies in light nuclei.

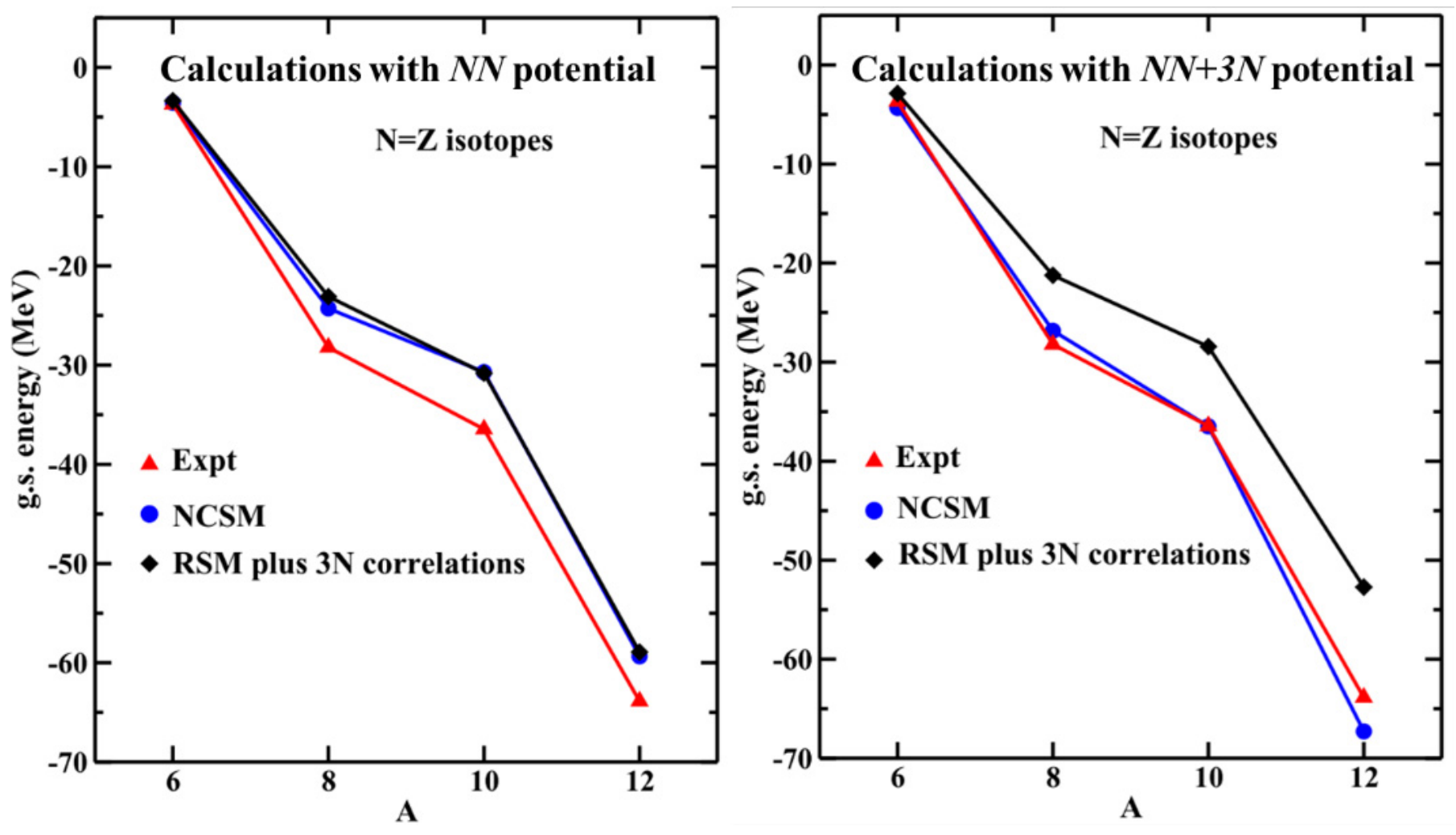

Fig. 31. From Ref. 327. Ground-state energies calculated by NCSM and RSM, without (left) or with (right) $N^2LO$ $3N$ force. Experimental data are from Ref. 88.

Further, the contribution to the monopole component of the effective Hamiltonian from the three-body force was investigated in Ref. 328. Following the procedure in Ref. 327, the authors applied a two-body force at next-to-next-to-next-to leading order ($N^3LO$) and the $N^2LO$ three-body force to construct the realistic effective Hamiltonian in the *pf*-shell. Results of ESPEs showed that the three-body force was necessary for the monopole component, which is believed to govern the shell evolution of $N = 28$. A series of calculations of $S_{2n}$ also illustrated the importance of the three-body force. In Fig. 32, $H^{2NF}$ and $H^{3NF}$ represent the effective Hamiltonian obtained with only two-body force vertices, and that includes three-body force, respectively, while $H^{mon}$ is the multipole part of $H^{2NF}$ plus the monopole component of $H^{3NF}$. Since $H^{2NF}$ did not include the three-body force, it provided a non-satisfactory description of iron and nickel isotopes. Besides, the authors suggested including higher-order perturbative expansion terms and testing the validity of the proposed realistic effective Hamiltonian in heavier and exotic nuclei in future work.

The same framework was also used to study the calcium dripline.[329] With nice reproduction of the bound properties of Ca isotopes no heavier than 60, this work suggested that the dripline of Ca had a mass up to 70, which would drop to 60 without three-body force. The sensitivity of the $S_{2n}$ and energy level of the yrast $2^+$ state to the SPE value of the $0g_{9/2}$ orbit was also investigated to complement this prediction.

It is also of great interest to introduce the progress in the effective Hamiltonian starting from the chiral effective potential derived with *ab initio* approaches. It was proved that the *ab initio* calculations were equivalent to the valence space SM calculations with *ab initio* effective interactions.[310,330] The NCSM effective Hamiltonians without three-body force were compared with MBPT interactions and phenomenological interactions in the *sd*-shell model space in Ref. 330. As a result, all the realistic interactions had a more attractive $T = 1$ channel and, inversely, a more repulsive $T = 0$ channel than phenomenological interactions. This defect was suggested to be removed by including the three-body force.

Researchers investigated *ab initio* effective interactions, including the three-body force in Ref. 298. Within the chiral effective field, interactions from two *ab initio* theories, CC and NCSM, as well as $V_{\text{low-}k}$ interactions, were adopted and compared with the phenomenological USDB interaction. In particular, the importance of the two-body and the three-body force for the CC method and the effects of different cutoffs for $V_{\text{low-}k}$ interactions were investigated. The three microscopic methods, applying to the same potential, led to similar effective interactions, while the dependency of $V_{\text{low-}k}$ interactions on cutoffs was very weak. In contrast, as shown in Fig. 33, the three-body force for the CC method was important for the precision of effective interaction and essential for describing the ground state of neutron-rich oxygen isotopes, which was not the same as that in excited states. In addition, as a part of the two-body force, the spin-orbit (LS) force was proved to have explicit effects on those nuclei.

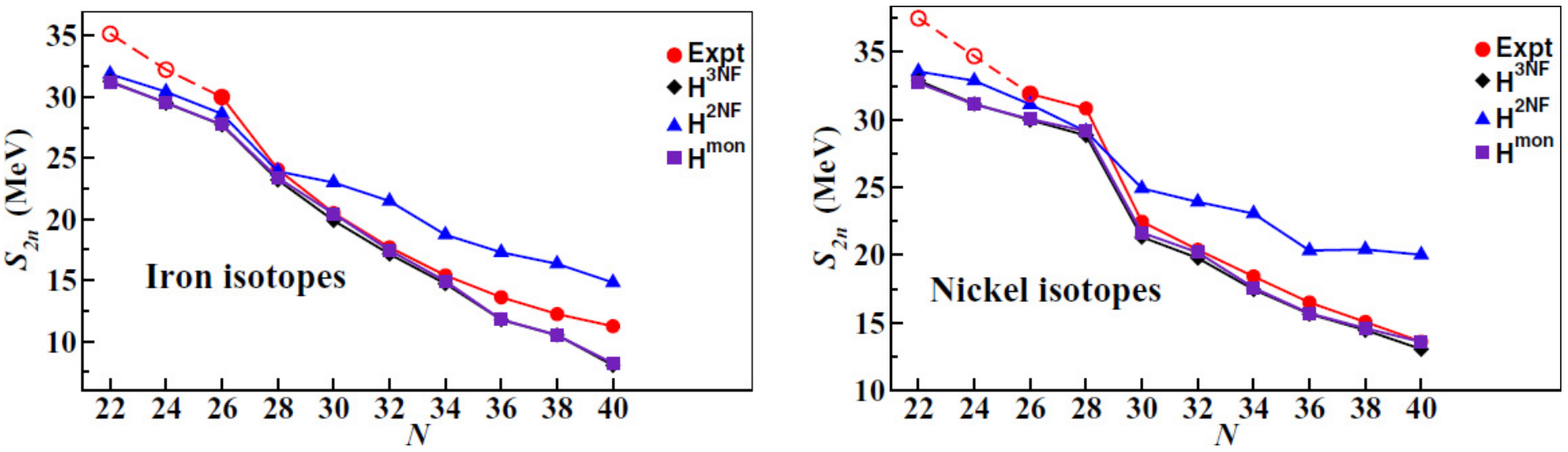

Fig. 32. From Ref. 328. Experimental and calculated $S_{2n}$ values for iron and nickel isotopes. See details in text.

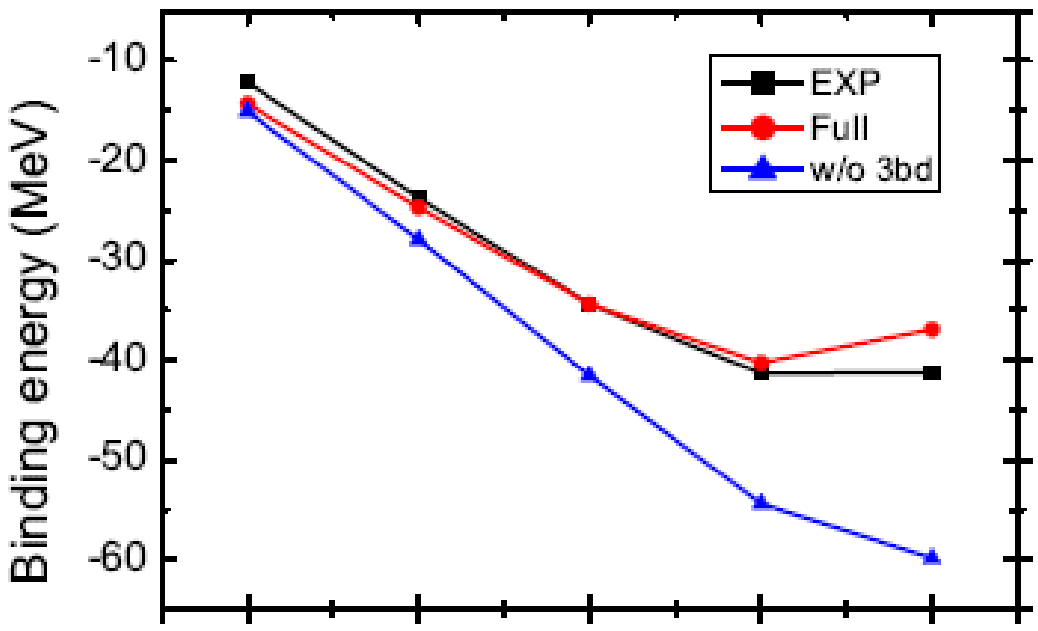

Fig. 33. From Ref. 298. Ground-state energies relative to $^{16}$O of O isotopes, derived by the CC method, with or without the three-body force.

In summary, the shell model study based on the high-precision realistic nuclear force has been flourishingly developed. Special attention has been paid to its description of nuclear structure and convergence rate to benchmark the newly-constructed realistic effective interactions. The inclusion of the three-body force is a crucial subject in the RSM framework, as it is in the *ab initio* studies considering all nucleons active. In the last decade, many-body perturbation theories have been extended to include the three-body force. In addition, *ab initio* methods have been adopted to construct effective interactions for valence-space shell model calculations, replacing direct calculations with *ab initio* theories and focusing on the three-body forces.

#### 3.2.3.2. GSM with realistic nuclear forces

The combination of the Gamow basis and the realistic nuclear force has been attempted many times in recent years to study exotic nuclei more strictly and precisely. The realistic Hamiltonian was, for the first time, combined with GSM in Ref. 42. The application of $^{6}$He and $^{18}$O proved the feasibility of these proposed methods. We briefly introduce some recent progress in the GSM using the realistic Hamiltonians in the following.

Compared with the standard CISM, treating bound, resonant and continuum states on equal footing brings merit and demerit to the Gamow basis, concretely, the weakly bound and unbound nuclear system can be better described, while very few nuclei can be treated. At present, the applicability of GSM is rather limited. It has been attempted to investigate the very few body systems in the GSM framework without a core. The no-core GSM starting from the chiral $N^3LO$ potential was proposed nearly a decade ago[301] and applied to $^3H$,[301] $^{4,5}He$,[301] tri-neutron[303] and tetra-neutron.[302,303] In Ref. 303, the *ab initio* no-core GSM based on the chiral $N^3LO$ nuclear force was used to figure out the existence of tri-neutron and tetra-neutron systems. Tri-neutron was predicted to be observed experimentally easier than tetra-neutron. The calculated energy and decay width of resonance in tetra-neutron were within the range of experimental error. As shown in Fig. 34, the predicted energy and resonance width of the tetra-neutron have been confirmed by the latest experimental study.[331]

While for heavier systems, one usually considers few valence particles in the model space, with minority valence particles occupying the Gamow (continuum) basis and the rest in the HO or WS basis. Actually, He, O and Ca isotopes near the shell closures are typical target nuclei. With the recent development of GSM, special attention was paid to the location of the neutron dripline of O and Ca isotopes. The realistic Hamiltonian was, for the first time, combined with GSM.[42] In Ref. 42, the single-particle basis was built from the renormalized $N^3LO$ nuclear force with the self-consistent Hartree–Fock method. The residual interaction was expressed with a finite set of HO functions in the built basis. The chiral $N^3LO$ potential is soften with the $V_{\text{low-}k}$ approach, and the effective interaction is obtained with MBPT (the degenerate $Q$-box approach). The proposed framework was applied to $^6He$ and $^{18}O$ with only two valence neutrons in the model space[42] and to nuclei with up to three valence nucleons beyond the $^{22}$O core.[333]

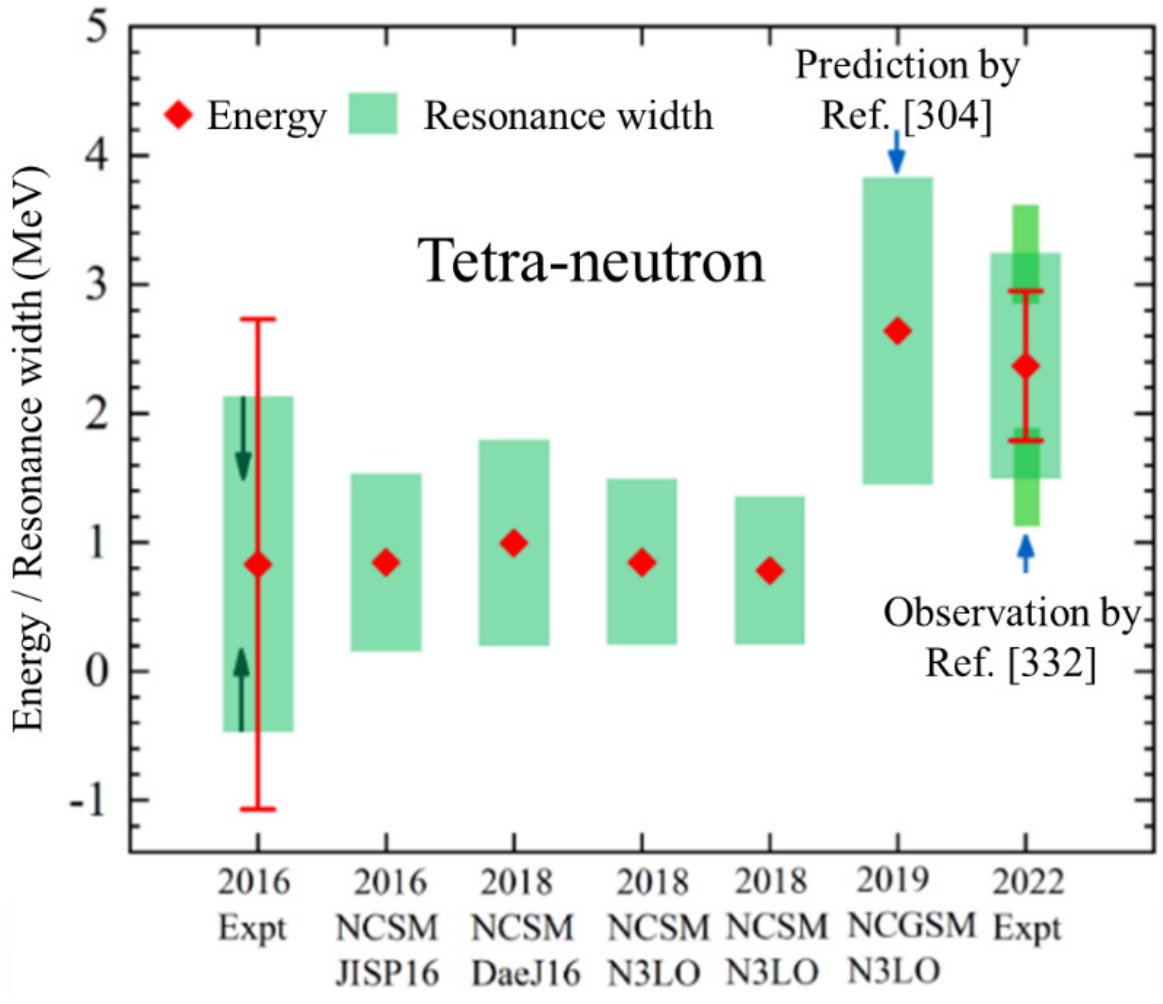

Fig. 34. From Ref. 332. Observed and theoretical energy and resonance width of tetra-neutron system.

Researchers have also developed a more sophisticated GSM[334] by exploiting the full $Q$-box folded-diagram (EKK method[335]) to the non-degenerate Berggren complex-$k$ space to construct the realistic effective interaction on the Gamow Wood–Saxon basis. The GSM calculations based on the $V_{\text{low-}k}$ CD-Bonn potential gave a nice description of both bound and resonant states in $^{24}$O and $^{25}$O, and a reasonable prediction of low-lying resonant states for $^{26}$O.[334] The developed method was also adopted to calculate calcium isotopes comprehensively in Ref. 336. The calculated binding energies agreed well with existing experimental results and predicted that 57 and 70 were, respectively, the mass number of the heaviest bound odd and even calcium isotopes. Furthermore, shell evolutions in the calcium chain were well explained, and resonant states were predicted. The continuum effect was believed to be significant in the neutron-rich Ca isotopes predicted to be weakly bound.

Later, the developed MBPT procedures were extended to construct both the SPEs and TBMEs of the multi-shell effective interaction within the Gamow Hartree–Fock basis.[337] The *sdpf*-shell interaction was constructed based on the optimized chiral potential $\text{NNLO}_{\text{opt}}$,[338] which nicely described the nuclear structure without resorting to three-body forces. The good convergence of the Gamow Hartree–Fock basis within the MBPT calculations was confirmed. With the proposed method, the bound properties of most oxygen and fluorine isotopes, as well as the spectra of $^{26}$O and $^{26}$F, were nicely reproduced. The description of the neutron-rich nuclei was significantly improved because of including the continuum coupling and *pf* shell, while more efforts remained to describe fluorine isotopes near the neutron driplines well.

Meanwhile, the effects of three-body forces were also phenomenologically included by adding a mass dependence. Based on such a method, in Ref. 339, the GSM with effective field theory type interaction was used to describe the proton-rich $^{16}$O isotones. A nice agreement between the experimental energy spectra and the GSM calculations was achieved. The long-range Coulomb force was well treated in GSM. And the Coulomb contributions of each low-lying state of $^{18}$Ne and $^{20}$Mg were calculated. In addition, the importance of three-body force and continuum coupling was confirmed. Moreover, the prediction ability for reaction properties of the GSM was suggested to be expectable with its good performance on the $^{18}$Ne $(p, p)$ reaction. The effects of mass dependence were also discussed for O isotopes near the neutron drip line.[340] The chiral nucleon–nucleon potential at leading-order (LO) was softened with different momentum cutoffs to construct the effective interaction in the *sdpf* model space.[340] It was found that the mass-dependence was needed mainly for the nuclear force to soften with a small momentum cutoff. Calculations with mass-independence and medium momentum cutoffs reproduced the location of the oxygen neutron drip line and suggested the weakly unbound properties of $^{28}$O. It is noteworthy that the calculated ground-state energies of $^{24-28}$O are of high precision, which was challenging for *ab initio* studies.

Table 6. Summary of the calculated resonance width (in MeV) in $^{24-26}$O with different effective Hamiltonians, compared with available experimental data.

| Work | Origin | Potential | Basis | Space |
|---|---|---|---|---|
| Exp | Refs. 88 and 344 | | | |
| Th1 | Ref. 334 | CD-Bonn | WS | $1s_{1/2}$, $0d_{5/2}$, $0d_{3/2}+d_{3/2}$-continuum |
| Th2 | Ref. 337 | NNLO$_{\rm opt}$ | HF | (bound: $\pi 1s_{1/2}$, $\pi 0d_{3/2}$, $\pi 0d_{5/2}$, $\nu 1s_{1/2}$); (resonance: $\nu 0d_{3/2}$, $\nu 1p_{1/2}$, $\nu 1p_{3/2}$); (scattering continua: $\nu d_{3/2}$, $\nu p_{1/2}$, $\nu p_{3/2}$, $\nu f_{7/2}$) |
| Th3 | Ref. 340 | LO | WS | *sdpf*, with *f* partial waves in the HO basis |
| Th4 | Ref. 341 | N$^3$LO | WS | $0d_{5/2}$, $1s_{1/2}$, $0d_{3/2}$, $d_{3/2}$-continuum |
| Th5 | Ref. 341 | N$^3$LO (2N) + N$^2$LO (3N) | WS | $0d_{5/2}$, $1s_{1/2}$, $0d_{3/2}$, $d_{3/2}$-continuum |

| Work | $^{24}$O $(0^+)$ | $^{24}$O$(2^+)$ | $^{24}$O$(1^+)$ | $^{25}$O$(3/2^+)$ | $^{26}$O$(0^+)$ | $^{26}$O$(2^+)$ |
|---|---|---|---|---|---|---|
| Exp | 0 | $0.05^{+21}_{-5}$ | $0.03^{+12}_{-3}$ | 0.088 | 0 | 0 |
| Th1 | 0 | 0.11 | 0.12 | 0.12 | 0.230 | 0.25 |
| Th2 | | | | | 0.029 | 0.085 |
| Th3 | 0 | 0.108 | 0.216 | 0.084 | 0 | 0.098 |
| Th4 | 0 | 0.008 | 0.015 | 0.004 | 0.021 | 0.004 |
| Th5 | 0 | 0.02 | 0.049 | 0.058 | 0.015 | 0.097 |

Recently the chiral three-body force has been introduced in the GSM framework. Based on the same MBPT method extended in Ref. 334, the *sd*-shell effective interaction starting from chiral N$^3$LO two-body force and N$^2$LO three-body force has been constructed for the GSM calculations.[341] Considering the computational cost, the normal-ordering approach[342] was used, and only the three-body effects in the model space were included. The constructed interaction has been applied to $^{18-28}$O.[341] The three-body force was confirmed again to reproduce the oxygen neutron-drip line and nearby nuclei.[341] The effective interaction beyond the $^{14}$O core has been constructed in the same way to investigate the Borromean nucleus $^{17}$Ne.[343] Results indicated that the three-body force was essential to reproduce the ground-state energy and the Borromean structure of $^{17}$Ne, and the continuum effect was crucial for the density. While the effective interactions in Refs. 341 and 343 were constructed in the Gamow WS basis, and an *sd*-shell interaction has as well been constructed in the Gamow Hartree–Fock basis to compare the neutron-rich O isotopes with their mirror partners. Results indicated that including three-body force improved the calculation precision of the ground-state and excited energies. Particularly, the ground-state spins of $^{19}$O and $^{19}$Na were corrected by including the three-body force. The inclusion of the continuum coupling was crucial for the separation energies of $^{26}$O, which are employed to identify the unbound structure of $^{26}$O. The weakly bound effect of the $\pi s_{1/2}$ orbit was highlighted to understand the energy differences between O isotopes and their mirror partners, which is consistent with conclusions in Sec. 3.1.2.

In the past few years, with the development of the *ab initio* GSM methods, including the extension of MPBT methods and self-consistent approaches, the inclusion of cross-shell interactions and three-body forces, the overall precision of light and medium mass nuclei, especially the exotic oxygen isotopes, have been further improved. Table 6 summarizes part of the calculated results, comparing the calculated resonance width of states in $^{24-26}$O.

To conclude, two statements have been highlighted through the recent development of *ab initio* GSM studies: On the one hand, the continuum coupling plays a significant role in weakly bound nuclei, as indicated in the GSM studies with phenomenological interactions; on the other hand, the three-body force is crucial to improve the theoretical predictions, as suggested in other *ab initio* approaches. The inclusion of continuum coupling and the three-body forces helps study exotic nuclei more strictly and precisely.

## 4. Conclusion

The nuclear shell model stands as one of the most significant theoretical frameworks for nuclear structure studies. However, the incomplete knowledge of the nuclear force and the limited computation ability constrain the application of the shell model, especially for exotic nuclei or nuclei far from the doubly magic nuclei. The CISM has been developed as a modern shell model to solve such problems partially. It includes the residual interaction, calculates in the truncated model space and uses the configuration mixing to express each state.

This review aims to provide an overview of the recent progress of the CISM and its related approaches. We shed light on the phenomenological effective Hamiltonians which have been proposed or improved to understand better the nuclear structure, the importance-truncation-related methods including NPA approach and the MCSM, the generalized shell models using different basis including the PSM and the GSM and finally the extension of shell model to the *ab initio* theories.

In recent decades, several phenomenological effective Hamiltonians have been constructed or (and) improved so that the applicability and accuracy of CISM have been improved. In the light-mass neutron-rich side, cross-shell excitation has been systematically investigated. The YSOX interaction, more exactly considering cross-shell excitation, has partially solved the puzzle of C, N and O driplines as well as the long lifetime of $^{14}$C.

In the *sd* region, the weakly-bound effect of the $1s_{1/2}$ orbit has been emphasized. Modified USD Hamiltonians, which include this effect, have been successfully applied to proton-rich nuclei, highlighting the difference from their mirror counterparts. These modified Hamiltonians have nicely interpreted the latest experimental progress achieved by the RIBLL collaboration, particularly regarding the $^{22}\mathrm{Si} \to {}^{22}\mathrm{Al}$, $^{26}\mathrm{P} \to {}^{26}\mathrm{Si}$, and $^{27}\mathrm{S} \to {}^{27}\mathrm{P}$ decays.

In the medium and heavy mass regions where few Hamiltonians were constructed a decade ago, a series of Hamiltonians based on $\mathrm{V_{MU}}$ have been developed and improved. They help investigate the systematics and the new isomers (nuclides). As a result, physical mechanisms, such as the shell evolution in In and Ag isotopes and the seniority conservation in $(9/2)^4$ systems and Sn and Te isotopic chain, have been better understood. In addition, an exploration of the unified effective nuclear force to describe the medium and heavy nuclei near $^{132}$Sn and $^{208}$Pb has been primarily employed. The $\mathrm{V_{MU}}$ plus M3Y spin-orbit force results in a root mean square error of around 0.2 MeV for 825 energy levels.

However, some areas of nuclear interaction warrant further investigation, such as the cross-shell interaction, the off-diagonal TBMEs, the proton-neutron interaction, and the pairing interaction. Deeper studies are planned on the unified effective nuclear force in the medium and heavy mass regions.

Meanwhile, related shell models have been developed to reduce computation dimensions, investigate largely deformed nuclei, include continuum effects, or start more fundamentally. The combination of advantages from several of these methods could be of great interest to future research.

## Acknowledgments

The authors appreciate fruitful discussions from Takaharu Otsuka, Toshio Suzuki, Furong Xu, Noritaka Shimizu, Chong Qi, Guanjian Fu, Yang Lei, Hankui Wang, Changfeng Jiao, Zhonghao Sun, Baishan Hu, Yuanzhuo Ma, Jianguo Li and many other collaborators. This research was funded by the Guangdong Major Project of Basic and Applied Basic Research under Grant No. 2021B0301030006, the National Natural Science Foundation of China under Grant No. 11775316, the computational resources from SYSU, the National Supercomputer Center in Guangzhou and the Key Laboratory of High Precision Nuclear Spectroscopy, Institute of Modern Physics, Chinese Academy of Sciences.

## ORCID

Menglan Liu https://orcid.org/0000-0001-8028-1487
Cenxi Yuan https://orcid.org/0000-0002-3495-3614

## References

1. M. G. Mayer, *Phys. Rev.* **75** (1949) 1969.
2. O. Haxel, J. H. D. Jensen and H. E. Suess, *Phys. Rev.* **75** (1949) 1766.
3. E. Caurier, G. Martínez-Pinedo, F. Nowacki, A. Poves and A. P. Zuker, *Rev. Mod. Phys.* **77** (2005) 427.
4. J. Rainwater, *Phys. Rev.* **79** (1950) 432.
5. J. P. Elliott and P. H. Flowers, *Proc. R. Soc. Lond. Ser. A. Math. Phys. Sci.* **229** (1954) 536.
6. J. P. Elliott, *Proc. R. Soc. Lond. Ser. A. Math. Phys. Sci.* **245** (1958) 128.
7. K. Heyde, P. Van Isacker, R. F. Casten and J. L. Wood, *Phys. Lett. B* **155** (1985) 303.
8. E. K. Warburton, J. A. Becker and B. A. Brown, *Phys. Rev. C* **41** (1990) 1147.
9. P. G. Hansen and B. Jonson, *Europhys. Lett.* **4** (1987) 409.
10. L. Gaudefroy *et al.*, *Phys. Rev. Lett.* **109** (2012) 202503.
11. T. Nakamura *et al.*, *Phys. Rev. Lett.* **103** (2009) 262501.
12. C. Thibault, R. Klapisch, C. Rigaud, A. M. Poskanzer, R. Prieels, L. Lessard and W. Reisdorf, *Phys. Rev. C* **12** (1975) 644.
13. D. D. Warner, M. A. Bentley and P. Van Isacker, *Nat. Phys.* **2** (2006) 311.
14. M. A. Bentley and S. M. Lenzi, *Prog. Part. Nucl. Phys.* **59** (2007) 497.
15. C. Yuan, C. Qi, F. Xu, T. Suzuki and T. Otsuka, *Phys. Rev. C* **89** (2014) 044327.
16. R. V. Janssens, *Nature* **459** (2009) 1069.
17. R. Kanungo *et al.*, *Phys. Rev. Lett.* **102** (2009) 152501.
18. C. R. Hoffman *et al.*, *Phys. Lett. B* **672** (2009) 17.
19. M. Hjorth-Jensen, T. T. S. Kuo and E. Osnes, *Phys. Rep.* **261** (1995) 125.
20. B. A. Brown, *Prog. Part. Nucl. Phys.* **47** (2001) 517.
21. S. Cohen and D. Kurath, *Nucl. Phys.* **73** (1965) 1.
22. T. T. S. Kuo and G. E. Brown, *Nucl. Phys. A* **114** (1968) 241 (1968).
23. B. A. Brown, W. A. Richter, R. E. Julies and B. H. Wildenthal, *Ann. Phys.* **182** (1988) 191.
24. D. S. Ahn *et al.*, *Phys. Rev. Lett.* **123** (2019) 212501.
25. C. X. Yuan, T. Suzuki, T. Otsuka, F. Xu and N. Tsunoda, *Phys. Rev. C* **85** (2012) 064324.
26. R. Han *et al.*, *Phys. Lett. B* **772** (2017) 529.
27. A. Revel *et al.*, *Phys. Rev. Lett.* **124** (2020) 152502.
28. C. X. Yuan, M. L. Liu and Y. L. Ge, *Nucl. Phys. R* **37** (2020) 447.
29. T. Mizusaki, *Riken Acce. Prog. Rep.* **33** (2000) 14.

30. S. Noritaka, M. Takahiro, U. Yutaka and T. Yusuke, *Comput. Phys. Commun.* **244** (2019) 372.
31. N. Shimizu, arXiv:1310.5431 (2013).
32. C. Yuan, Y. Ge, M. Liu, G. Chen and B. Cai, *EPJ Web Conf.* **239** (2020) 04002.
33. M. M. Zhang *et al.*, *Phys. Lett. B* **800** (2020) 135102.
34. C. Yuan, M. Liu, N. Shimizu, Z. Podolyák, T. Suzuki, T. Otsuka and Z. Liu, *Phys. Rev. C* **106** (2022) 044314.
35. J. Erler, N. Birge, M. Kortelainen, W. Nazarewicz, E. Olsen, A. M. Perhac and M. Stoitsov, *Nature* **486** (2012) 509.
36. A. Arima and F. Iachello, *Ann. Phys.* **99** (1976) 253.
37. M. Honma, T. Mizusaki and T. Otsuka, *Phys. Rev. Lett.* **75** (1995) 1284.
38. M. Honma, T. Mizusaki and T. Otsuka, *Phys. Rev. Lett.* **77** (1996) 3315.
39. T. Mizusaki, M. Honma and T. Otsuka, *Phys. Rev. C Nucl. Phys.* **53** (1996) 2786.
40. T. Otsuka, M. Honma, T. Mizusaki, N. Shimizu and Y. Utsuno, *Prog. Part. Nucl. Phys.* **47** (2001) 319.
41. K. Hara and Y. Sun, *Int. J. Mod. Phys. E* **4** (1995) 637.
42. G. Hagen, M. Hjorth-Jensen and N. Michel, *Phys. Rev. C* **73** (2006) 064307.
43. J. F. Dawson, I. Talmi and J. D. Walecka, *Ann. Phys.* **18** (1962) 339.
44. L. Coraggio, A. Covello, A. Gargano, N. Itaco and T. T. S. Kuo, *Prog. Part. Nucl. Phys.* **62** (2008) 135.
45. M. Goeppert and J. J. Hans. D., *Angew. Chem.* **68** (1956) 360.
46. P. Doornenbal *et al.*, *Phys. Rev. C* **95** (2017) 041301(R).
47. Y. Utsuno, T. Otsuka, T. Mizusaki and M. Honma, *Phys. Rev. C* **60** (1999) 054315.
48. M. Liu and C. Yuan, *Symmetry* **13** (2021) 2167.
49. S. Cohen and D. Kurath, *Nucl. Phys. A* **101** (1967) 1.
50. S. Cohen and D. Kurath, *Nucl. Phys. A* **141** (1970) 145.
51. B. A. Brown and B. H. Wildenthal, *Annu. Rev. Nucl. Part. Sci.* **38** (1988) 29.
52. A. Poves, J. Sánchez-Solano, E. Caurier and F. Nowacki, *Nucl. Phys. A* **694** (2001) 157.
53. A. Poves and A. Zuker, *Phys. Rep.* **70** (1981) 235.
54. M. Honma, T. Otsuka, B. A. Brown and T. Mizusaki, *Eur. Phys. J. A* **25** (2005) 499.
55. E. K. Warburton and B. A. Brown, *Phys. Rev. C* **46** (1992) 923.
56. H. A. Bethe, *Phys. Rev.* **57** (1940) 260.
57. H. A. Bethe, *Phys. Rev.* **57** (1940) 390.
58. L. Eisenbud and E. P. Wigner, *Proc. Natl. Acad. Sci. USA* **27** (1941) 281.
59. T. Otsuka, T. Suzuki, R. Fujimoto, H. Grawe and Y. Akaishi, *Phys. Rev. Lett.* **95** (2005) 232502.
60. R. K. Bansal and J. B. French, *Phys. Lett.* **11** (1964) 145.
61. T. Otsuka, A. Gade, O. Sorlin, T. Suzuki and Y. Utsuno, *Rev. Mod. Phys.* **92** (2020) 015002.
62. M. Dufour and A. P. Zuker, *Phys. Rev. C* **54** (1996) 1641.
63. B. A. Brown and W. A. Richter, *Phys. Rev. C* **74** (2006) 034315.
64. N. C. Francis and K. M. Watson, *Phys. Rev.* **92** (1953) 291.
65. K. M. Watson, *Phys. Rev.* **89** (1953) 575.
66. D. J. Millener and D. Kurath, *Nucl. Phys. A* **255** (1975) 315.
67. T. Suzuki, R. Fujimoto and T. Otsuka, *Phys. Rev. C* **67** (2003) 044302.
68. Y. Utsuno, T. Otsuka, T. Mizusaki and M. Honma, *Phys. Rev. C* **60** (1999) 054315.
69. T. Otsuka, T. Suzuki, M. Honma, Y. Utsuno, N. Tsunoda, K. Tsukiyama and M. Hjorth-Jensen, *Phys. Rev. Lett.* **104** (2010) 012501.
70. G. Bertsch, J. Borysowicz, H. McManus and W. G. Love, *Nucl. Phys. A* **284** (1977) 399.

71. Y. Utsuno, T. Otsuka, B. A. Brown, M. Honma, T. Mizusaki and N. Shimizu, *Phys. Rev. C* **86** (2012) 051301(R).
72. F. Nowacki and A. Poves, *Phys. Rev. C* **79** (2009) 014310.
73. M. Hasegawa and K. Kaneko, *Phys. Rev. C* **59** (1999) 1449.
74. H.-K. Wang, S. K. Ghorui, K. Kaneko, Y. Sun and Z. H. Li, *Phys. Rev. C* **96** (2017) 054313.
75. H.-K. Wang, K. Kaneko and Y. Sun, *Phys. Rev. C* **89** (2014) 064311.
76. H.-K. Wang, K. Kaneko and Y. Sun, *Phys. Rev. C* **91** (2015) 021303(R).
77. H.-K. Wang, K. Kaneko, Y. Sun, Y.-Q. He, S.-F. Li and J. Li, *Phys. Rev. C* **95** (2017) 011304(R).
78. E. K. Warburton, *Phys. Rev. C* **44** (1991) 233.
79. E. K. Warburton and B. A. Brown, *Phys. Rev. C* **43** (1991) 602.
80. C. Lanczos, *J. Res. Natl. Bur. Stand.* **45** (1950) 255.
81. E. R. Davidson, *J. Comput. Phys.* **17** (1975) 87.
82. E. Caurier, *Nucl. Phys. A* **704** (2002) 60.
83. E. Caurier, G. Martínez-Pinedo, F. Nowacki, A. Poves, J. Retamosa and A. P. Zuker, *Phys. Rev. C* **59** (1999) 2033.
84. Oxbash for Windows PC, www.nscl.msu.edu/∼brown.
85. NuShellX, http://www.garsington.eclipse.co.uk/.
86. C. W. Johnson, W. Erich Ormand, K. S. McElvain and H. Shan, BIGSTICK: A flexible configuration-interaction shell-model code, arXiv:1801.08432v1.
87. T. Otsuka, T. Suzuki, J. D. Holt, A. Schwenk and Y. Akaishi, *Phys. Rev. Lett.* **105** (2010) 032501.
88. NNDC Databases, NuDat 3.0, https://www.nndc.bnl.gov/nudat2/.
89. A. G. Artukh, V. V. Avdeichikov, L. P. Chelnokov, G. F. Gridnev, V. L. Mikheev, V. I. Vakatov, V. V. Volkov and J. Wilczynski, *Phys. Lett. B* **32** (1970) 43.
90. H. Sakurai *et al.*, *Phys. Lett. B* **448** (1999) 180.
91. G. Audi, A. H. Wapstra and C. Thibault, *Nucl. Phys. A* **729** (2003) 337.
92. N. Kobayashi *et al.*, *Phys. Rev. C* **86** (2012) 054604.
93. T. Suzuki, T. Otsuka, C. Yuan and N. Alahari, *Phys. Lett. B* **753** (2016) 199.
94. J. Chen *et al.*, *Phys. Rev. C* **100** (2019) 064314.
95. C. R. Hoffman *et al.*, *Phys. Rev. C* **88** (2013) 044317.
96. Y. Jiang *et al.*, *Phys. Rev. C* **101** (2020) 024601.
97. M. Stanoiu *et al.*, *Phys. Rev. C* **78** (2008) 034315.
98. D. Sohler *et al.*, *Phys. Rev. C* **77** (2008) 044303.
99. C. X. Yuan, C. Qi and F. R. Xu, *Chin. Phys. C* **33** (2009) 55.
100. C. X. Yuan, C. Qi and F. R. Xu, *Nucl. Phys. A* **883** (2012) 25.
101. D. R. Tilley, H. R. Weller and C. M. Cheves, *Nucl. Phys. A* **564**, (1993) 1.
102. D. R. Tilley, H. R. Weller, C. M. Cheves and R. M. Chasteler, *Nucl. Phys. A* **595** (1995) 1.
103. T. Aumann *et al.*, *Phys. Rev. Lett.* **84** (2000) 35.
104. J. Chen *et al.*, *Phys. Lett. B* **781** (2018) 412.
105. J. Chen *et al.*, *Phys. Rev. C* **98** (2018) 014616.
106. R. Kanungo *et al.*, *Phys. Lett. B* **682** (2010) 391.
107. C. X. Yuan, *Chin. Phys. C* **41** (2017) 104102.
108. C. X. Yuan, *Nucl. Phys. R* **33** (2016) 246.
109. Y. Kanada-En'yo and T. Suhara, *Phys. Rev. C* **89** (2014) 044313.
110. S. Aroua, P. Navrátil, L. Zamick, M. S. Fayache, B. R. Barrett, J. P. Vary and K. Heyde, *Nucl. Phys. A* **720** (2003) 71.
111. F. Ajzenberg-Selove, *Nucl. Phys. A* **523** (1991) i.
112. S. Raman, C. W. Nestor and P. Tikkanen, *At. Data Nucl. Data Tables* **78** (2001) 1.

113. A. Negret *et al.*, *Phys. Rev. Lett.* **97** (2006) 062502.
114. R. Lică *et al.*, *Phys. Rev. C* **95** (2017) 021301(R).
115. R. Han *et al.*, *Sci. China Phys. Mech.* **60** (2017) 042021.
116. C.-B. Moon and C. Yuan, *Chin. Phys. C* **46** (2022) 124102.
117. D. Bazin *et al.*, *Phys. Rev. C* **103** (2021) 064318.
118. L.-C. Tao *et al.*, *Chin. Phys. Lett.* **36** (2019) 062101.
119. Y.-F. Lv *et al.*, *Chin. Phys. C* **43** (2019) 104102.
120. S. Y. Jin *et al.*, *Phys. Rev. C* **104** (2021) 024302.
121. H.-K. Wang, Z.-H. Li, C.-X. Yuan, Z.-Q. Chen, N. Wang, W. Qin and Y.-Q. He, *Chin. Phys. C* **43** (2019) 054101.
122. Y. Ichikawa *et al.*, *Phys. Rev. C* **80** (2009) 044302.
123. B. Blank and M. J. G. Borge, *Prog. Part. Nucl. Phys.* **60** (2008) 403.
124. Y. B. Wang *et al.*, *Phys. Rev. C* **103** (2021) L011301.
125. C. G. Wu *et al.*, *Phys. Rev. C* **104** (2021) 044311.
126. M. D. Cable, J. Honkanen, R. F. Parry, H. M. Thierens, J. M. Wouters, Z. Y. Zhou and J. Cerny, *Phys. Rev. C* **26** (1982) 1778.
127. M. D. Cable, J. Honkanen, R. F. Parry, S. H. Zhou, Z. Y. Zhou and J. Cerny, *Phys. Rev. Lett.* **50** (1983) 404.
128. N. L. Achouri *et al.*, *Eur. Phys. J. A* **27** (2006) 287.
129. S. Czajkowski *et al.*, *Nucl. Phys. A* **616** (1997) 278.
130. X. X. Xu *et al.*, *Phys. Lett. B* **766** (2017) 312.
131. L. J. Sun *et al.*, *Phys. Rev. C* **99** (2019) 064312.
132. J. Lee *et al.*, *Phys. Rev. Lett.* **125** (2020) 192503.
133. P. F. Liang *et al.*, *Phys. Rev. C* **101** (2020) 024305.
134. L. J. Sun *et al.*, *Phys. Lett. B* **802** (2020) 135213.
135. H. Jian *et al.*, *Symmetry* **13** (2021) 2278.
136. G. Z. Shi *et al.*, *Phys. Rev. C* **103** (2021) L061301.
137. J. J. Liu *et al.*, *Phys. Rev. Lett.* **129** (2022) 242502.
138. Z. Sun, W. L. Zhan, Z. Y. Guo, G. Xiao and J. X. Li, *Nucl. Instrum. Meth. Phys. Res. A, Accel. Spectrom. Detect. Assoc. Equip.* **503** (2003) 496.
139. M. Wang, W. J. Huang, F. G. Kondev, G. Audi and S. Naimi, *Chin. Phys. C* **45** (2021) 030003.
140. M. Wang, G. Audi, F. G. Kondev, W. J. Huang, S. Naimi and X. Xu, *Chin. Phys. C* **41** (2017) 030003.
141. L. Weissman *et al.*, *J. Phys. G Nucl. Part. Phys.* **31** (2005) 553.
142. M. S. Basunia and A. M. Hurst, *Nucl. Data Sheets* **134** (2016) 1.
143. M. Shamsuzzoha Basunia, *Nucl. Data Sheets* **112** (2011) 1875.
144. B. Moon *et al.*, *Phys. Rev. C* **95** (2017) 044322.
145. B. Moon *et al.*, *Phys. Rev. C* **96** (2017) 014325.
146. V. E. Viola and G. T. Seaborg, *J. Inorg. Nucl. Chem.* **28** (1966) 697.
147. B. Cai, G. Chen, J. Xu, C. Yuan, C. Qi and Y. Yao, *Phys. Rev. C* **101** (2020) 054304.
148. C. Qi, A. N. Andreyev, M. Huyse, R. J. Liotta, P. Van Duppen and R. A. Wyss, *Phys. Rev. C* **81** (2010) 064319.
149. H. Watanabe *et al.*, *Phys. Rev. Lett.* **111** (2013) 152501.
150. G. S. Simpson *et al.*, *Phys. Rev. Lett.* **113** (2014) 132502.
151. P. Lee *et al.*, *Phys. Rev. C* **92** (2015) 044320.
152. Y. M. Xing *et al.*, *Phys. Rev. C* **107** (2023) 014304.
153. H. B. Zhou *et al.*, *Phys. Rev. C* **103** (2021) 044314.
154. X. Xu *et al.*, *Phys. Rev. C* **100** (2019) 051303(R).
155. Z. Q. Chen *et al.*, *Phys. Rev. Lett.* **122** (2019) 212502.

156. H. Watanabe *et al.*, *Phys. Lett. B* **823** (2021) 136766.
157. B. Moon, C. B. Moon, G. D. Dracoulis, R. A. Bark, A. P. Byrne, P. A. Davidson, G. J. Lane, T. Kibédi, A. N. Wilson and C. Yuan, *Phys. Rev. C* **100** (2019) 024319.
158. B. Moon *et al.*, *Phys. Lett. B* **782** (2018) 602.
159. T. Otsuka and Y. Tsunoda, *J. Phys. G Nucl. Part. Phys.* **43** (2016) 024009.
160. J. K. Hwang *et al.*, *Phys. Rev. C* **65** (2002) 054314.
161. C. X. Yuan *et al.*, *Phys. Lett. B* **762** (2016) 237.
162. V. H. Phong *et al.*, *Phys. Rev. C* **100** (2019) 011302(R).
163. Z. Y. Zhang *et al.*, *Phys. Rev. C* **89** (2014) 014308.
164. M. H. Huang *et al.*, *Phys. Lett. B* **834** (2022) 137484.
165. H. B. Yang *et al.*, *Phys. Rev. C* **105** (2022) L051302.
166. L. Ma *et al.*, *Phys. Rev. C* **91** (2015) 051302(R).
167. H. B. Yang *et al.*, *Eur. Phys. J. A* **51** (2015) 88.
168. Z. Y. Zhang *et al.*, *Phys. Rev. Lett.* **126** (2021) 152502.
169. Z. Y. Zhang *et al.*, *Phys. Rev. Lett.* **122** (2019) 192503.
170. M. D. Sun *et al.*, *Phys. Lett. B* **771** (2017) 303.
171. T. H. Huang *et al.*, *Phys. Rev. C* **98** (2018) 044302.
172. C. B. Li *et al.*, *Phys. Rev. C* **101** (2020) 044313.
173. A. E. Champagne, C. Iliadis and R. Longland, *AIP Adv.* **4** (2014) 041006.
174. V. Artisyuk, C. Broeders, E. Gonzalez-Romero, W. Gudowski, A. Ignatyuk, A. Konobeyev, Y. Korovin, G. Pilnov, A. Stankovskiy and Y. Titarenko, *Prog. Nucl. Energy* **50** (2008) 341.
175. E. Sartori, *Ann. Nucl. Energy* **62** (2013) 579.
176. S. M. Qaim, *J. Nucl. Med. Biol.* **44** (2017) 31.
177. C. Yuan, *EPJ Web Conf.* **178** (2018) 02016.
178. C. Yuan, in *Recent Nuclear Structure Study through Large Scale Shell Model, from Light to Heavy Nuclei - Proc. Ito Int. Research Center Symp.* "*Perspectives of the Physics of Nuclear Structure*" (JPS, Japan, 2018) 012026.
179. C. Yuan, *Phys. Rev. C* **93**, 034310 (2016).
180. M. Liu, Y. Gao and N. Wang, *Chin. Phys. C* **41** (2017) 114101.
181. B. Cai, G. Chen, C. Yuan and H. Jian-Jun, *Chin. Phys. C* **46** (2022) 084104.
182. G. Racah, *Phys. Rev.* **63** (1943) 367.
183. J. J. Ressler *et al.*, *Phys. Rev. C* **69** (2004) 034317.
184. I. Talmi, *Nucl. Phys. A* **172** (1971) 1.
185. C. Qi, *Phys. Rev. C* **81** (2010) 034318.
186. A. Escuderos and L. Zamick, *Phys. Rev. C* **73** (2006) 044302.
187. C. Qi, Z. X. Xu and R. J. Liotta, *Nucl. Phys. A* **884–885** (2012) 21.
188. C. Qi, *Phys. Rev. C* **83** (2011) 014307.
189. Y. Qian and C. Qi, *Phys. Rev. C* **98** (2018) 061303(R).
190. C. Qi, *Phys. Lett. B* **773** (2017) 616.
191. H. Mach *et al.*, *Phys. Rev. C* **95** (2017) 014313.
192. B. Das *et al.*, *Phys. Rev. C* **105** (2022) L031304.
193. R. M. Perez-Vidal *et al.*, *Phys. Rev. Lett.* **129** (2022) 112501.
194. A. Jungclaus *et al.*, *Phys. Lett. B* **695** (2011) 110.
195. I. O. Morales, P. Van Isacker and I. Talmi, *Phys. Lett. B* **703** (2011) 606.
196. T. Bäck, C. Qi, B. Cederwall, R. Liotta, F. Ghazi Moradi, A. Johnson, R. Wyss and R. Wadsworth, *Phys. Rev. C* **87** (2013) 031306(R).
197. C. Qi and Z. X. Xu, *Phys. Rev. C* **86** (2012) 044323.
198. A. Bohr and B. R. Mottelson, *Nuclear Structure Volume II: Nuclear Deformations* (World Scientific, 1975).

199. J. Cederkäll *et al.*, *Phys. Rev. Lett.* **98** (2007) 172501.
200. C. Vaman *et al.*, *Phys. Rev. Lett.* **99** (2007) 162501.
201. A. Ekstrom *et al.*, *Phys. Rev. Lett.* **101** (2008) 012502.
202. P. Doornenbal *et al.*, *Phys. Rev. C* **78** (2008) 031303(R).
203. C. Qi, *Phys. Rev. C* **94** (2016) 034310.
204. T. Bäck *et al.*, *Phys. Rev. C* **84** (2011) 041306(R).
205. M. Doncel *et al.*, *Phys. Rev. C* **91** (2015) 061304(R).
206. B. Cederwall *et al.*, *Nature* **469** (2011) 68.
207. C. Qi, J. Blomqvist, T. Bäck, B. Cederwall, A. Johnson, R. J. Liotta and R. Wyss, *Phys. Rev. C* **84** (2011) 021301(R).
208. Y. M. Zhao and A. Arima, *Phys. Rep.* **545** (2014) 1.
209. J. Chen, *Nucl. Phys. A* **626** (1997) 686.
210. Y. M. Zhao, N. Yoshinaga, S. Yamaji, J. Q. Chen and A. Arima, *Phys. Rev. C* **62** (2000) 014304.
211. B. C. He, L. Li, Y. A. Luo, Y. Zhang, F. Pan and J. P. Draayer, *Phys. Rev. C* **102** (2020) 024304.
212. M. A. Caprio, F. Q. Luo, K. Cai, V. Hellemans and C. Constantinou, *Phys. Rev. C* **85** (2012) 034324.
213. Y.-Y. Cheng, H. Wang, J.-J. Shen, X.-R. Zhou, Y.-M. Zhao and A. Arima, *Phys. Rev. C* **100** (2019) 024321.
214. Y. Lei, Z. Y. Xu, Y. M. Zhao and A. Arima, *Phys. Rev. C* **82** (2010) 034303.
215. H. Jiang, G. J. Fu, Y. M. Zhao and A. Arima, *Phys. Rev. C* **84** (2011) 034302.
216. Y. Y. Cheng, C. Qi, Y. M. Zhao and A. Arima, *Phys. Rev. C* **94** (2016) 024321.
217. H. Jiang, Y. Lei, C. Qi, R. Liotta, R. Wyss and Y. M. Zhao, *Phys. Rev. C* **89** (2014) 014320.
218. L. Y. Jia, H. Zhang and Y. M. Zhao, *Phys. Rev. C* **75** (2007) 034307.
219. Z. Y. Xu, Y. Lei, Y. M. Zhao, S. W. Xu, Y. X. Xie and A. Arima, *Phys. Rev. C* **79** (2009) 054315.
220. Y. Y. Cheng, Y. Lei, Y. M. Zhao and A. Arima, *Phys. Rev. C* **92** (2015) 064320.
221. Y. Lei and Z. Y. Xu, *Phys. Rev. C* **92** (2015) 014317.
222. Y. Lei, H. Jiang and S. Pittel, *Phys. Rev. C* **92** (2015) 024321.
223. Y. Lei, *Phys. Rev. C* **93** (2016) 024319.
224. Y. Lei, H. Jiang and S. Pittel, *Phys. Rev. C* **102** (2020) 024310.
225. G. J. Fu, Y. Lei, Y. M. Zhao, S. Pittel and A. Arima, *Phys. Rev. C* **87** (2013) 044310.
226. G. J. Fu, J. J. Shen, Y. M. Zhao and A. Arima, *Phys. Rev. C* **87** (2013) 044312.
227. G. J. Fu, Y. M. Zhao and A. Arima, *Phys. Rev. C* **90** (2014) 054333.
228. G. J. Fu, Y. M. Zhao and A. Arima, *Phys. Rev. C* **91** (2015) 054318.
229. G. J. Fu, Y. M. Zhao and A. Arima, *Phys. Rev. C* **91** (2015) 054322.
230. G. J. Fu, Y. M. Zhao and A. Arima, *Phys. Rev. C* **97** (2018) 024337.
231. G. J. Fu, Y. Y. Cheng, Y. M. Zhao and A. Arima, *Phys. Rev. C* **94** (2016) 024336.
232. G. J. Fu and C. W. Johnson, *Phys. Lett. B* **809** (2020) 135705.
233. G. J. Fu, C. W. Johnson, P. Van Isacker and Z. Ren, *Phys. Rev. C* **103** (2021) L021302.
234. T. Otsuka, M. Honma and T. Mizusaki, *Phys. Rev. Lett.* **81** (1998) 1588.
235. N. Shimizu, Y. Utsuno, T. Mizusaki, T. Otsuka, T. Abe and M. Honma, *Phys. Rev. C* **82** (2010) 061305(R).
236. N. Shimizu, Y. Utsuno, T. Mizusaki, M. Honma, Y. Tsunoda and T. Otsuka, *Phys. Rev. C* **85** (2012) 054301.
237. N. Shimizu, T. Abe, Y. Tsunoda, Y. Utsuno, T. Yoshida, T. Mizusaki, M. Honma and T. Otsuka, *Prog. Theor. Exp. Phys.* **2012** (2012) 1.

238. T. Abe, P. Maris, T. Otsuka, N. Shimizu, Y. Utsuno and J. P. Vary, *Phys. Rev. C* **86** (2012) 054301.
239. N. Shimizu, T. Abe, M. Honma, T. Otsuka, T. Togashi, Y. Tsunoda, Y. Utsuno and T. Yoshida, *Phys. Scr.* **92** (2017) 063001.
240. Y. Tsunoda, T. Otsuka, N. Shimizu, M. Honma and Y. Utsuno, *Phys. Rev. C* **89** (2014) 031301(R).
241. S. Suchyta *et al.*, *Phys. Rev. C* **89** (2014) 021301(R).
242. C. J. Chiara *et al.*, *Phys. Rev. C* **91** (2015) 044309.
243. F. Flavigny *et al.*, *Phys. Rev. C* **91** (2015) 034310.
244. J. M. Régis *et al.*, *Phys. Rev. C* **95** (2017) 054319.
245. T. Togashi, Y. Tsunoda, T. Otsuka and N. Shimizu, *Phys. Rev. Lett.* **117** (2016) 172502.
246. B. A. Marsh *et al.*, *Nat. Phys.* **14** (2018) 1163.
247. S. Sels *et al.*, *Phys. Rev. C* **99** (2019) 044306.
248. R. Broda *et al.*, *Phys. Rev. C* **86** (2012) 064312.
249. F. Recchia *et al.*, *Phys. Rev. C* **88** (2013) 041302(R).
250. R. Roth and P. Navratil, *Phys. Rev. Lett.* **99** (2007) 092501.
251. R. Roth, *Phys. Rev. C* **79** (2009) 064324.
252. L. F. Jiao, Z. H. Sun, Z. X. Xu, F. R. Xu and C. Qi, *Phys. Rev. C* **90** (2014) 024306.
253. L. Y. Jia, *J. Phys. G Nucl. Part. Phys.* **42** (2015) 115105.
254. L. Y. Jia, *Phys. Rev. C* **96** (2017) 034313.
255. C. Qi, L. Y. Jia and G. J. Fu, *Phys. Rev. C* **94** (2016) 014312.
256. N. Michel, W. Nazarewicz, M. Płoszajczak and T. Vertse, *J. Phys. G Nucl. Part. Phys.* **36** (2009) 013101.
257. M. Nicolas and P. Marek, *Gamow Shell Model.* (Springer, Cham, 2021).
258. Y. Sun, *Proc. Sci.* **085** (2007) 1.
259. Z. Sun, Q. Wu and F. Xu, *Chin. Sci. Bull.* **61** (2016) 2793.
260. J. Chen *et al.*, *Phys. Rev. C* **103** (2021) L031302.
261. Z. H. Yang *et al.*, *Phys. Rev. Lett.* **126** (2021) 082501.
262. Y. Sun, Y. C. Yang, H. Jin, K. Kaneko and S. Tazaki, *Phys. Rev. C* **85** (2012) 054307.
263. G. Dong, X. Wang, F. Xu and S. Yu, *Chin. Sci. Bull.* **59** (2014) 3847.
264. G. X. Dong, X. B. Wang, H. L. Liu and F. R. Xu, *Phys. Rev. C* **88** (2013) 024328.
265. Q.-J. Zhi and Z.-Z. Ren, *Phys. Lett. B* **638** (2006) 166.
266. L.-J. Wang, Y. Sun, T. Mizusaki, M. Oi and S. K. Ghorui, *Phys. Rev. C* **93** (2016) 034322.
267. X.-Y. Wu, S. K. Ghorui, L.-J. Wang, K. Kaneko and Y. Sun, *Nucl. Phys. A* **957** (2017) 208.
268. X. Y. Wu, S. K. Ghorui, L. J. Wang, Y. Sun, M. Guidry and P. M. Walker, *Phys. Rev. C* **95** (2017) 064314.
269. T. Berggren, *Nucl. Phys. A* **109** (1968) 265.
270. R. Id Betan, R. J. Liotta, N. Sandulescu and T. Vertse, *Phys. Rev. Lett.* **89** (2002) 042501.
271. N. Michel, W. Nazarewicz, M. Ploszajczak and K. Bennaceur, *Phys. Rev. Lett.* **89** (2002) 042502.
272. G. X. Dong, N. Michel, K. Fossez, M. Płoszajczak, Y. Jaganathen and R. M. I. Betan, *J. Phys. G Nucl. Part. Phys.* **44** (2017) 045201.
273. N. Michel, J. G. Li, F. R. Xu and W. Zuo, *Phys. Rev. C* **101** (2020) 031301(R).
274. R. B. Wiringa, V. G. Stoks and R. Schiavilla, *Phys. Rev. C* **51** (1995) 38.
275. V. G. Stoks, R. A. Klomp, C. P. Terheggen and J. J. de Swart, *Phys. Rev. C Nucl. Phys.* **49** (1994) 2950.
276. R. Machleidt, *Phys. Rev. C* **63** (2001) 024001.
277. R. Machleidt, F. Sammarruca and Y. Song, *Phys. Rev. C* **53** (1996) R1483.

278. U. van Kolck, *Phys. Rev. C* **49** (1994) 2932.
279. C. Ordóñez, L. Ray and U. van Kolck, *Phys. Rev. Lett.* **72** (1994) 1982.
280. R. Machleidt and D. R. Entem, *Phys. Rep.* **503** (2011) 1.
281. S. Bogner, T. T. S. Kuo, L. Coraggio, A. Covello and N. Itaco, *Phys. Rev. C* **65** (2002) 051301.
282. S. K. Bogner, T. T. S. Kuo and A. Schwenk, *Phys. Rep.* **386** (2003) 1.
283. S. Bonger, T. T. S. Kuo and L. Coraggio, *Nucl. Phys. A* **684** (2001) 432.
284. S. K. Bogner, R. J. Furnstahl and R. J. Perry, *Phys. Rev. C* **75** (2007) 061001.
285. R. Roth, H. Hergert, P. Papakonstantinou, T. Neff and H. Feldmeier, *Phys. Rev. C* **72** (2005) 034002.
286. P. Navrátil, S. Quaglioni, I. Stetcu and B. R. Barrett, *J. Phys. G Nucl. Part. Phys.* **36** (2009) 083101.
287. F. Coester, *Nucl. Phys.* **7** (1958) 421.
288. G. Hagen, M. Hjorth-Jensen, G. R. Jansen, R. Machleidt and T. Papenbrock, *Phys. Rev. Lett.* **108** (2012) 242501.
289. H. Hergert, S. K. Bogner, T. D. Morris, A. Schwenk and K. Tsukiyama, *Phys. Rep.* **621** (2016) 165.
290. A. Carbone, A. Cipollone, C. Barbieri, A. Rios and A. Polls, *Phys. Rev. C* **88** (2013) 054326.
291. Z. Sun, Q. Wu and F. Xu, *Sci. China Phys. Mech.* **59** (2016) 692013.
292. B. S. Hu, Q. Wu and F. R. Xu, *Chin. Phys. C* **41** (2017) 104101.
293. B. S. Hu, F. R. Xu, Z. H. Sun, J. P. Vary and T. Li, *Phys. Rev. C* **94** (2016) 014303.
294. L. Coraggio, N. Itaco, A. Covello, A. Gargano and T. T. S. Kuo, *Phys. Rev. C* **68** (2003) 034320.
295. B. S. Hu, F. R. Xu, Q. Wu, Y. Z. Ma and Z. H. Sun, *Phys. Rev. C* **95** (2017) 034321.
296. Y. N. Zhang, S. K. Bogner and R. J. Furnstahl, *Phys. Rev. C* **98** (2018) 064306.
297. K. T. R. Davies and R. J. McCarthy, *Phys. Rev. C* **4** (1971) 81.
298. X.-B. Wang, G.-X. Dong, H.-L. Wang, C.-X. Yuan, Y.-J. Chen and Y. Tu, *Chin. Phys. C* **42** (2018) 114103.
299. L. Liu, T. Otsuka, N. Shimizu, Y. Utsuno and R. Roth, *Phys. Rev. C* **86** (2012) 014302.
300. T. Abe, P. Maris, T. Otsuka, N. Shimizu, Y. Utsuno and J. P. Vary, *Phys. Rev. C* **104** (2021) 054315.
301. G. Papadimitriou, J. Rotureau, N. Michel, M. Płoszajczak and B. R. Barrett, *Phys. Rev. C* **88** (2013) 044318.
302. K. Fossez, J. Rotureau, N. Michel and M. Płoszajczak, *Phys. Rev. Lett.* **119** 032501.
303. J. G. Li, N. Michel, B. S. Hu, W. Zuo and F. R. Xu, *Phys. Rev. C* **100** (2019) 054313.
304. B. S. Hu, Q. Wu, Z. H. Sun and F. R. Xu, *Phys. Rev. C* **99** (2019) 061302(R).
305. Q. Yuan, S. Q. Fan, B. S. Hu, J. G. Li, S. Zhang, S. M. Wang, Z. H. Sun, Y. Z. Ma and F. R. Xu, *Phys. Rev. C* **105** (2022) L061303.
306. J. D. Holt, J. Menéndez, J. Simonis and A. Schwenk, *Phys. Rev. C* **90** (2014) 024312.
307. H. Hergert, S. K. Bogner, T. D. Morris, S. Binder, A. Calci, J. Langhammer and R. Roth, *Phys. Rev. C* **90** (2014) 041302(R).
308. G. Hagen, M. Hjorth-Jensen, G. R. Jansen, R. Machleidt and T. Papenbrock, *Phys. Rev. Lett.* **109** (2012) 032502.
309. A. F. Lisetskiy, B. R. Barrett, M. K. G. Kruse, P. Navratil, I. Stetcu and J. P. Vary, *Phys. Rev. C* **78** (2008) 044302.
310. E. Dikmen, A. F. Lisetskiy, B. R. Barrett, P. Maris, A. M. Shirokov and J. P. Vary, *Phys. Rev. C* **91** (2008) 064301.
311. G. R. Jansen, J. Engel, G. Hagen, P. Navratil and A. Signoracci, *Phys. Rev. Lett.* **113** (2014) 142502.

312. S. K. Bogner, H. Hergert, J. D. Holt, A. Schwenk, S. Binder, A. Calci, J. Langhammer and R. Roth, *Phys. Rev. Lett.* **113** (2014) 142501.
313. S. C. Pieper and R. B. Wiringa, *Annu. Rev. Nucl. Part. Sci.* **51** (2001) 53.
314. P. Navratil, V. G. Gueorguiev, J. P. Vary, W. E. Ormand and A. Nogga, *Phys. Rev. Lett.* **99** (2007) 042501.
315. P. Maris, J. P. Vary, P. Navratil, W. E. Ormand, H. Nam and D. J. Dean, *Phys. Rev. Lett.* **106** (2011) 202502.
316. A. P. Zuker, *Phys. Rev. Lett.* **90** (2003) 042502.
317. T. T. S. Kuo and G. E. Brown, *Nucl. Phys.* **85** (1966) 40.
318. T. Hamada and I. D. Johnston, *Nucl. Phys.* **34** (1962) 382.
319. X.-B. Wang, Y.-H. Meng, Y. Tu and G.-X. Dong, *Chin. Phys. C* **43** (2019) 124106.
320. C. Qi and F. R. Xu, *Nucl. Phys. A* **800** (2008) 47.
321. C. Qi and F. R. Xu, *Nucl. Phys. A* **814** (2008) 48.
322. M. A. Bentley *et al.*, *Phys. Rev. C* **73** (2006) 024304.
323. D. R. Entem and R. Machleidt, *Phys. Rev. C* **68** (2003) 041001(R).
324. P. Maris, J. P. Vary and P. Navrátil, *Phys. Rev. C* **87** (2013) 014327.
325. T. T. S. Kuo and E. Osnes, *Folded-Diagram Theory of the Effective Interaction in Nuclei, Atoms and Molecules* (Springer, Berlin, Heidelberg, 1990).
326. L. Coraggio, A. Covello, A. Gargano, N. Itaco and T. T. S. Kuo, *Ann. Phys.* **327** (2012) 2125.
327. T. Fukui, L. De Angelis, Y. Z. Ma, L. Coraggio, A. Gargano, N. Itaco and F. R. Xu, *Phys. Rev. C* **98** (2018) 044305.
328. Y. Z. Ma, L. Coraggio, L. De Angelis, T. Fukui, A. Gargano, N. Itaco and F. R. Xu, *Phys. Rev. C* **100** (2019) 034324.
329. L. Coraggio, G. De Gregorio, A. Gargano, N. Itaco, T. Fukui, Y. Z. Ma and F. R. Xu, *Phys. Rev. C* **102** (2020) 054326.
330. X. Wang, G. Dong, Q. Li, C. Shen and S. Yu, *Sci. China Phys. Mech.* **59** (2016) 692011.
331. M. Duer *et al.*, *Nature* **606** (2022) 678.
332. https://www.phy.pku.edu.cn/info/1031/8116.htm.
333. K. Tsukiyama, M. Hjorth-Jensen and G. Hagen, *Phys. Rev. C* **80** (2009) 051301(R).
334. Z. H. Sun, Q. Wu, Z. H. Zhao, B. S. Hu, S. J. Dai and F. R. Xu, *Phys. Lett. B* **769** (2017) 227.
335. K. Takayanagi, *Nucl. Phys. A* **852** (2011) 61.
336. J. G. Li, B. S. Hu, Q. Wu, Y. Gao, S. J. Dai and F. R. Xu, *Phys. Rev. C* **102** (2020) 034302.
337. B. S. Hu, Q. Wu, J. G. Li, Y. Z. Ma, Z. H. Sun, N. Michel and F. R. Xu, *Phys. Lett. B* **802** (2020) 135206.
338. A. Ekstrom *et al.*, *Phys. Rev. Lett.* **110** (2013) 192502.
339. N. Michel, J. G. Li, F. R. Xu and W. Zuo, *Phys. Rev. C* **100** (2019) 064303.
340. J. G. Li, N. Michel, W. Zuo and F. R. Xu, *Phys. Rev. C* **103** (2021) 034305.
341. Y. Z. Ma, F. R. Xu, L. Coraggio, B. S. Hu, J. G. Li, T. Fukui, L. De Angelis, N. Itaco and A. Gargano, *Phys. Lett. B* **802** (2020) 135257.
342. R. Roth, S. Binder, K. Vobig, A. Calci, J. Langhammer and P. Navrátil, *Phys. Rev. Lett.* **109** (2012) 052501.
343. Y. Z. Ma, F. R. Xu, N. Michel, S. Zhang, J. G. Li, B. S. Hu, L. Coraggio, N. Itaco and A. Gargano, *Phys. Lett. B* **808** (2020) 135673.
344. Y. Kondo *et al.*, *Phys. Rev. Lett.* **116** (2016) 102503.
345. Y. Lei, Y. Lu and Y. M. Zhao, *Chinese Phys. C* **45** (2021) 054103
346. C. Ma, X. Yin and Y.M. Zhao, *Phys, Rev. C* **108** (2023) 034308.
347. J. Z. Han, S. Xu, A. Jalili and H. K. Wang, *Nucl. Sci. Tech.* **34** (2023) 85.

348. L. Zhou, S.-M. Wang, D.-Q. Fang and Y.-G. Ma, *Nucl. Sci. Tech.* **33** (2022) 105.
349. Z.-X. Fang, M. Yu, Y.-G. Huang, J.-B. Chen, J. Su and L. Zhu, *Nucl. Sci. Tech.* **32** (2021) 72.
350. Q.-F. Song, L. Zhu, H. Guo and J. Su, *Nucl. Sci. Tech.* **34** (2023) 32.
351. Y.-F. Gao, B.-S. Cai and C.-X. Yuan, *Nucl. Sci. Tech.* **34** (2023) 9.